\documentclass[a4,usenatbib]{mnras}
\usepackage{amsmath}
\usepackage{amssymb} 
\usepackage{multirow}
\usepackage{textcomp}
\usepackage{microtype}
\usepackage{subcaption}
\usepackage[dvipsnames]{xcolor}
\usepackage{ulem}
\usepackage{graphicx}
\usepackage{float}
\usepackage{mathptmx}
\usepackage{tabularx}
\usepackage{caption}

\newcommand{\email}[1]{\mbox{\href{mailto:#1}{#1}}}

\newcommand{\Mc}{\mathcal{M}}
\newcommand{\Msol}{M_{\odot}}

\newcommand{\Adet}{A_{\mathrm{det}}}

\newcommand{\Atrial}{A_{\mathrm{trial}}}

\newcommand{\sigdet}{\sigma_{\mathrm{det}}}

\def\ltsima{$\; \buildrel < \over \sim \;$}
\def\simlt{\lower.5ex\hbox{\ltsima}}
\def\gtsima{$\; \buildrel > \over \sim \;$}
\def\simgt{\lower.5ex\hbox{\gtsima}}

\def\aj{Astronomical Journal}                 
\def\apj{Astrophysical Journal}                
\def\apjl{Astrophysical Journal}             

\def\apjs{ApJS}              
\def\mnras{MNRAS}            
\def\prd{Phys.~Rev.~D}       
\def\araa{ARA\&A}             
\def\nat{Nature}              
\def\aap{A\&A}                

\title[Astrophysical observables from PTAs]{Constraining astrophysical observables of Galaxy and Supermassive Black Hole Binary Mergers using Pulsar Timing Arrays}
\author[Chen et al.]{
Siyuan Chen,$^{1,2,3,4}$\thanks{E-mail: \email{schen@star.sr.bham.ac.uk; siyuan.chen@cnrs-orleans.fr}}
Alberto Sesana,$^{1,5}$
and Christopher J. Conselice$^6$
\\
$^1$Institute of Gravitational Wave Astronomy and School of Physics \& Astronomy, University of Birmingham, Birmingham, B15 2TT, UK \\
$^2$Station de Radioastronomie de Nan\c{c}ay, Observatoire de Paris, PSL University, CNRS, Universit\'{e} d'Orl\'{e}ans, 18330 Nan\c{c}ay, France and \\
$^3$FEMTO-ST Institut de recherche, Department of Time and Frequency, UBFC and CNRS, ENSMM, 25030 Besan\c{c}on, France and \\
$^4$Laboratoire de Physique et Chimie de l'Environnement et de l'Espace, LPC2E UMR7328, Universit\'{e} d'Orl\'{e}ans, CNRS, 45071 Orl\'{e}ans, France \\
$^5$Dipartimento di Fisica "G. Occhialini", Universita` degli Studi di Milano-Bicocca, Piazza della Scienza 3, 20126 Milano, Italy \\
$^6$Centre for Astronomy and Particle Theory, University of Nottingham, Nottingham, NG7 2RD, UK
}

\date{Accepted \dots Received \dots; in original form \dots}

\pubyear{2019}

\begin{document}
\label{firstpage}
\pagerange{\pageref{firstpage}--\pageref{lastpage}}
\maketitle

\begin{abstract}

We present an analytic model to describe the supermassive black hole binary (SMBHB) merger rate in the Universe with astrophysical observables: galaxy stellar mass function, pair fraction, merger timescale and black hole - host galaxy relations. We construct observational priors and compute the allowed range of the characteristic spectrum $h_c$ of the gravitational wave background (GWB) to be $10^{-16}<h_c<10^{-15}$ at a frequency of $f=1/{\rm yr}$. We exploit our parametrization to tackle the problem of astrophysical inference from Pulsar Timing Array (PTA) observations. We simulate a series of upper limits and detections and use a nested sampling algorithm to explore the parameter space. Corroborating previous results, we find that the current PTA non-detection does not place significant constraints on any observables; however, either future upper limits or detections will significantly enhance our knowledge of the SMBHB population. If a GWB is not detected at a level of $h_c(f=1/{\rm yr})=10^{-17}$, our current understanding of galaxy and SMBHB mergers is disfavoured at a $5\sigma$ level, indicating a combination of severe binary stalling, over-estimating of the SMBH -- host galaxy relations, and extreme dynamical properties of merging SMBHBs. Conversely, future detections of a Square Kilometre Array (SKA)-type array will allow to constrain the normalization of the SMBHB merger rate in the Universe, the time between galaxy pairing and SMBHB merging, the normalization of the SMBH -- host galaxy relations and the dynamical binary properties, including their eccentricity and density of stellar environment.

\end{abstract}


\section{Introduction}
\label{sec:Introduction}

It is well established that supermassive black holes (SMBHs) reside at the centre of massive galaxies \citep[see e.g.][]{2019ApJ...875L...1E}, and that their masses correlate with several properties of the hosts \citep[see][and references therein]{2013ARA&A..51..511K}. In the hierarchical clustering scenario of structure formation \citep{1978MNRAS.183..341W}, the SMBHs hosted in merging galaxies sink to the centre of the merger remnant because of dynamical friction, eventually forming a bound SMBH binary (SMBHB) at parsec scales \citep{1980Natur.287..307B}. The binary subsequently hardens because of (hydro)dynamical interaction with the dense background of stars and gas \citep[see][for a review]{2012AdAst2012E...3D}, until gravitational wave (GW) emission takes over at sub-parsec separations, leading to the final coalescence of the system \citep{PetersMathews:1963}. Upon coalescence, the frequency emitted by a SMBHB of mass $M$ at the last stable orbit is $f_{\rm LSO}\approx 4/M_9$ $\mu$Hz, where $M_9=M/10^9$M$_\odot$, making inspiralling SMBHBs the loudest sources in the Universe of sub-$\mu$Hz GWs. This frequency regime is accessible via precise timing of millisecond pulsars (MSPs). The most massive systems closest to Earth might be powerful enough to be detected as individual deterministic sources at nano-Hz frequencies \citep{SesanaVecchioVolonteri:2009,2017NatAs...1..886M}. There are, however, $\approx 10^{10}$ massive galaxies in the Universe. If each of them experienced one or more major merger in its lifetime and if the resulting SMBHB emits GWs at nano-Hz frequencies for $\approx 1$ Myr \citep{2017NatAs...1..886M}, then there are, at any time $\approx 1$ million SMBHBs emitting GWs in the frequency band probed by pulsar observations, resulting in an unresolved stochastic GW background \citep[GWB, see e.g.][]{1995ApJ...446..543R,2003ApJ...583..616J,SesanaVecchioColacino:2008,RaviEtAl:2012}. 

A GWB affects the time of arrivals (TOAs) of radio pulses emitted by an ensemble of MSPs in a characteristic and correlated fashion \citep{HellingsDowns:1983}. Pulsar timing arrays \citep[PTA][]{1990ApJ...361..300F} search for GWs using this Hellings \& Downs correlation. Although a GWB has not been detected yet, the three currently leading PTAs -- the European Pulsar Timing Array \citep[EPTA][]{2016MNRAS.458.3341D}, the North American Nanohertz Observatory for Gravitational Waves \citep[NANOGrav][]{2018ApJS..235...37A}, and the Parkes Pulsar Timing Array \citep[PPTA][]{2016MNRAS.455.1751R}  -- already produced stringent upper limits \citep{2015Sci...349.1522S,2015MNRAS.453.2576L,2018ApJS..235...37A}. The three PTAs work together under the aegis of the International Pulsar Timing Array \citep[IPTA][]{2016IPTA}, with the goal of building a larger TOA dataset to improve sensitivity. With the contribution of emerging PTAs in India, China and South Africa, a detection is expected within the next decade \citep{2015MNRAS.451.2417R,2016ApJ...819L...6T,2017MNRAS.471.4508K}.

Besides detecting the low frequency GWB, the final goal of PTAs is to extract useful astrophysical information from their data to address the 'inverse problem'. Since the GWB shape and normalization depends on the statistical properties of the SMBHB population and on the dynamics of individual binaries in their late inspiral \citep[see][for a general discussion of the relevant processes]{2013CQGra..30v4014S}, stringent upper limits, and eventually a detection, will allow to gain invaluable insights in the underlying relevant physical processes. In fact, the most stringent upper limits to date have been already used to place tentative constrains on the population of SMBHBs \citep{2016ApJ...826...11S,2017PhRvL.118r1102T,2018NatCo...9..573M}. Generally speaking, the normalization of the GWB depends on the cosmic SMBHB merger rate, and its shape on the typical SMBHB eccentricity and on the effectiveness of energy and angular momentum loss to the dense environment of gas and stars surrounding the binary. \cite{2016ApJ...821...13A} investigated the implications of the NANOGrav nine-year upper limits on several astrophysical ingredients defining the underlying SMBHB population model. They found that the data prefer low SMBHB merger rate normalization, light SMBHs for a given galaxy mass (i.e. a low normalization of the SMBH -- host galaxy scaling relation), eccentric binaries and dense stellar environment. Their analysis should be taken as a proof of concept, since each parameter was investigated separately, keeping all the other fixed. In a subsequent extension of the work, \cite{2016ApJ...826...11S} have shown that it is possible to use that limit to constrain simultaneously the parameters describing the $M_{BH} - M_{\rm bulge}$ and the typical SMBHB merger timescale, but still keeping other relevant parameters within a narrow prior range and assuming a GWB characteristic strain, $h_c$, described by a $f^{-2/3}$ power-law, appropriate for circular, GW driven binaries (thus not considering the detailed SMBHB dynamics). \cite{2017PhRvL.118r1102T} focused on the determination of the parameters driving the dynamical evolution of individual binaries, showing with detailed GWB simulations interpolated by means of Gaussian processes, that eccentricity and density of the stellar environment can be constrained for a specific choice of the SMBHB merger rate. Finally, a more sophisticated astrophysical inference investigation was conducted in \cite{2018ApJS..235...37A}, including model selection between different SMBHB population models from the literature, and constrains on the SMBHB eccentricity and environment density for different $M_{BH} - M_{\rm bulge}$ scaling relations. 

In \cite{2016MNRAS.455L..72M} we started a long-term project of creating a general framework for astrophysical inference from PTA data. In \cite{2017MNRAS.470.1738C} we presented a fast and flexible way to compute the stochastic GWB shape for a general parametrization of the SMBHB merger rate and the relevant properties defining the SMBHB dynamics. We demonstrated the versatility of our model on synthetic simulations in \cite{2017MNRAS.468..404C} and eventually applied it to the most stringent PTA limit to date in \cite{2018NatCo...9..573M}. This latter study, in particular, was instrumental in demonstrating that PTA upper limits are not in tension with our current understanding of the cosmic galaxy and SMBH build-up.

Although previous work has focussed on particular bits of physics contributing to the amplitude and shape of the GWB spectrum, we combine here all ingredients of the SMBHB merger rate into one overall model to simultaneously contrain the entire parameter space without keeping certain aspects fixed. We make in this paper an important step towards this goal by re-writing our model of the SMBHB merger rate as a parametric function of astrophysical observables rather then considering a purely phenomenological form. In fact, as shown in \cite{Sesana:2013}, the SMBHB merger rate can be derived from the galaxy stellar mass function (GSMF), the galaxy pair fraction, the SMBHB merger timescale and the scaling relation connecting SMBHs and their hosts. By expressing the SMBHB merger rate as a function of simple analytical parametrizations of these ingredients -- constrained by independent observations --, we build a GWB model that allows to use PTA observations to constrain a number of extremely relevant astrophysical observables.

The paper is organized as follows. Section \ref{sec:Model} summarizes the model to compute the characteristic strain of the GWB and highlights the changes introduced in this paper. Section \ref{sec:Galaxy} derives the parametric formulation of the SMBHB merger rate as a function of all the relevant observational parameters describing the properties of merging galaxies and their SMBHs. In section \ref{sec:Observations} we briefly described how the PTA signal is constructed, the simulation set-up of the different investigated PTAs, and the Bayesian method used in the analysis. Observationally motivated prior distributions for all model parameters are given in section \ref{sec:Prior}. Detailed results are presented and discussed in section \ref{sec:Results} and in section \ref{sec:Conclusions} we summarize our main findings and outline future research directions.

Unless stated otherwise, we use the standard Lambda CDM as our cosmology with the Hubble parameter $h_0 = 0.7$ and constant $H_0 = 70$ km\,Mpc$^{-1}$s$^{-1}$ and energy density ratios $\Omega_M = 0.3$, $\Omega_k = 0$ and $\Omega_\Lambda = 0.7$.

\section{GWB strain model}
\label{sec:Model}
Deviations from an unperturbed spacetime arising from an incoherent superposition of GW sources (i.e. a stochastic GWB) are costumarily described in terms of characteristic strain $h_c$, which represents the amplitude of the perturbation per unit logarithmic frequency interval. We compute $h_c$ following \cite{2017MNRAS.470.1738C} (paper I hereafter). The model allows for the quick calculation of $h_c$ given the chirp mass $\cal M$, redshift $z$ and eccentricity $e$ at decoupling of any individual binary. The total strain of the GWB can then be computed by integrating over the population $\frac{d^2n}{dzd{\cal M}}$, giving the main equation of paper I:
\begin{equation}
\begin{split}
  h_c^2(f) = & \int dz \int d{\cal M} \frac{d^2n}{dzd{\cal M}} h_{c,{\rm fit}}^2\Big(f\frac{f_{p,0}}{f_{p,t}}\Big) \\ & \times \Big(\frac{f_{p,t}}{f_{p,0}}\Big)^{-4/3} \Big(\frac{\mathcal{M}}{\mathcal{M}_0}\Big)^{5/3} \Big(\frac{1+z}{1+z_0}\Big)^{-1/3}
  \label{eqn:hsquared}
\end{split}
\end{equation}
where $h_{c,{\rm fit}}$ is the strain of a reference binary with chirp mass $\mathcal{M}_0$, redshift $z_0$ and eccentricity $e_0$ and $f_{p,0}$ is the peak frequency of the spectrum (see equation 13 in paper I and relative discussion therein). The main concept of equation (\ref{eqn:hsquared}) is to use the self-similarity of the characteristic strain of a purely GW emission driven binary to go from the reference spectrum $h_{c,{\rm fit}}$ with fixed parameters to the emitted spectrum of a binary $h_c$ with arbitrary parameters via shifts in frequency, chirp mass and redshift.

As in paper I, we assume that the evolution of the binary is driven by hardening in a stellar environment before GW emission takes over at a transition frequency given by (equation 21 in paper I):
\begin{equation}
  f_t = 0.356\, {\rm nHz}\, \left(\frac{1}{F(e)}\frac{\rho_{i,100}}{\sigma_{200}}\zeta_0\right)^{3/10}\mathcal{M}_9^{-2/5},
  \label{eq:fdec}
\end{equation}
where
\begin{equation}
F(e) = \frac{1+(73/24)e^2+(37/96)e^4}{(1-e^2)^{7/2}}
\end{equation}
\citep{PetersMathews:1963}, $\mathcal{M}_9 = \mathcal{M}/(10^9 \Msol)$ is the rescaled chirp mass, $\rho_{i,100}=\rho_i/(100\,\Msol{\rm pc}^{-3})$ is the density of the stellar environment at the influence radius of the SMBHB, $\sigma_{200}=\sigma/(200\,{\rm km\,s}^{-1})$ is the velocity dispersion of stars in the galaxy and $\zeta_0$ is an additional multiplicative factor absorbing all systematic uncertainties in the estimate of $\rho_{i,100}$. In fact, as extensively described in Paper I, the stellar density of the host galaxy bulge follows a Dehnen profile \citep{1993MNRAS.265..250D} with a fiducial inner density slope $\gamma = 1$. This specific profile choice, together with an empirical estimate of scale radius, fixes $\rho_i$ for a given stellar bulge mass. Galaxies can, however, be more/less compact and have steeper/shallower density profiles, thus resulting in $\rho_i$ that can be different by orders of magnitude from this value. We thus capture this possibility by introducing the multiplicative factor $\zeta_0$. If, for example, $\zeta_0$ is measured to be $\approx 10$, this means that massive galaxies have on average higher central densities than implied by a standard Dehnen profile. Note that both $\rho_i$ and $\sigma$ enter in the calculation to the 3/10 power. Although difference in $\rho_i$ can be significant, massive galaxies have generally $250\,$km\,s$^{-1}<\sigma<350\,$km\,s$^{-1}$. We thus keep $\sigma$ constant in our calculation, since we found that such small range does have a negligible impact on the shape of the spectrum. Note however that $\zeta_0$ can be considered as a multiplicative factor absorbing systematics in the determination of $\rho_i$ and $\sigma$.

Finally, the spectrum described by equation (\ref{eqn:hsquared}) is corrected by including an a high frequency drop related to an upper mass limit calculated, at each frequency, via (equation 39 paper I)
\begin{equation}
  N_{\Delta{f}}=\int_{f-\Delta f/2}^{f+\Delta f/2}df \int_{\bar\Mc}^{\infty}d\Mc \int_{0}^{\infty}dz \frac{d^3 N}{df dz d\mathcal{M}} = 1,
  \label{eq:Nf}
\end{equation}
This upper mass limit $\bar\Mc$ takes into account that, particularly at high frequencies, there is less than 1 binary above $\bar\Mc$ contributing to the signal within a frequency bin $\Delta f = 1/T$. Statistically, this means that in a given realization of the universe, there will be either one or zero loud sources contributing to the signal. In the case the source is present, it can be removed from the GWB computation since it will be likely resolvable as an individual deterministic GW source \citep[see discussion in][]{SesanaVecchioColacino:2008}.

In paper I, we used a phenomenological parametric function to describe the SMBHB merger rate $d^2n/(dzd{\cal M})$, and introduced an extra parameter $e_0$ to allow for eccentric binaries at $f_t$. The quantity $d^2n/(dzd{\cal M})$, however, cannot be directly measured from observations. It can be either computed theoretically from galaxy and SMBH formation and evolution models \citep[e.g.][]{SesanaVecchioColacino:2008,RaviEtAl:2012,2017MNRAS.471.4508K} or it can be indirectly inferred from observations of other astrophysical quantities, such as the galaxy mass function, pair fraction, typical merger timescales, and the SMBH -- host galaxy relation. Parametrizing the SMBHB merger rate as a function of astrophysical observables would therefore allow to reverse engineer the outcome of current and future PTA observations to obtain useful constrains on those observables. With this goal in mind, in this paper we expand the model from paper I in two ways:
\begin{enumerate}
\item we introduce an extra parameter $\zeta_0$, see equation \eqref{eq:fdec}, to allow for variations from the fiducial values of the density of the stellar environment;
\item we cast the phenomenological SMBHB merger rate $d^2n/(dzd{\cal M})$ in terms of astrophysical observables, such as galaxy mass function and pair fraction, galaxy - black hole relations, etc., as we detail next in Section \ref{sec:Galaxy}.
\end{enumerate}


\section{Parametric model of the SMBHB merger rate}
\label{sec:Galaxy}

As detailed in \cite{2013CQGra..30v4014S} and \cite{2016MNRAS.463L...6S}, the differential galaxy merger rate per unit redshift, mass and mass ratio, can be written as
\begin{equation}
\frac{d^3 n_G}{dz'dMdq} = \frac{\Phi(M,z)}{M\ln10} \frac{\mathcal{F}(M,z,q)}{\tau(M,z,q)} \frac{dt}{dz}
\label{ng}
\end{equation}
where $\Phi(M,z)$ is the redshift dependent galaxy stellar mass function (GSMF), $\mathcal{F}(M,z,q)$ is the differential pair fraction with respect to the mass ratio $q$ (see equation (\ref{dfdq}) below) and $\tau(M,z,q)$ is the merger timescale. $M$ is the mass of the primary galaxy, $z$ is the redshift of the {\it galaxy pair} and $q$ is the mass ratio between the two galaxies. It is important to note that a pair of galaxies at redshift $z$ will merge at redshift $z'<z$. The timescale $\tau(M,z,q)$ is used to convert the pair fraction of galaxies at $z$ into the galaxy merger rate at $z'<z$ \citep{2017MNRAS.470.3507M}. The merger redshift is obtained by solving for $z'$ the implicit equation
\begin{equation}
\int_{z'}^z \frac{dt}{d \tilde{z}} d \tilde{z} = \tau(M,z,q),
\end{equation}
where, assuming a flat Lambda CDM model,
\begin{equation}
\frac{dt}{d \tilde{z}} = \frac{1}{H_0 (1+\tilde{z}) \sqrt{\Omega_M(1+\tilde{z})^3 + \Omega_k(1+\tilde{z})^2 + \Omega_\Lambda}}.
\end{equation}

\begin{figure}
\centering
\includegraphics[width=8.cm,clip=true,angle=0]{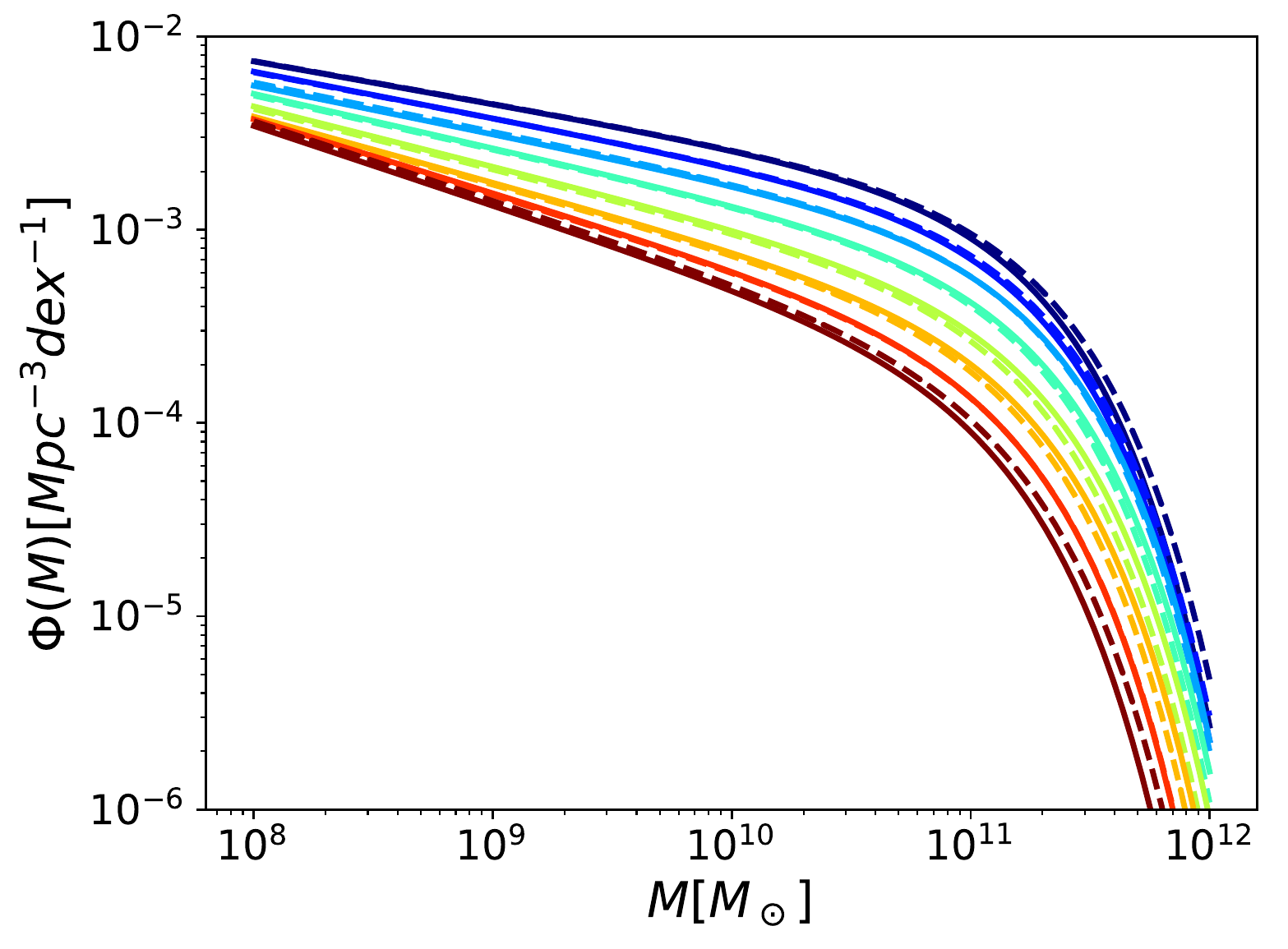}
\caption{Comparison between measured and computed GSMFs for 8 redshift bins in the range $0.4 < z < 3$, where blue represents lower and red higher redshift values. The solid lines represent the original mass functions reported by \protect\cite{2016ApJ...830...83C} and \protect\cite{2015MNRAS.447....2M}, the dashed lines are best fits obtained using equation (\ref{gsmfeq}) with the parametric functions $\Phi_0(z), M_0(z), \alpha(z)$ for the central values of the 8 corresponding redshift bins.}
\label{fig:gsmf}
\end{figure}

The galaxy stellar mass function $\Phi(M,z)$ can be written as a single Schechter function \citep{2016ApJ...830...83C}
\begin{equation}
\Phi(M,z) = \frac{d n_G}{d \log_{10} M} = \ln 10 \ \Phi_0(z) \Big( \frac{M}{M_0(z)} \Big)^{1+\alpha_0(z)} \exp \Big( -\frac{M}{M_0(z)} \Big),
\label{gsmfeq}
\end{equation}
where $\Phi_0(z), \ M_0(z), \ \alpha_0(z)$ are phenomenological functions of redshift of the form \citep{2015MNRAS.447....2M}:
\begin{align}
\log_{10} \Phi_0(z) & = \Phi_0 + \Phi_I z
\\
M_0(z) & = M_0
\\
\alpha_0(z) & = \alpha_0 + \alpha_I z
\end{align}
The 5 parameters $\Phi_0, \Phi_I, M_0, \alpha_0, \alpha_I$ are sufficient to fit the original Schechter functions at any redshift; an example is shown in figure \ref{fig:gsmf}. To simplify the notation, in the following $\Phi_0, \ M_0, \ \alpha_0$ will implicitly denote their corresponding redshift dependent functions $\Phi_0(z), \ M_0(z), \ \alpha_0(z)$.

The differential pair fraction as a function of $q$ is given by
\begin{equation}
\mathcal{F}(M,z,q) = \frac{df_{\rm pair}}{dq} = f_0' \Big( \frac{M}{a M_0} \Big)^{\alpha_f} (1+z)^{\beta_f} q^{\gamma_f}
\label{dfdq}
\end{equation}
where $a M_0 = 10^{11} M_\odot$ is an arbitrary reference mass. Note that, in the literature, pair fractions are usually given as a function of primary galaxy mass and redshift only \citep[e.g.][]{2017MNRAS.470.3507M}, such that
\begin{equation}
  f_{\rm pair} = f_0 \Big( \frac{M}{a M_0} \Big)^{\alpha_f} (1+z)^{\beta_f},
  \label{eq:pf_lit}
\end{equation}
i.e. integrated over the mass ratio of the pairs. The integral of equation (\ref{dfdq}) over $q$ gives 
\begin{equation}
f_{\rm pair} = f_0' \Big( \frac{M}{a M_0} \Big)^{\alpha_f} (1+z)^{\beta_f} \int q^{\gamma_f} dq,
\end{equation}
which becomes equivalent to equation (\ref{eq:pf_lit}) by setting
\begin{equation}
  f_0 = f_0' \int q^{\gamma_f} dq.
  \label{eq:f_f'}
\end{equation}
Equation (\ref{eq:f_f'}) allows to map an observational prior of the form of equation (\ref{eq:pf_lit}) into the four parameters of our model $f_0', \alpha_f, \beta_f, \gamma_f$.

We use an analogue parametrization for the merger timescale:
\begin{equation}
\tau(M,z,q) = \tau_0 \Big( \frac{M}{b M_0} \Big)^{\alpha_\tau} (1+z)^{\beta_\tau} q^{\gamma_\tau}
\label{tau}
\end{equation}
where $b M_0 = 0.4/h_0 \times 10^{11} M_\odot$, and $\tau_0, \alpha_\tau, \beta_\tau, \gamma_\tau$ are four further model parameters. Equation \eqref{tau} has originally been derived to describe the {\it galaxy} merger timescale \citep{2017MNRAS.468..207S}. A further delay is, however, expected between the galaxy merger and the SMBHB final coalescence. In fact, after dynamical friction has merged the two galaxies and has brought the two SMBHs in the nuclear region, the newly formed SMBHB has to harden via energy and angular momentum losses mediated by either stars or gas, before GW emission eventually takes over \citep[see][for a review]{2012AdAst2012E...3D}. Depending on the details of the environment, this process can take up to several Gyrs, and even cause the binary to stall \citep{2015MNRAS.454L..66S,2015ApJ...810...49V,2017MNRAS.464.3131K}. For simplicity, we assume here that this further delay can be re-absorbed in equation (\ref{tau}), which we then use to describe the time elapsed between the observed galaxy pair and the final {\it SMBHB} coalescence.

Substituting equations \eqref{gsmfeq}, \eqref{dfdq} and \eqref{tau} into \eqref{ng} gives
\begin{equation}
\frac{d^3 n_G}{dz'dMdq} = n_\mathrm{eff} \Big( \frac{M}{M_0} \Big)^{\alpha_\mathrm{eff}} e^{-M/M_0} (1+z)^{\beta_\mathrm{eff}} q^{\gamma_\mathrm{eff}} \frac{dt}{dz},
\label{ngq}
\end{equation}
where the effective parameters are
\begin{align}
\label{eq:eff_param}
n_\mathrm{eff} = \frac{\Phi_0 f_0'}{M_0 \tau_0} \frac{b^{\alpha_\tau}}{a^{\alpha_f}} & = \frac{\Phi_0 f_0'}{M_0 \tau_0} \Big( \frac{0.4}{h_0} \Big)^{\alpha_\tau} \Big( \frac{10^{11}}{M_0} \Big)^{\alpha_\tau-\alpha_f}\nonumber
\\
\alpha_\mathrm{eff} & = \alpha_0 + \alpha_f - \alpha_\tau\nonumber
\\
\beta_\mathrm{eff} & = \beta_f - \beta_\tau\nonumber
\\
\gamma_\mathrm{eff} & = \gamma_f - \gamma_\tau
\end{align}

Equation \eqref{ngq} is still a function of the merging galaxy stellar masses, which needs to be translated into SMBH masses. The total mass of a galaxy $M$ can be converted into its bulge mass $M_{\rm bulge}$, using assumptions on the ellipticity of the galaxy: more massive galaxies are typically elliptical and have higher bulge to total stellar mass ratio. We use a phenomenological fitting function \citep{2014MNRAS.443..874B,2016MNRAS.463L...6S} to link the bulge mass to the total stellar mass of a galaxy:

\begin{equation}
\frac{M_{\rm bulge}}{M} = 
\begin{cases}
\frac{\sqrt{6.9}}{(\log M - 10)^{1.5}} \exp \Big( \frac{-3.45}{\log M - 10} \Big) + 0.615 & \text{if $\log M > 10$}\\
0.615 & \text{if $\log M < 10$}.
\label{eq:mstar_mbulge}
\end{cases}
\end{equation}
Note that this fit is appropriate for ellipticals and spheroidals, whereas spiral galaxies usually have smaller bulge to total mass ratio. In \cite{2013CQGra..30v4014S} different scaling relations were used for blue and red galaxy pairs (under the assumption that blue pairs are predominantly spirals and red pairs predominantly elliptical). The result was that the GW signal is completely dominated by red pairs. We have checked on \cite{2013CQGra..30v4014S} data that approximating all galaxies as spheroidals affects the overall signal by less than $0.05$dex. We therefore apply equation \eqref{eq:mstar_mbulge} to all galaxies, independent on their colour or morphology.

We can then apply a scaling relation between the galaxy bulge mass $M_{\rm bulge}$ and black hole mass $M_{\rm BH}$ of the form \citep[see, e.g., ][]{2013ARA&A..51..511K}
\begin{equation}
M_{\rm BH} = {\cal N}\left\{M_* \Big( \frac{M_{\rm bulge}}{10^{11} M_\odot} \Big)^{\alpha_*}, \epsilon\right\},
\label{eqnmbulge}
\end{equation}
where ${\cal N}\{x,y\}$ is a log normal distribution with mean value $x$ and standard deviation $y$, to translate galaxy mass $M$ into black hole mass $M_{\rm BH}$. Note that the galaxy mass ratio $q$ is in general different from the black hole mass ratio $q_{\rm BH} = q^{\alpha_*}$. Finally, the galaxy merger rate $n_G$ \eqref{ngq} can be converted into the SMBHB merger rate $n$:
\begin{equation}
\frac{d^3 n}{dz'dM_{\rm BH}dq_{\rm BH}} = \frac{d^3 n_G}{dz'dMdq} \frac{dM}{dM_{\rm BH}} \frac{dq}{dq_{\rm BH}}.
\end{equation}
Equation \eqref{eqnmbulge} adds three further parameters to the model: $M_*,\alpha_*, \epsilon$.
Lastly, it is convenient to map $M_{\rm BH}, q_{\rm BH}$  into the SMBHB chirp mass $\mathcal{M} = M_{\rm BH} q_{\rm BH}^{3/5} / (1+q_{\rm BH})^{1/5}$, by performing a variable change and integrate over the black hole mass ratio to produce a SMBHB merger rate as a function of chirp mass and redshift only:
\begin{equation}
\frac{d^2 n}{dz'd \mathcal{M}} = \int \frac{d^3 n_G}{dz'dM_{\rm BH}dq_{\rm BH}} \frac{dM_{\rm BH}}{d\mathcal{M}} dq_{\rm BH}.
\label{nchirp}
\end{equation}

Summarizing, the SMBHB merger rate $d^2 n/(dz'd \mathcal{M})$ is described as a function of 16 empirical parameters that are related to astrophysical observables: $(\Phi_0, \Phi_I, M_0, \alpha_0, \alpha_I)$ for the GSMF, $(f_0', \alpha_f, \beta_f, \gamma_f)$ for the pair fraction, $(\tau_0, \alpha_\tau, \beta_\tau, \gamma_\tau)$ for the merger timescale, and $(M_*,\alpha_*, \epsilon)$ for the galaxy -- SMBH scaling relation. Further, the first three sets of parameters can be grouped into the four effective parameters given by equation \eqref{eq:eff_param}. The two extra parameters $(e_0, \zeta_0)$ enter the computation of the shape of the GW spectrum via the transition frequency $f_t$ given in equation \eqref{eq:fdec}. We can therefore express the stochastic GWB in equation \eqref{eqn:hsquared} as a function of 18 phenomenological parameters, listed in table \ref{table:prior}.


\section{GWB simulations and analysis}
\label{sec:Observations}

As in \cite{2017MNRAS.468..404C} (paper II hereafter), we compute the signal-to-noise-ratio S/N $\mathcal{S}$ of a detection of a GWB in the frequency domain as \citep{2015CQGra..32e5004M,2015MNRAS.451.2417R}:
\begin{equation}
\mathcal{S}^2=2\sum_{i=1,N}\sum_{j>i}T_{ij}\int \frac{\Gamma_{ij}^2S_h^2}{(S_n^2)_{ij}}df,
\label{eqrho}
\end{equation}
where $\Gamma_{ij}$ are the Hellings-Downs coefficients \citep{HellingsDowns:1983}:
\begin{equation}
\label{eq:gamma}
\Gamma_{ij}=\frac{3}{2}\gamma_{ij}\ln \left(\gamma_{ij}\right)-\frac{1}{4}\gamma_{ij}+\frac{1}{2}+\frac{1}{2}\delta_{ij},
\end{equation}
where $\gamma_{ij}=[1-\cos(\theta_{ij})]/2$ and $\theta_{ij}$ is the relative angle between pulsars $i$ and $j$. $S_h$ and $S_n$ in equation \eqref{eqrho} are spectral densities of the signal and noise respectively, and $S_n$ includes the 'self noise' contribution of the pulsar term (see equation 11 in paper II for details).

We can simplify equation \eqref{eqrho} by assuming that all pulsars are identical (except for their position in the sky), i.e. all pulsars have the same properties: rms $\delta_i = \delta$, observation time $T_{ij} = T$ and observation cadence $\Delta t$. Furthermore, we also assume that there is a sufficient number of pulsars $N$, uniformly distributed in the sky, so that each individual coefficient $\Gamma_{ij}$ can be replaced by the rms computed across the sky $\Gamma = \sqrt{\langle\Gamma_{ij}^2\rangle} = 1/(4\sqrt{3})$, and the double sum over all pairs of pulsars $\sum_{i=1,N}\sum_{j>i}$ becomes $N(N-1)$. For an observation time $T$ the spectrum of the GWB is resolved into a Fourier series of frequencies $1/T, 2/T, ..., (k+1)/T$ with an equal bin width $\Delta f=1/T$ and central frequencies $f_k=(2k+1)/(2T)$. The total S/N in equation \eqref{eqrho} can thus be split into frequency bin components $\mathcal{S}_k$:
\begin{equation}
\mathcal{S}^2_k \approx T \Gamma^2 N(N-1) \frac{S_h^2}{S_n^2} \Delta f.
\label{eqrhobin}
\end{equation}
In the strong signal regime ($S_h \gg S_n$) equation \eqref{eqrhobin} can further be reduced to the approximate total S/N of a strong detection in $M$ frequency bins
\begin{equation}
\mathcal{S}=\left(\sum_k \mathcal{S}_k^2\right)^{1/2}=\left(\frac{\Gamma^2}{1+\Gamma^2}MN(N-1)\right)^{1/2}\approx \Gamma N M^{1/2},
\label{totrhosimple}
\end{equation}
where we used the fact that $\Gamma\ll1$ and $N\gg1$. Equation (\ref{totrhosimple}) is a drastic simplification, still it provides the relevant scaling between $\mathcal{S}$, number of pulsars in the array, and frequency range in which the signal is resolved. For $\Gamma \approx 0.14$, to achieve a S/N $\approx$ 5 in the lowest few frequency bins, an array of about 20 equally good pulsars is needed \citep[see also][]{JenetEtAl:2006}.

PTA data are simulated as in paper II. For a signal $h_c$ with amplitude $A_{\mathrm{det},k}$ in the $k$-th frequency bin, the detection S/N $\mathcal{S}_k$ is related to the detection uncertainty $\sigma_{\mathrm{det},k}$ via (see equation 18 in paper II)
\begin{equation}
\sigma_{\mathrm{det},k}=\frac{1}{\mathcal{S}_k}.
\label{eqrhotot}
\end{equation}

\begin{figure}
\centering
\includegraphics[width=8.cm,clip=true,angle=0]{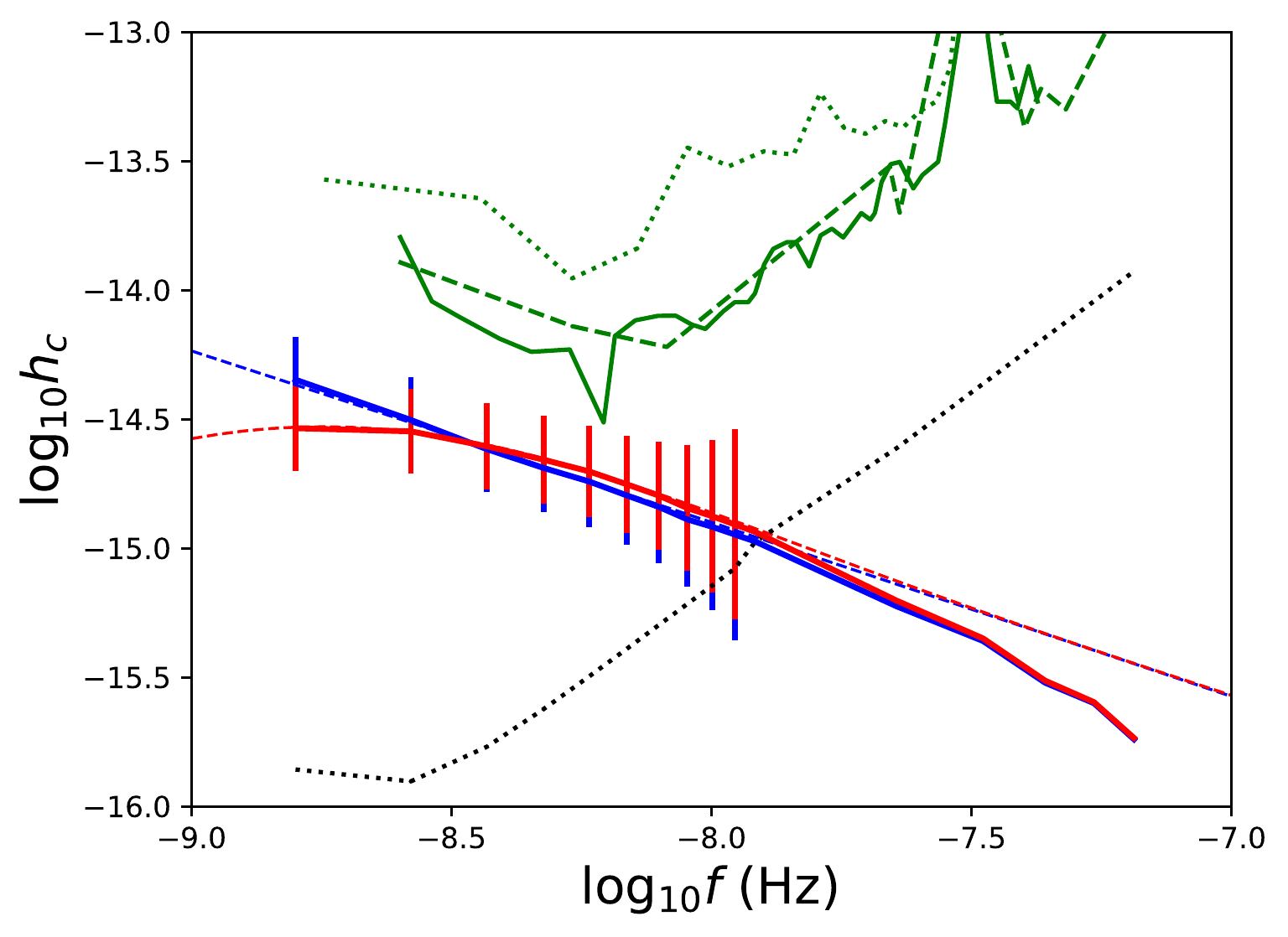}
\caption{Examples of simulated detections for two different spectral shapes. Signal models correspond to the default SMBHB population with parameters defined in Section \ref{sec:Prior} and high eccentricity ($e_t=0.9$, red) and almost circular ($e_t=0.01$, blue). For each model, solid lines are the theoretical spectra including the high frequency steepening due to the mass upper limit defined by equation (\ref{eq:Nf}), dashed lines depict spectra excluding this feature (therefore with $h_c\propto f^{-2/3}$ at high frequency) for comparison. Error bars centred around the model value are the observed amplitudes with associated uncertainties when $\mathcal{S}_k>1$. The black dotted line represents the nominal $1\sigma$ model sensitivity curve of the PTA for the \textit{IPTA30} case ($h_n$ as computed from equation 20 in paper II). Green lines in the upper part of the figure are current EPTA (dotted), NANOGrav (dashed) and PPTA (solid) upper limits.}
\label{fig:hc}
\end{figure}

\subsection{Simulated datasets}
\label{sec:Simdata}

Besides adding a future and an ideal upper limit, we use the same simulation setup as in paper II, with the simplifying assumptions that all pulsars are observed with the same cadence $\Delta t$ for the same duration of $T$ and have the same rms of $\delta$. These assumptions only affect the S/N of the detection, thus setting the error bars $\sigma_{\mathrm{det},k}$. This is purely a choice of convenience that does not affect the general validity of our results. We expand upon the 4 cases from paper II by adding 2 more upper limit cases to get a total of 6 fiducial cases (3 upper limits and 3 detections): 
\begin{enumerate}
\item case {\it PPTA15}: we use the upper limit curve of the most recent PPTA analysis, as given by \cite{2015Sci...349.1522S}, which is representative of current PTA capabilities and results in a GWB upper limit of $A=10^{-15}$;\footnote{$A$ represents the amplitude of the GWB at a reference frequency of $f=1/{\rm yr}$ under the assumption that its spectrum is described by a single power law with $h_c\propto f^{-2/3}$, appropriate for circular, GW-driven binaries.}
\item case {\it PPTA16}: we shift the {\it PPTA15} curve down by one order of magnitude, which is representative of an upper limit of $A=10^{-16}$, reachable in the SKA era;
\item case {\it PPTA17}: we shift the {\it PPTA15} curve down by two orders of magnitude, which is representative of an upper limit of $A=10^{-17}$. Although a two orders of magnitude leap in sensitivity might require decades of timing with the full SKA, we use this scenario to infer what conclusions can be drawn by a non-detection at a level well below currently predicted GWB values;
\item case {\it IPTA30}: $N=20$, $\delta=100$ns, $T=30$yr, $\Delta t=1$ week. This PTA results in a detection S/N$\approx 5-10$ and is based on a future extrapolation of the current IPTA, without the addition of new telescopes;
\item case {\it SKA20}: $N=100$, $\delta=50$ns, $T=20$yr, $\Delta t=1$ week. This PTA results in a high significance detection with S/N$\approx 30-40$, which will be technically possible in the SKA era;
\item case {\it ideal}: $N=500$, $\delta=1$ns, $T=30$yr, $\Delta t=1$ week. This theoretically possible ideal PTA provides useful insights of what might be achievable in principle.
\end{enumerate}

\subsection{Data analysis method}
\label{sec:DAmethod}

We apply Bayes' theorem to perform inference on our model $M$, given some data $d$ and a set of parameters $\theta$:
\begin{equation}
 p(\theta|d,M) = \frac{p(\theta| M) p(d|\theta,M)}{p(d|M)}\,,
 \label{eqn:bayes}
\end{equation}
where $p(\theta|d,M)$ is the posterior distribution coming from the analysis of the PTA measurement, $p(\theta| M)$ is the prior distribution and accounts for any beliefs on the constraints of the model parameters (prior to the PTA measurement), $p(d|\theta,M)$ is the likelihood of producing the data for a given model and parameter set, and $p(d|M)$ is the evidence, which is a measure of how likely the model is to produce the data.

To simulate detections we apply the likelihood from paper II
\begin{equation}
p_\mathrm{det}\left(d|\Atrial(f) \right) \propto \exp\left\{- \frac{\left[ \log_{10}\Atrial(f)-\log_{10}\Adet(f)\right]^2}{2\sigdet(f)^2} \right\} \,,
\label{eq:likedet}
\end{equation}
to each frequency bin for which $\mathcal{S}_k>1$, and then sum over the frequency bins to obtain the total likelihood. For the upper limit analyses, we use the directly derived likelihood from the PPTA upper limit, as described in Appendix A.3 of \cite{2018NatCo...9..573M}.

Prior distributions are taken from independent theoretical and observational constrains, as described in Section \ref{sec:Prior}. The parameter space is sampled using cpnest \citep{cpnest}, which is a parallel implementation of the nested sampling algorithm in the spirit of \cite{LALInference} and \cite{Skilling2004a}. Nested sampling algorithms do not only provide posterior distributions, but also the total evidence. This allows us to compute Bayes factors for model comparisons.  Each simulation has been run with 1000 livepoints, producing $\sim 2500$ independent posterior samples.


\section{Defining the prior ranges of the model parameters}
\label{sec:Prior}

There is a vast literature dedicated to the measurement of the GSMF, galaxy pair fraction, merger timescale and SMBH -- host galaxy scaling relations. We now described how independent observational and theoretical work translates into constrained prior distributions of the 18 parameters of our model. A summary of all the prior ranges is given in table \ref{table:prior}.

\subsection{Galaxy stellar mass function}
\label{sec:gsmfprior}

At any given redshift, the GSMF is usually described as a Schechter function with three parameters $(\Phi_0, M_0, \alpha_0)$. The parameters, however, are independently determined at any redshift. Depending on the number of redshift bins $n$ to be considered in the computation, this can easily lead to a very large number of parameters $3n$. To reduce the dimensionality of the problem from $3n$ to five, we note that the parameters $(\Phi_0, \alpha_0)$ show clear linear trends with redshift, whilst $M_0$ is fairly constant (see \cite{2015MNRAS.447....2M}). This allows for a re-parametrisation as a function of the 5 parameters $(\Phi_0, \Phi_I, M_0, \alpha_0, \alpha_I)$  performed in Section \ref{sec:Galaxy}.

\begin{figure}
\centering
\includegraphics[width=8.cm,clip=true,angle=0]{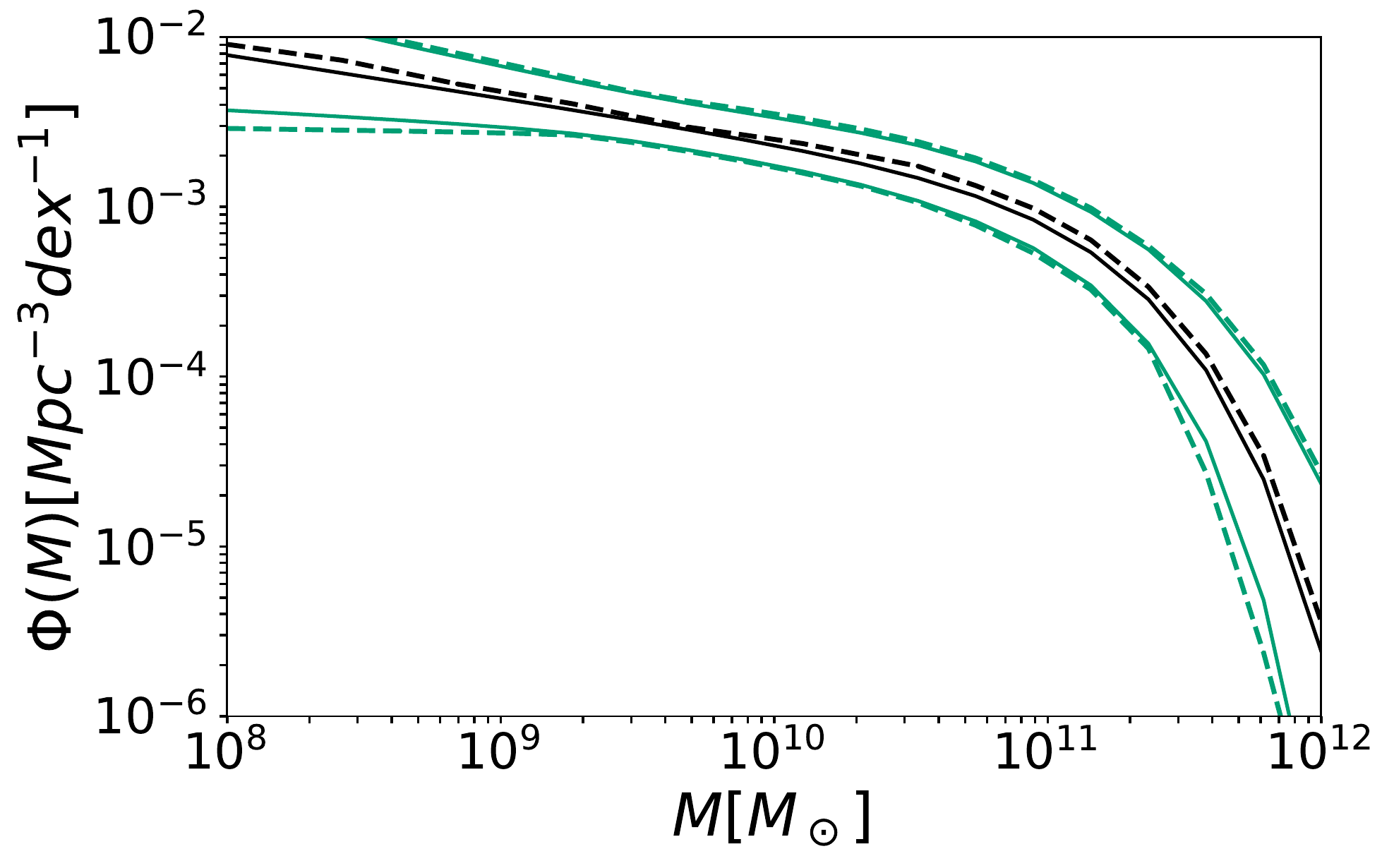} \\
\includegraphics[width=8.cm,clip=true,angle=0]{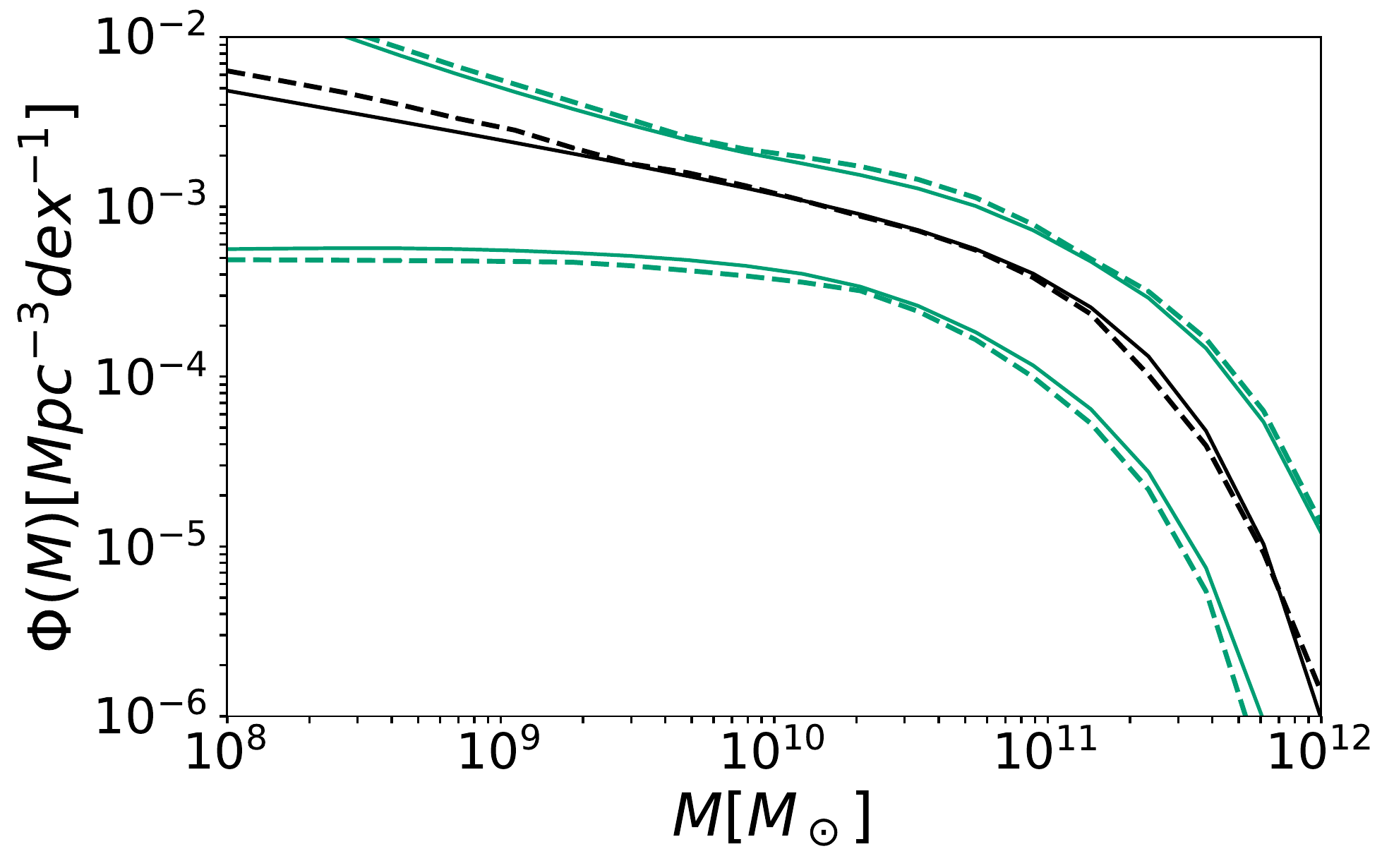}
\caption{Comparison between the allowed region of GSMFs for different redshift bins computed from \protect\cite{2016ApJ...830...83C} (3 dashed lines) and the region of GSMFs recovered by using the five GSMF parameters $(\Phi_0, \Phi_I, M_0, \alpha_0, \alpha_I)$ (3 solid lines). Black lines represent the median, whilst green lines represent the upper and lower bounds of the 99\% confidence region. The top and bottom panel show the GSMF in the $0<z<1$ and $1<z<3$ redshift bins respectively.}
\label{fig:gsmfregion}
\end{figure}

\begin{figure}
\centering
\includegraphics[width=8.cm,clip=true,angle=0]{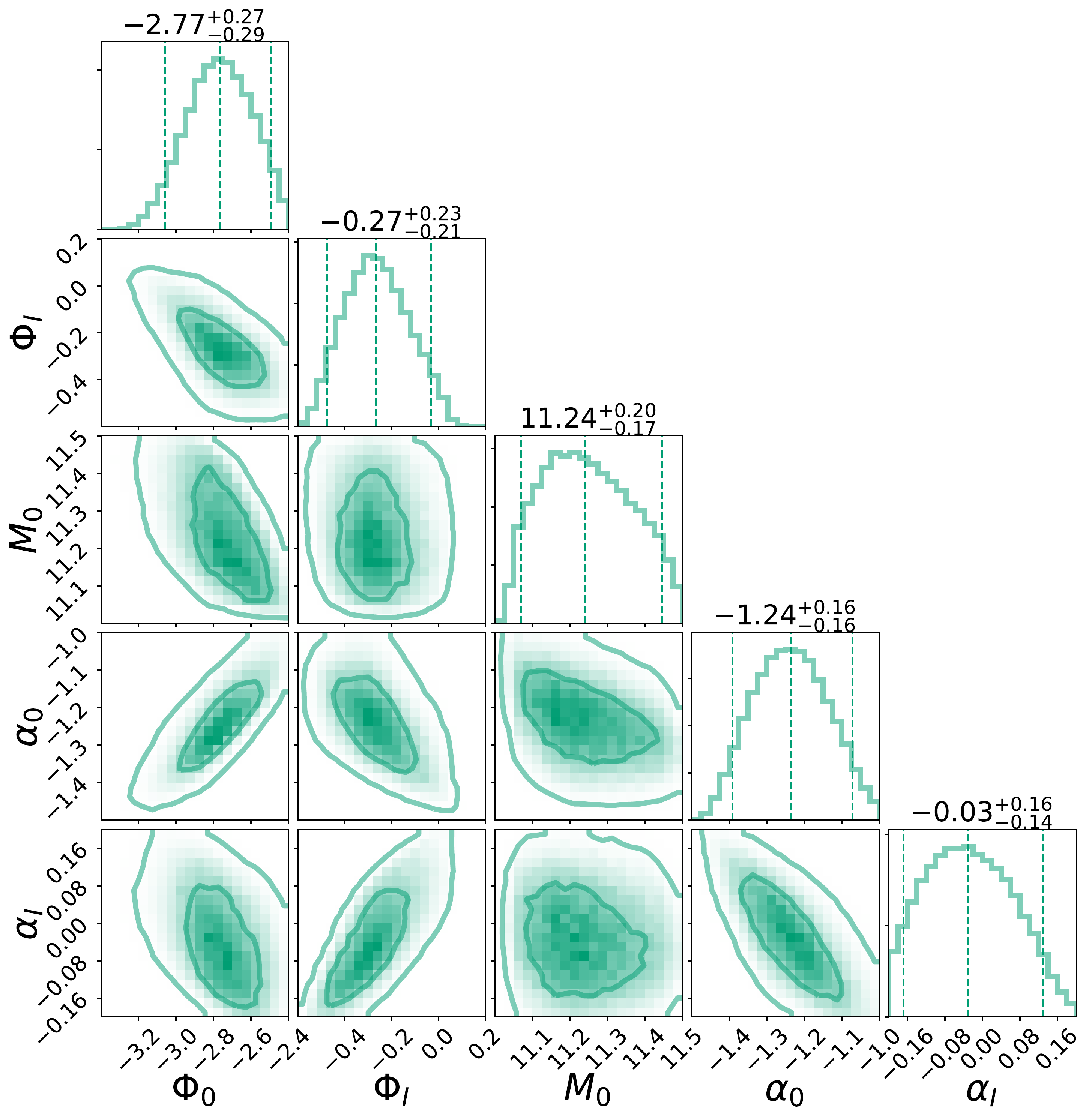}
\caption{Corner plot showing the prior distributions of the five GSMF parameters $(\Phi_0, \Phi_I, M_0, \alpha_0, \alpha_I)$ used in this work.}
\label{fig:gsmfprior}
\end{figure}

A comprehensive list of published values for the parameters $(\Phi_0, M_0, \alpha_0)$ for various redshift bins can be found in \cite{2016ApJ...830...83C}, which forms the basis of our prior distribution. We compute $\Phi(M,z)$ between $10^9 M_\odot$ and $10^{12} M_\odot$ for all sets of $(\Phi_0, M_0, \alpha_0)$, dividing the sample in two redshift bins: $0<z<1$ and $1<z<3$. This gives a range of values for $\Phi(M,z)$, shown in figure \ref{fig:gsmfregion}. We then take uniform distributions of $(\Phi_0 \in [-3.4,-2.4], \Phi_I \in [-0.6,0.2], M_0 \in [11,11.5], \alpha_0 \in [-1.5,-1.], \alpha_I \in [-0.2,0.2])$, compute the $\Phi(M,z)$ for each sample and redshift bins and compare them against the allowed range. If the value is within the range, the sample is accepted, otherwise it is rejected. The resulting prior distributions are shown in figure \ref{fig:gsmfprior}.


\subsection{Pair fraction}

Constraints on the pair fraction have been derived by counting the numbers of paired and merged galaxies in various surveys with a number of different photometric and spectroscopic techniques \citep[see, e.g., ][]{2003AJ....126.1183C,xu12,2014MNRAS.444.3986R,2014ApJ...795..157K,2019ApJ...876..110D}. Recently, \cite{2017MNRAS.470.3507M} have combined data from several surveys to produce an overall up-to-date constrain. We base our prior range on the results reported in table 3 of their paper, for the All+GAMA+D17 survey combination and galaxy separation of $5-30$ kpc:
\begin{equation}
f_{\rm pair} = 
\begin{cases}
0.028 \pm 0.002 \times (1+z)^{0.80 \pm 0.09} & \text{for $\log M > 10$}\\
0.024 \pm 0.004 \times (1+z)^{0.78 \pm 0.20} & \text{for $\log M > 11$}.
\label{eq:mstar_mbulge}
\end{cases}
\end{equation}
This is one of the flatter redshift dependences within the \cite{2017MNRAS.470.3507M} compilation. It is, however, likely the more accurate measurement, coming from a combination of deep surveys. Moreover, while stronger redshift dependences are common for Milky Way-size galaxies, most $f_{\rm pair}$ measurements of galaxies with $M>10^{11} M_\odot$ have a relatively flat redshift dependence. Most of the GWB will come from SMBHBs hosted in those massive galaxies, this justifies our choice. Noting that both sets of parameters and uncertainties in equation \eqref{eq:mstar_mbulge} are similar, we use flat priors for $f_0 \in [0.02,0.03]$ and $\beta_f \in [0.6,1]$ for all galaxy masses. Steeper redshift dependences are allowed in our set of 'extended priors', introduced in section \ref{sec:ext} below. \cite{2017MNRAS.470.3507M} also find no significant dependency on galaxy mass, thus we pick $\alpha_f, \gamma_f \approx 0$, adding the possibility of a mild deviation by imposing a flat prior $\alpha_f, \gamma_f \in [-0.2,0.2]$.

\subsection{Merger timescale}
\label{sec:mtime}

We define the merger timescale, as the time elapsed between the observation of a galaxy pair at a given projected separation (usually 20 or 30 kpc) and the final coalescence of the SMBHB, thus including the time that it takes for the two galaxies to effectively merge, plus the time required for the SMBHs to sink to the center,
form a binary and harden via stellar scattering. Galaxy merger timescales have been computed both for simulations of isolated galaxy mergers \citep{2011ApJ...742..103L} and from ensemble of halos and galaxies extracted from large cosmological simulations \citep{kit08,2017MNRAS.468..207S}, resulting in a large dynamical range, typically between 0.1 and 1Gyr. On the other hand, the SMBHB merger timescale has been estimated by means of N-body and special purpose Monte-Carlo codes \citep[e.g.][]{2012ApJ...756...30K,2015ApJ...810...49V,2015MNRAS.454L..66S}. All studies show that three-body scattering is efficient in driving the binary to final coalescence withing a Gyr.

We therefore choose the parametrisation given by equation \eqref{tau} with wide uniform prior ranges $\tau_0 \in [0.1,2]$ Gyr and $\beta_\tau \in [-2,1]$, which is sufficiently generic to cover the observation based range of possible effects influencing the total merger time. The mass dependencies are generally found to be milder, playing a minor role. We therefore choose flat prior ranges $\alpha_\tau, \gamma_\tau \in [-0.2,0.2]$. These conservative prior ranges are extended in section \ref{sec:ext} to include possibe physical effect further delaying the merger timescale, such as inefficient replenishment of the loss cone slowing down the binary hardening. Note that in this latter case all SMBHBs forming in galaxy pairs observed at $z< \sim 2$ would not merge by $z=0$, thus being effectively 'stalled' for the sake of our analysis.

\subsection{$M_{\rm bulge} - M_{\rm BH}$ relation}

Since SMBHs are thought to have an important impact on the formation and evolution of their host galaxy and vice versa, the relation between their mass and several properties of the host galaxy has been studied and constrained by a number of authors  \citep[see][for a comprehensive review]{2013ARA&A..51..511K}. Here we use the tight relation between the SMBH mass and the stellar mass of the spheroidal component (i.e. the bulge) of the host galaxy, which has been described as a power-law of the form of equation \eqref{eqnmbulge} with some intrinsic scattering. Although non-linear functions have been proposed in the literature \citep[see, e.g., ][]{grahamscott12,2016MNRAS.460.3119S}, the non-linearity is mostly introduced to describe the (observationally very uncertain) low mass end of the relation. Since the vast majority of the GWB is produced by SMBH with masses above $10^8 M_\odot$ \citep{SesanaVecchioColacino:2008}, we do not consider here those alternative parametrizations.

Similarly, we do also not consider the possibility of a redshift dependent $M_{\rm bulge} - M_{\rm BH}$ relation \citep[see][and references therein]{2011ApJ...742...33L}. Recent findings strenghten the view that there is no evidence for a cosmic evolution \citep{2014MNRAS.438.3422S} or only a very weak one \citep{2015ApJ...799..173S}. This additional weak redshift dependence would likely not have a significant impact on our results and would be in any case covariant with other redshift dependences, and thus unlikely to be constrained by our analysis.

\begin{figure}
\centering
\includegraphics[width=8.cm,clip=true,angle=0]{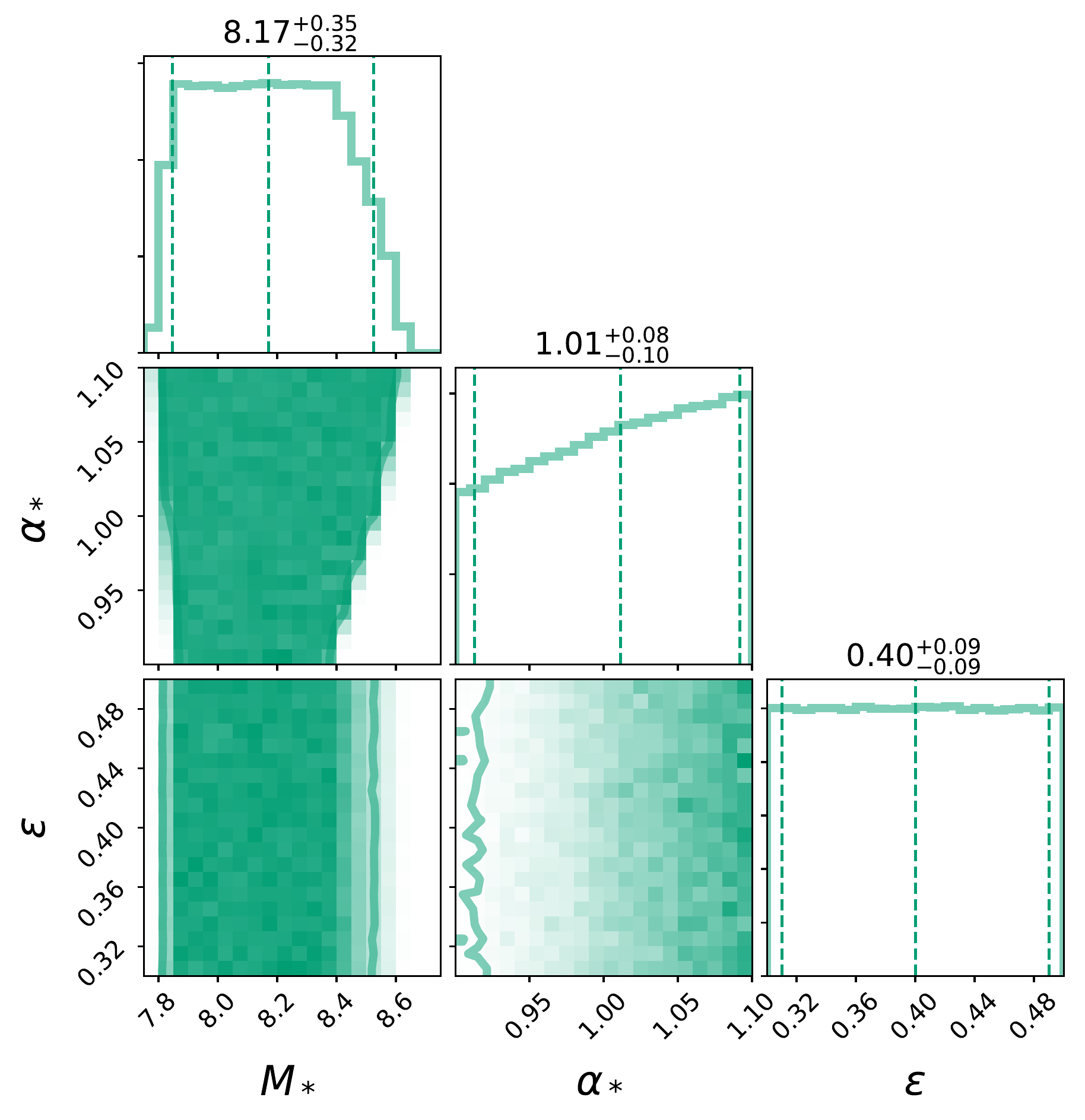}
\caption{Corner plot showing the prior distributions of the three $M_{\rm bulge} - M_{\rm BH}$ relation parameters $(M_*, \alpha_*, \epsilon)$ used in this work.}
\label{fig:mbulge}
\end{figure}

To construct the prior distributions, we apply the same method as in Section \ref{sec:gsmfprior}. We define the allow region of the $M_{\rm BH}-M_{\rm bulge}$ relation as the one enclosed within a compilation of relations collected from the literature in \cite{2018NatCo...9..573M}. We then draw relations from a uniform distribution of $\log_{10} M_* \in [7.75,8.75]$ and $\alpha_* \in [0.9,1.1]$ and accept them if they fall within the region allowed by observations.  Additionally, we assume a flat distribution for the scattering $\epsilon \in [0.3,0.5]$. Figure \ref{fig:mbulge} shows the obtained prior distributions for $(M_*, \alpha_*, \epsilon)$.

\subsection{Eccentricity and stellar density}

The last two parameters deal with the properties of the individual binary. As the eccentricity at decoupling is not well constrained \cite[see, e.g.][]{2015MNRAS.454L..66S,2017MNRAS.470..940M}, we choose an uninformative flat prior $e_0 \in [0.01,0.99]$. The other additional parameter describes the stellar density around the SMBHB (see section \ref{sec:Model}). $\zeta_0$ is a multiplicative factor added to the density at the SMBHB influence radius, $\rho_{i,100}$, calculated by using the fiducial Dehnen profile defined in paper I. This has an impact on the frequency of decoupling, as a higher density of stars in the galactic centre means more efficient scattering. The SMBHB thus experiences a faster evolution, reaching a higher $f_t$ before transitioning to the efficient GW emission stage. We choose to include densities that are between 0.01 and 100 times the fiducial value, aiming at covering the large variation of stellar densities observed in cusped vs cored galaxies \citep{2009ApJS..182..216K}. This translates into a flat prior $\log_{10} \zeta_0 \in [-2,2]$.

\begin{table}
\begin{center}
\def\arraystretch{1.5}
\begin{tabularx}{0.475\textwidth}{c|Xcc}
\hline
parameter & description & standard & extended \\
\hline
$\Phi_0$  & GSMF norm & $-2.77^{+0.27}_{-0.29}$ & $-2.77^{+0.27}_{-0.29}$ \\
$\Phi_I$  & GSMF norm redshift evolution & $-0.27^{+0.23}_{-0.21}$ & $-0.27^{+0.23}_{-0.21}$ \\
$\log_{10} M_0$  & GSMF scaling mass & $11.24^{+0.20}_{-0.17}$ & $11.24^{+0.20}_{-0.17}$ \\
$\alpha_0$  & GSMF mass slope & $-1.24^{+0.16}_{-0.16}$ & $-1.24^{+0.16}_{-0.16}$ \\
$\alpha_I$  & GSMF mass slope redshift evolution & $-0.03^{+0.16}_{-0.14}$ & $-0.03^{+0.16}_{-0.14}$ \\
$f_0$  & pair fraction norm & [0.02,0.03] & [0.01,0.05] \\
$\alpha_f$ & pair fraction mass slope & [-0.2,0.2] & [-0.5,0.5] \\
$\beta_f$  & pair fraction redshift slope & [0.6,1] & [0,2] \\
$\gamma_f$  & pair fraction mass ratio slope & [-0.2,0.2] & [-0.2,0.2] \\
$\tau_0$  & merger time norm & [0.1,2] & [0.1,10] \\
$\alpha_\tau$  & merger time mass slope & [-0.2,0.2] & [-0.5,0.5] \\
$\beta_\tau$  & merger time redshift slope & [-2,1] & [-3,1] \\
$\gamma_\tau$  & merger time mass ratio slope & [-0.2,0.2] & [-0.2,0.2] \\
$\log_{10} M_*$  & $M_{\rm bulge} - M_{\rm BH}$ relation norm & $8.17^{+0.35}_{-0.32}$ & $8.17^{+0.35}_{-0.32}$ \\
$\alpha_*$  & $M_{\rm bulge} - M_{\rm BH}$ relation slope & $1.01^{+0.08}_{-0.10}$ & $1.01^{+0.08}_{-0.10}$ \\
$\epsilon$ & $M_{\rm bulge} - M_{\rm BH}$ relation scatter & [0.3,0.5] & [0.2,0.5] \\
$e_0$ & binary eccentricity & [0.01,0.99] & [0.01,0.99] \\
$\log_{10} \zeta_0$  & stellar density factor & [-2,2] & [-2,2] \\
\hline
\end{tabularx}
\caption{List of the 18 parameters in the model, including their description, standard and extended prior distribution ranges. Squared brackets indicate flat uniform distributions, while $\pm$ signs indicate the median and 90\% credible intervals for the distributions, as shown in figures \ref{fig:gsmfprior} and \ref{fig:mbulge} of section \ref{sec:Prior}.}
\label{table:prior}
\end{center}
\end{table}

\subsection{Extended prior ranges}
\label{sec:ext}

Unless otherwise stated, the prior ranges just described are used in our analysis. However, we also consider 'extended' prior ranges for some of the parameters. Although observational determination of the galaxy mass function is fairly solid, identifying and counting galaxy pairs in large galaxy surveys is a delicate endeavour, especially beyond the local universe. We therefore also consider extended prior ranges $f_0 \in [0.01,0.05]$, $\alpha_f \in [-0.5,0.5]$ and $\beta_f \in [0,2]$, allowing for more flexibility in the overall normalization, redshift and mass evolution of the galaxy pair fraction. Likewise, SMBHB merger timescales are poorly constrained. The prior range adopted in Section \ref{sec:mtime} is rather wide, but notably does not allow for stalling of low redshift binaries (the maximum allowed merger timescale being 2 Gyrs). Also in this case we consider extended prior ranges $\tau_0 \in [0.1,10]$ Gyr, $\alpha_\tau \in [-0.5,0.5]$ and $\beta_\tau \in [-3,1]$, allowing the possibility of SMBHB stalling at any redshift. Finally we also consider a wider prior on the scatter of the $M_{\rm BH}-M_{\rm bulge}$ relation $\epsilon \in [0.2,0.5]$, mostly because several authors find $\epsilon \approx 0.3$, which is at the edge of our standard prior. All standard and extended priors are listed in table \ref{table:prior}.


\section{Results and discussion}
\label{sec:Results}

Having defined the mathematical form of the signal, the prior ranges of all the model parameters, the simulated data and the form of the likelihood function, we performed our analysis on the six limits and detections described in Section \ref{sec:Simdata}. In this section, we present the results of our simulations and discuss their astrophysical consequences in detail. We first present the implications of current and future upper limits and then move onto discussing the different cases of detection. Note that, although all 18 parameters are left free to vary within their respective priors, we will present posteriors only for the subset of parameters that can be significantly constrained via PTA observations. Those are the overall normalization of the merger rate $n_{\rm eff}$, the parameters defining the merger timescale $\tau_0, \alpha_\tau, \beta_\tau$, the parameters defining the $M_{\rm BH}-M_{\rm bulge}$ relation $M_*, \epsilon$, the eccentricity at the transition frequency $e_0$, and the normalization of the stellar density $\zeta_0$. Because the large number of parameters and the limited information enclosed in the GWB shape and normalization, other parameters are generally unconstrained. Corner plots including all 18 parameters for all the simulated upper limits and detections are presented in Appendix A, available in electronic form. All runs are performed using the standard prior distributions derived in Section \ref{sec:Prior}, unless stated otherwise.

\subsection{Predicted GWB Strain}

\begin{figure}
\centering
\includegraphics[width=8.cm,clip=true,angle=0]{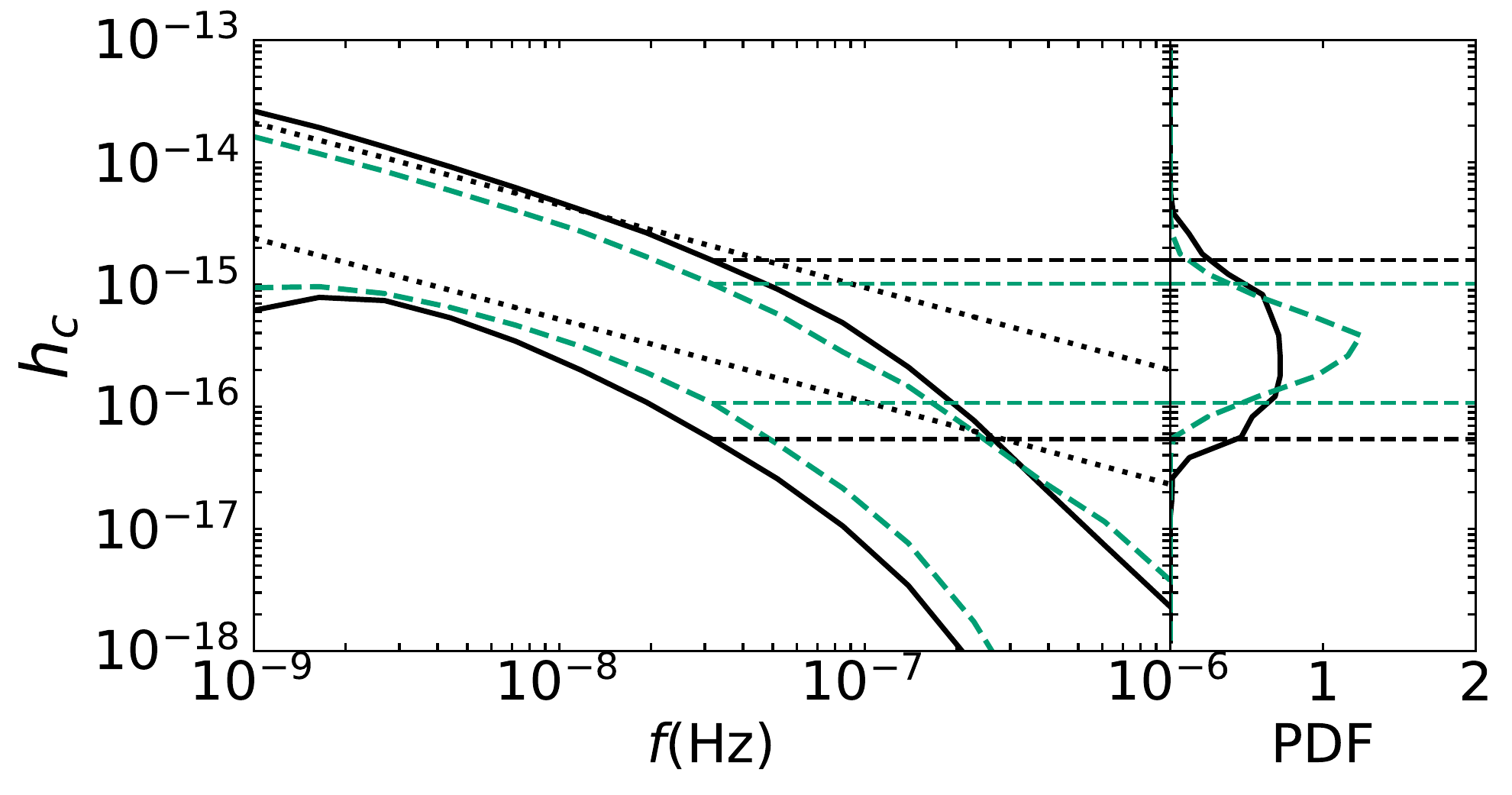}\\
\includegraphics[width=8.cm,clip=true,angle=0]{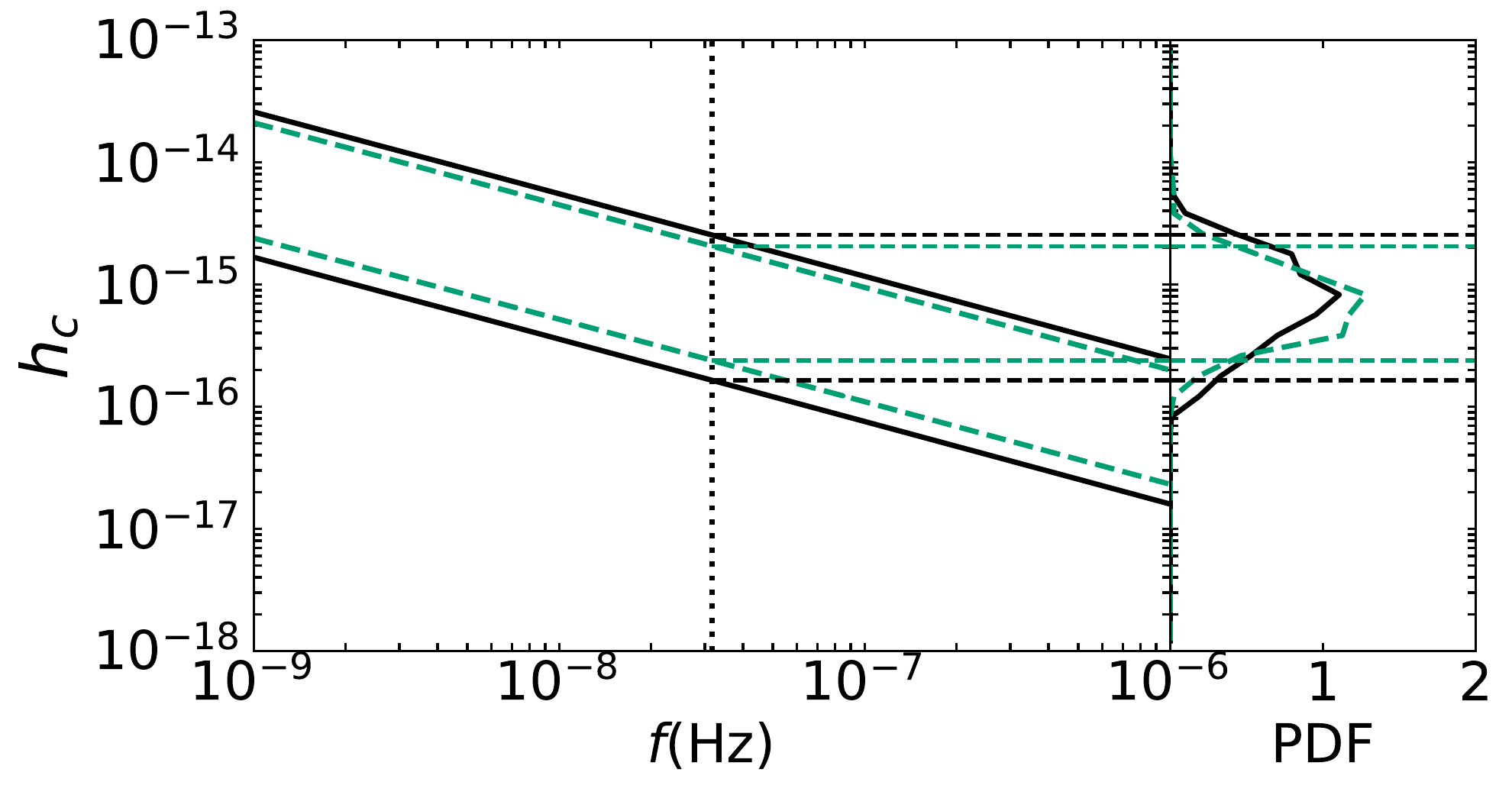}
\caption{Comparison of the predicted characteristic strain, $h_c$, predicted by our new model (green dashed lines) compared to the \textit{ALL} model from \protect\cite{2018NatCo...9..573M} (solid black lines). The top panel shows the predicted strain from the full model, while the bottom panel restricts the model to circular SMBHBs without the drop at high frequencies. For comparison between the top and bottom panel, the equivalent of the bottom panel black solid lines are plotted in the top panel as black dotted lines. The left panels show the frequency - strain plot, while the right panels show the posterior density function (PDF) at $f=1/{\rm yr}$.}
\label{fig:hccomp}
\end{figure}
\begin{figure*}
\includegraphics[width=0.45\textwidth]{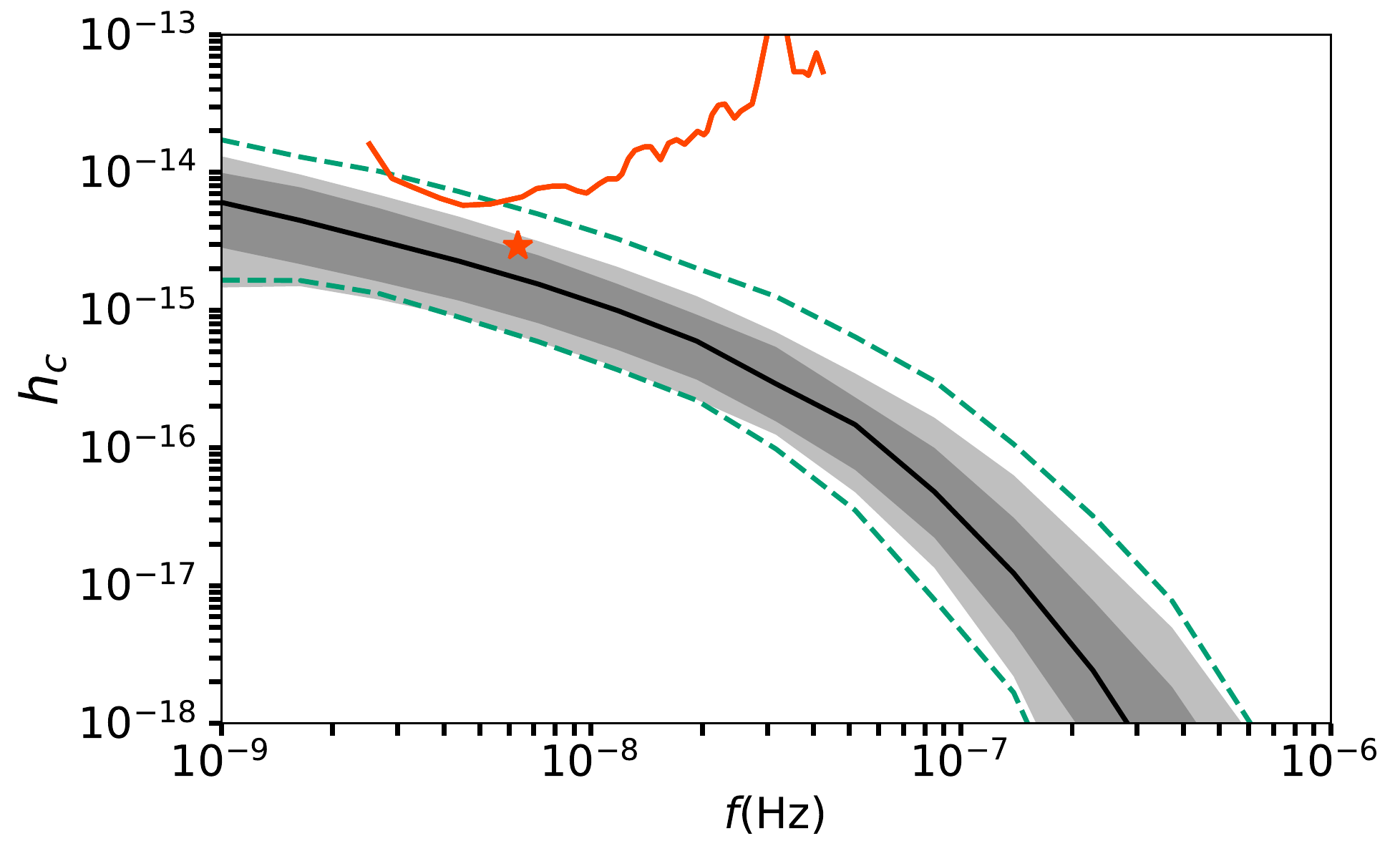} \hspace{1cm}
\includegraphics[width=0.45\textwidth]{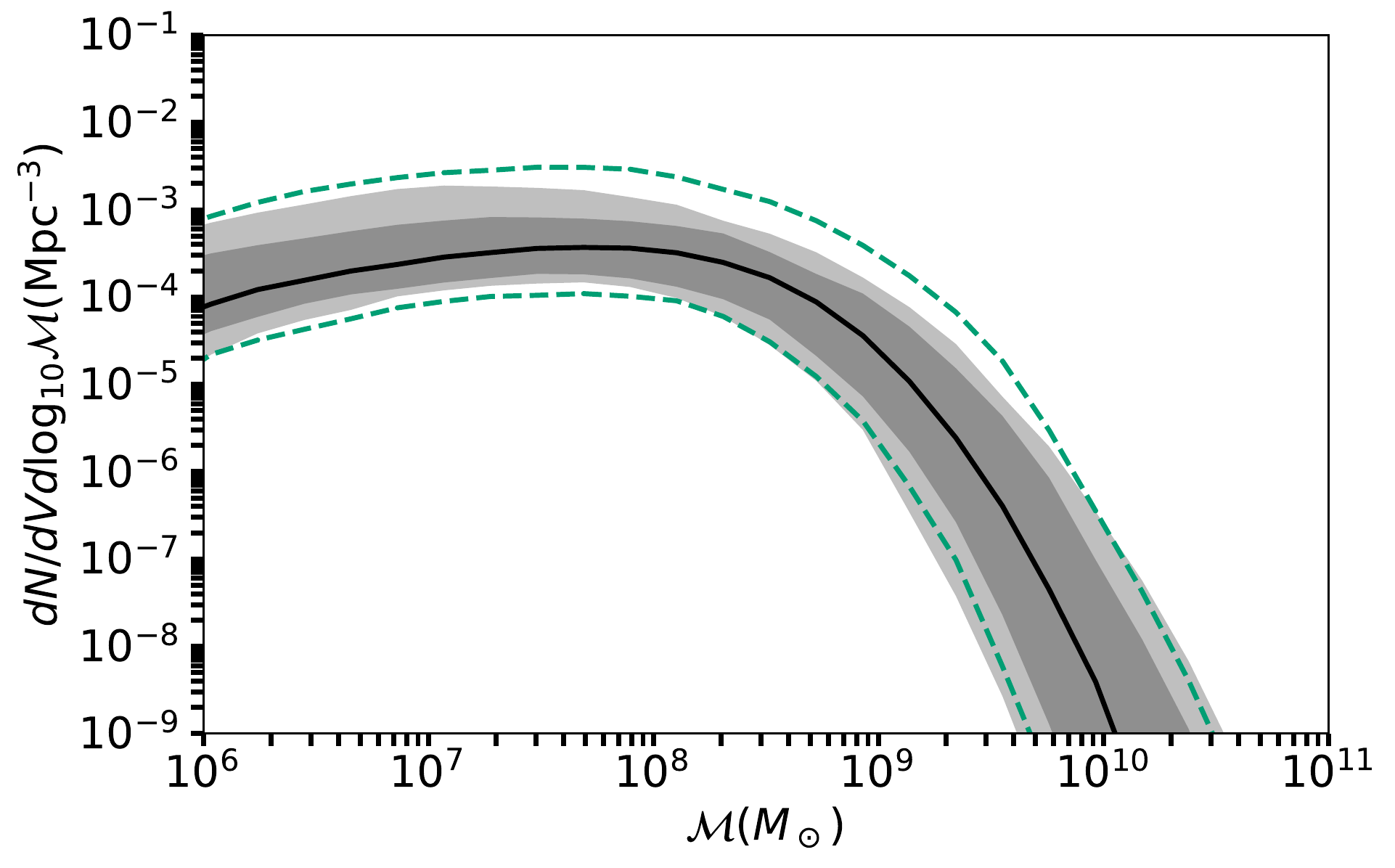}\\
\includegraphics[width=0.45\textwidth]{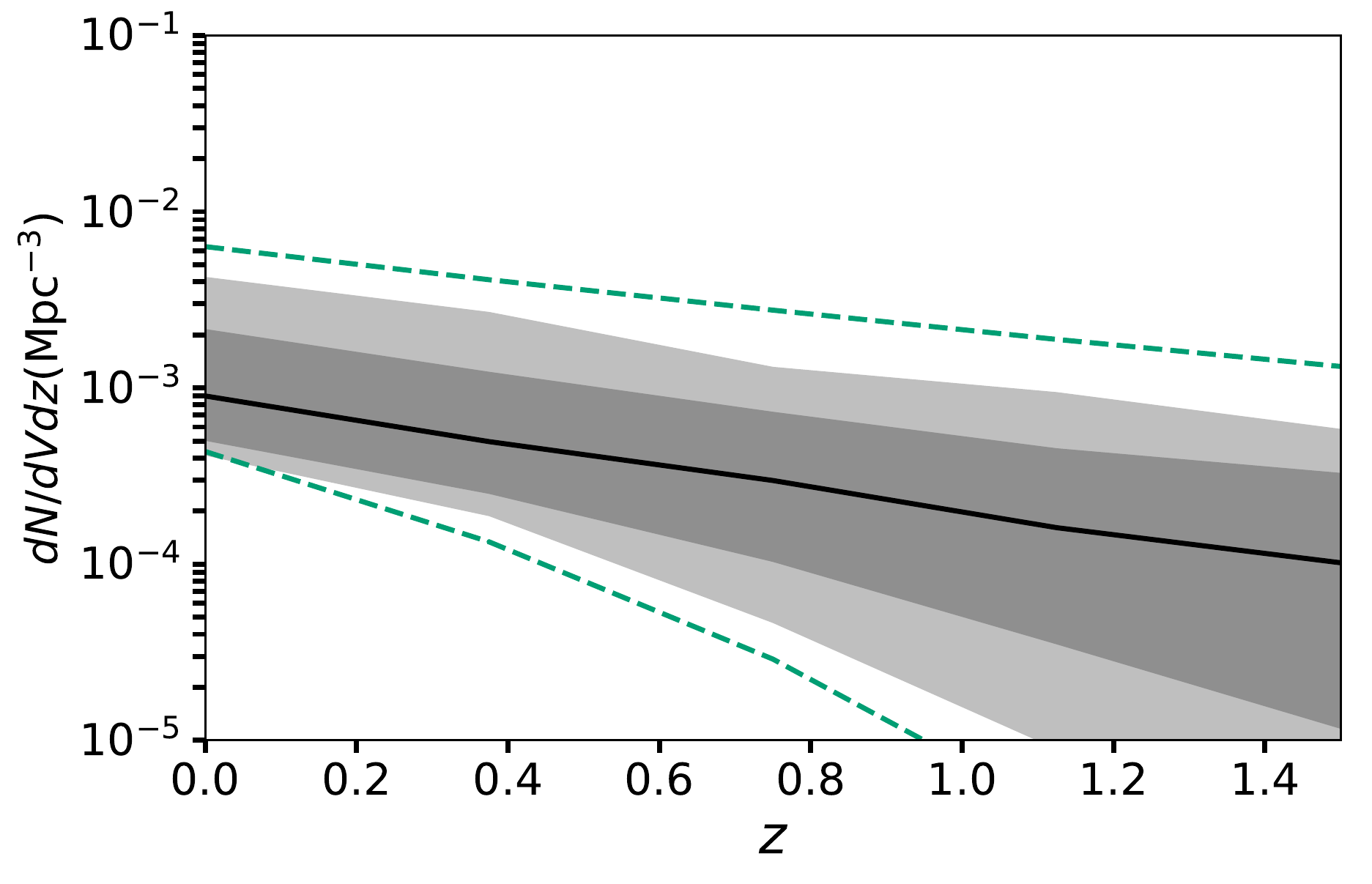} \hspace{1cm}
\includegraphics[width=0.45\textwidth]{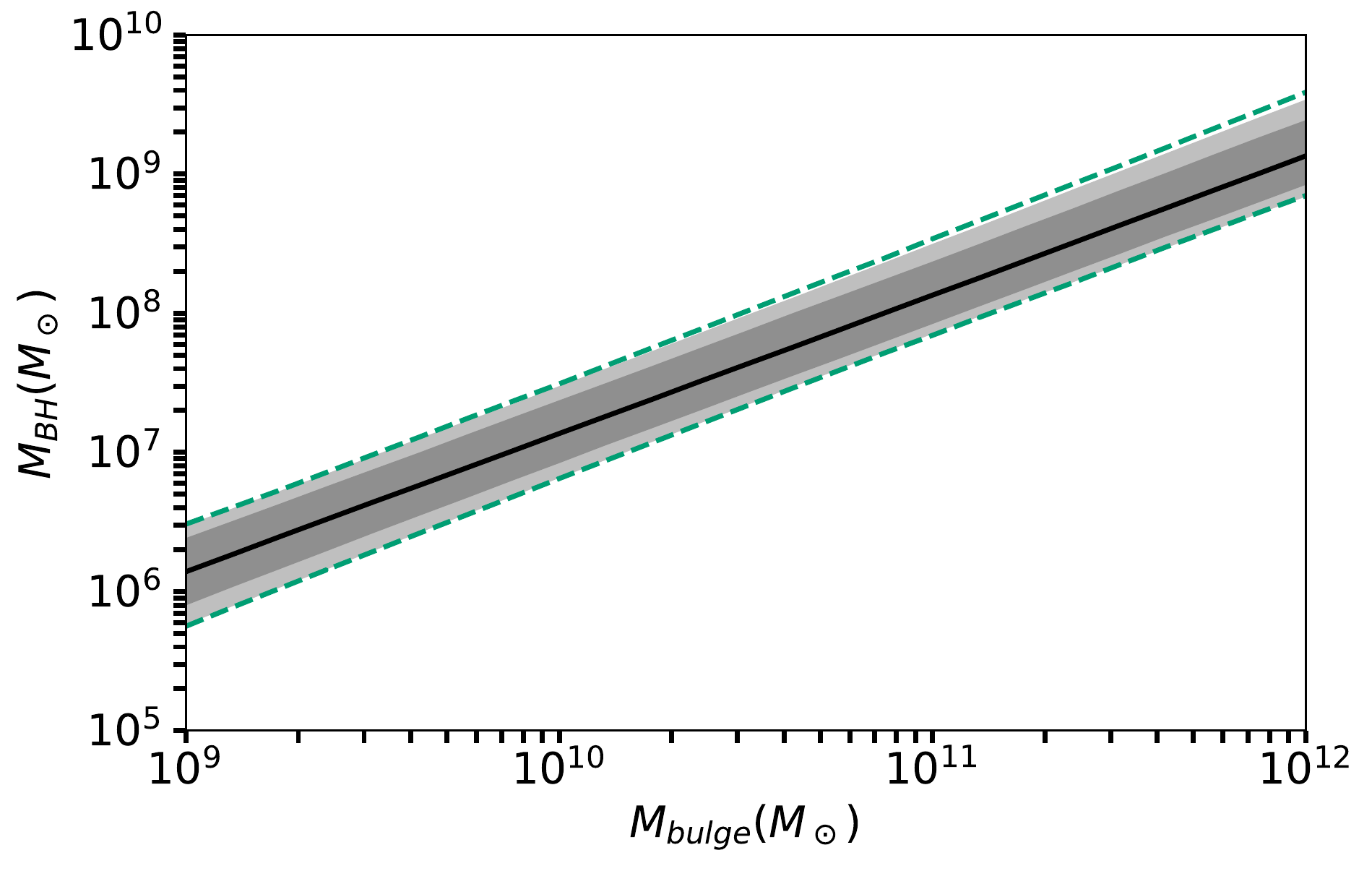}\\
\includegraphics[width=0.9\textwidth]{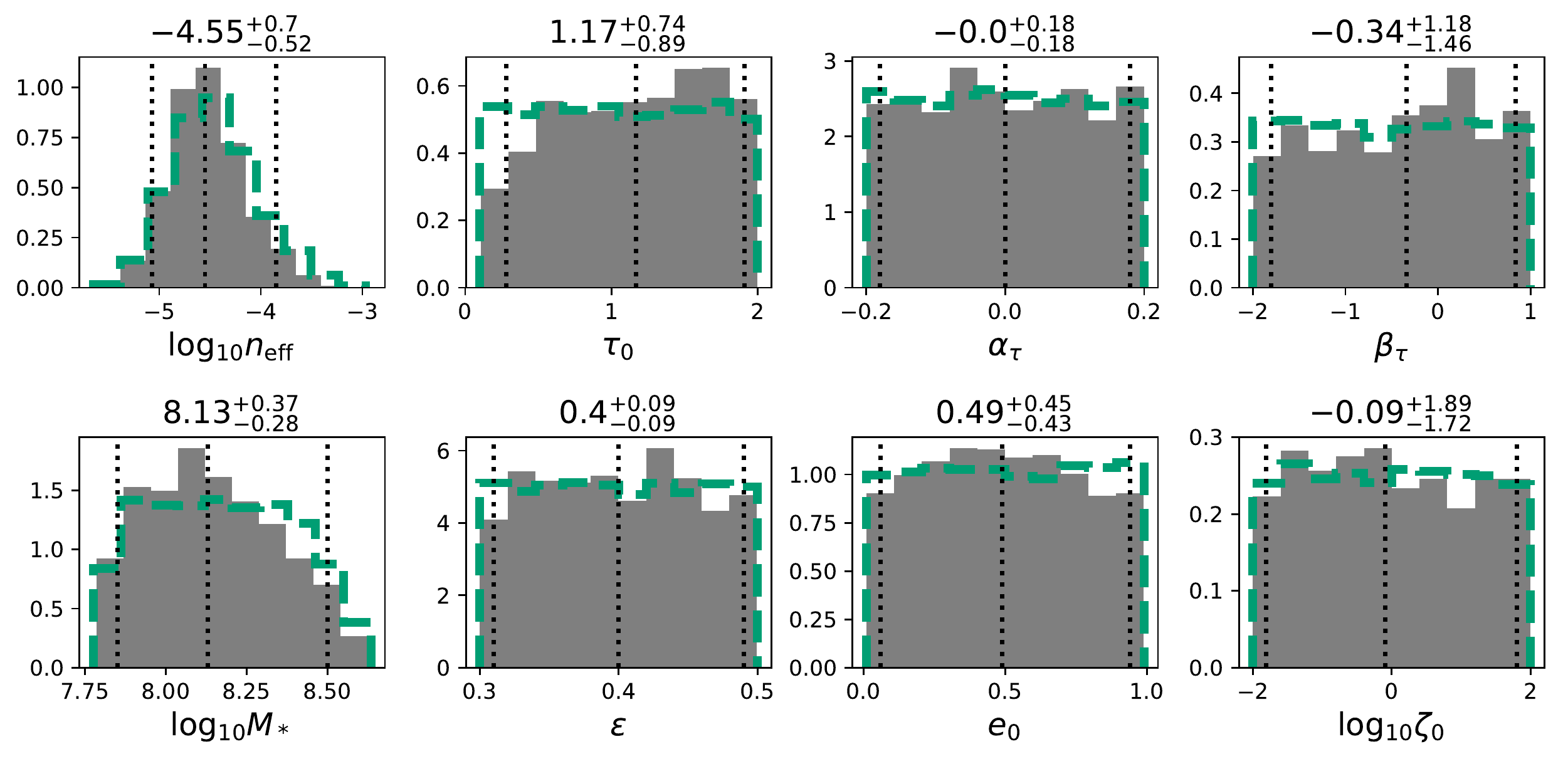} \\
\caption{Implication of a 95\% upper-limit of $A(f=1/{\rm yr})= 1\times 10^{-15}$, corresponding to the most stringent PTA upper limit to date. The posterior for the spectrum (top left), mass (top right), redshift functions (centre left) and $M_{\rm bulge} - M_{\rm BH}$ relations (centre right) are shown as shaded areas, with the central 68\% and 90\% confidence regions indicated by progressively lighter shades of grey, and the solid black line marking the median of the posterior. While the prior is indicated by green dashed lines. The solid orange line and star in the top left panel indicate the frequency dependent and nominal frequency integrated 95\% upper limit from \protect\cite{2015Sci...349.1522S} respectively. The bottom row shaded histograms show the marginalized posteriors for selected model parameters with the prior distributions indicated with green dashed lines, see Section \ref{sec:Prior} and table \ref{table:prior}.}
\label{fig:uplim15}
\end{figure*}
\begin{figure*}
\includegraphics[width=0.45\textwidth]{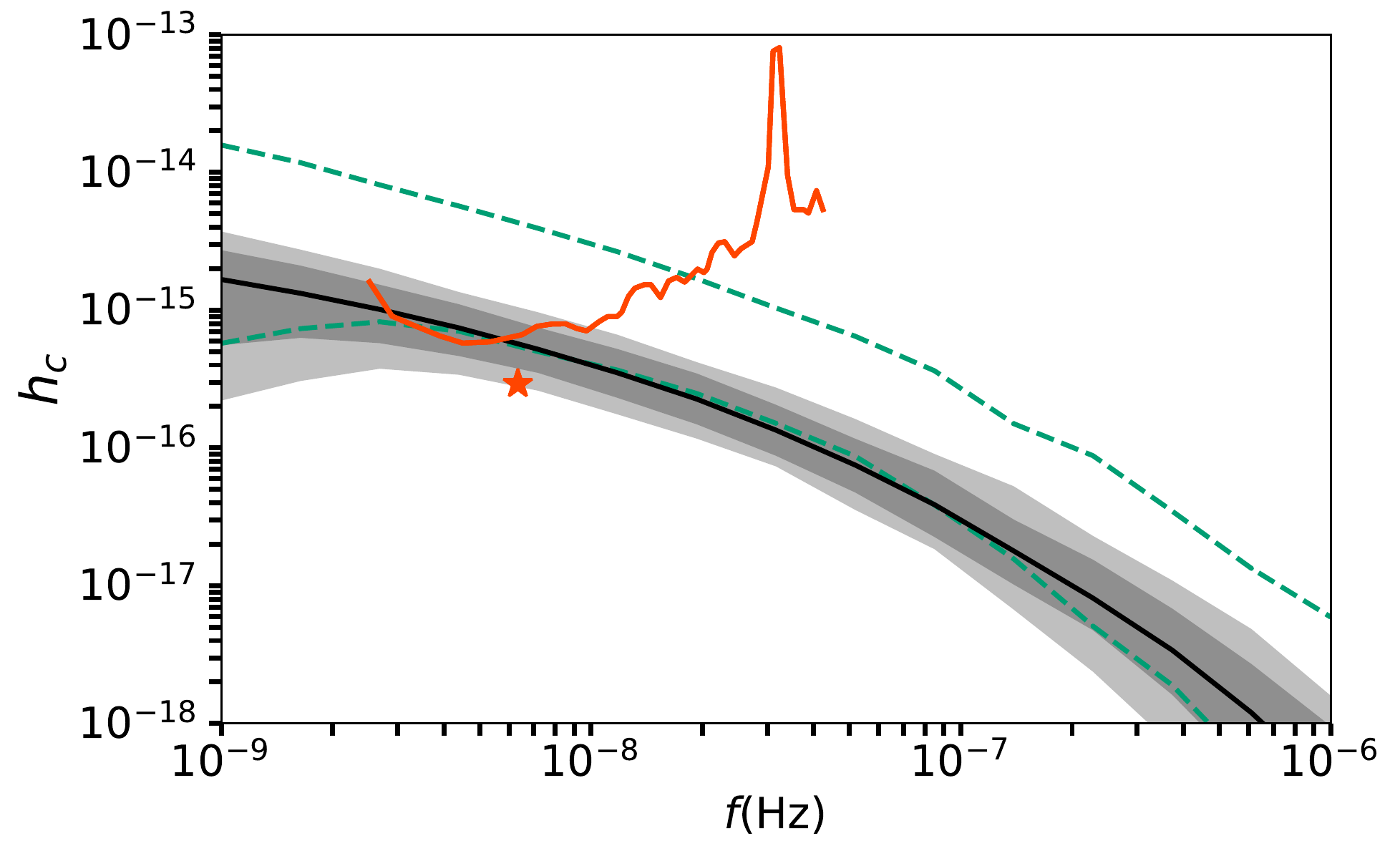} \hspace{1cm}
\includegraphics[width=0.45\textwidth]{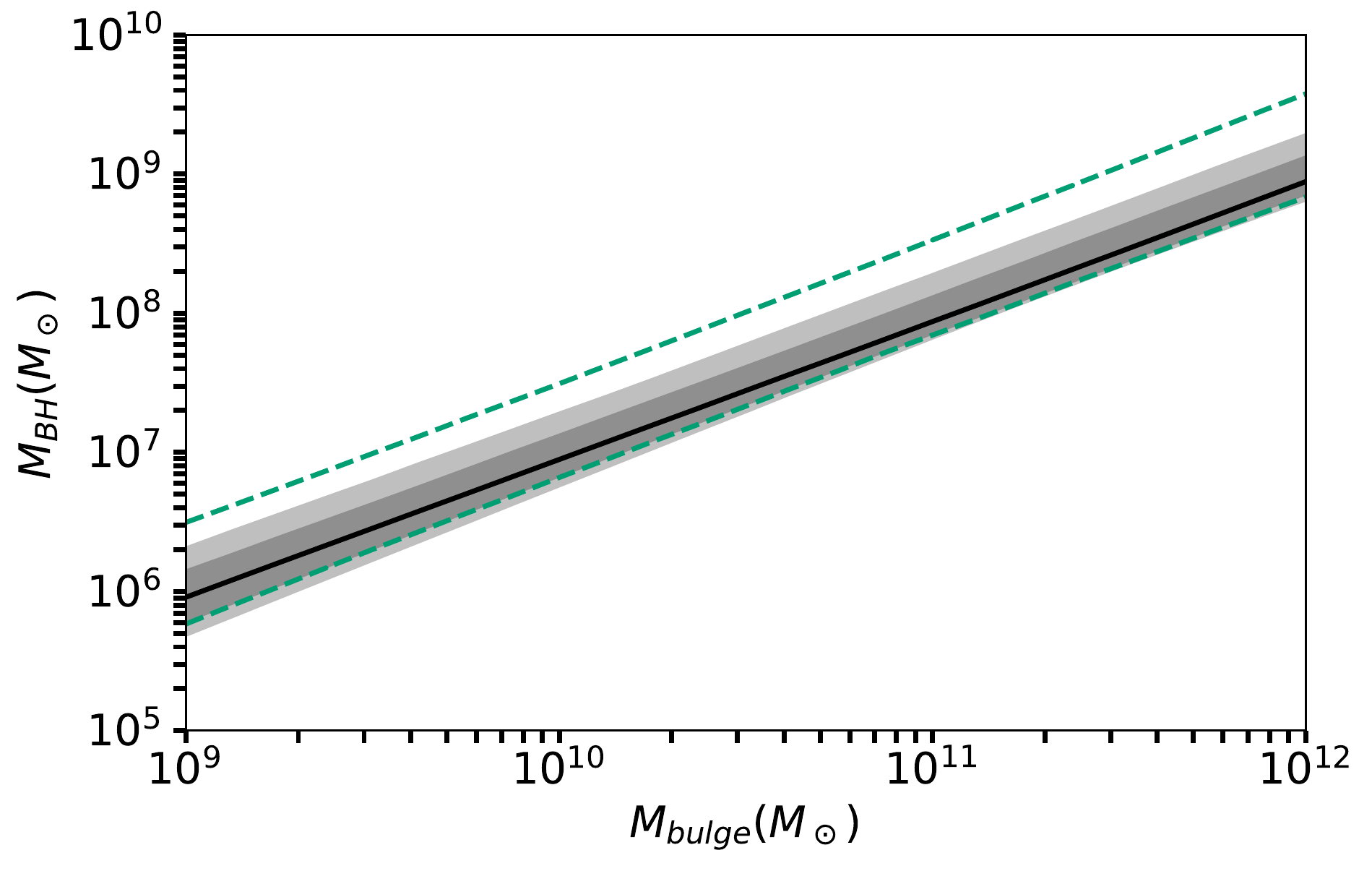} \\
\includegraphics[width=0.9\textwidth]{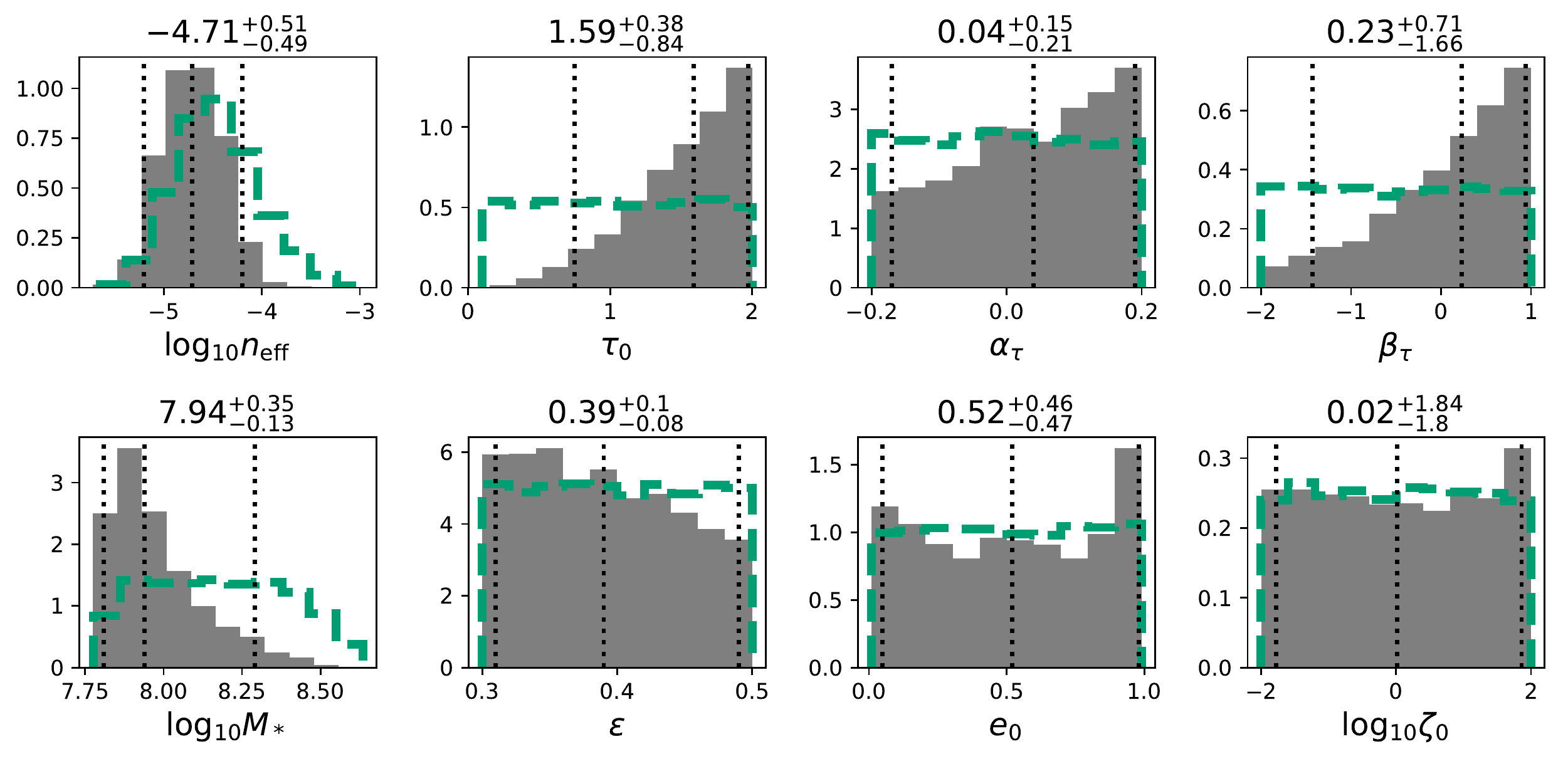} \\
\caption{Same as figure \ref{fig:uplim15} without the posterior distributions for the mass and redshift functions, but for an upper limit of $A(f=1/{\rm yr})= 1\times 10^{-16}$}
\label{fig:uplim16}
\end{figure*}
\begin{figure*}
\includegraphics[width=0.45\textwidth]{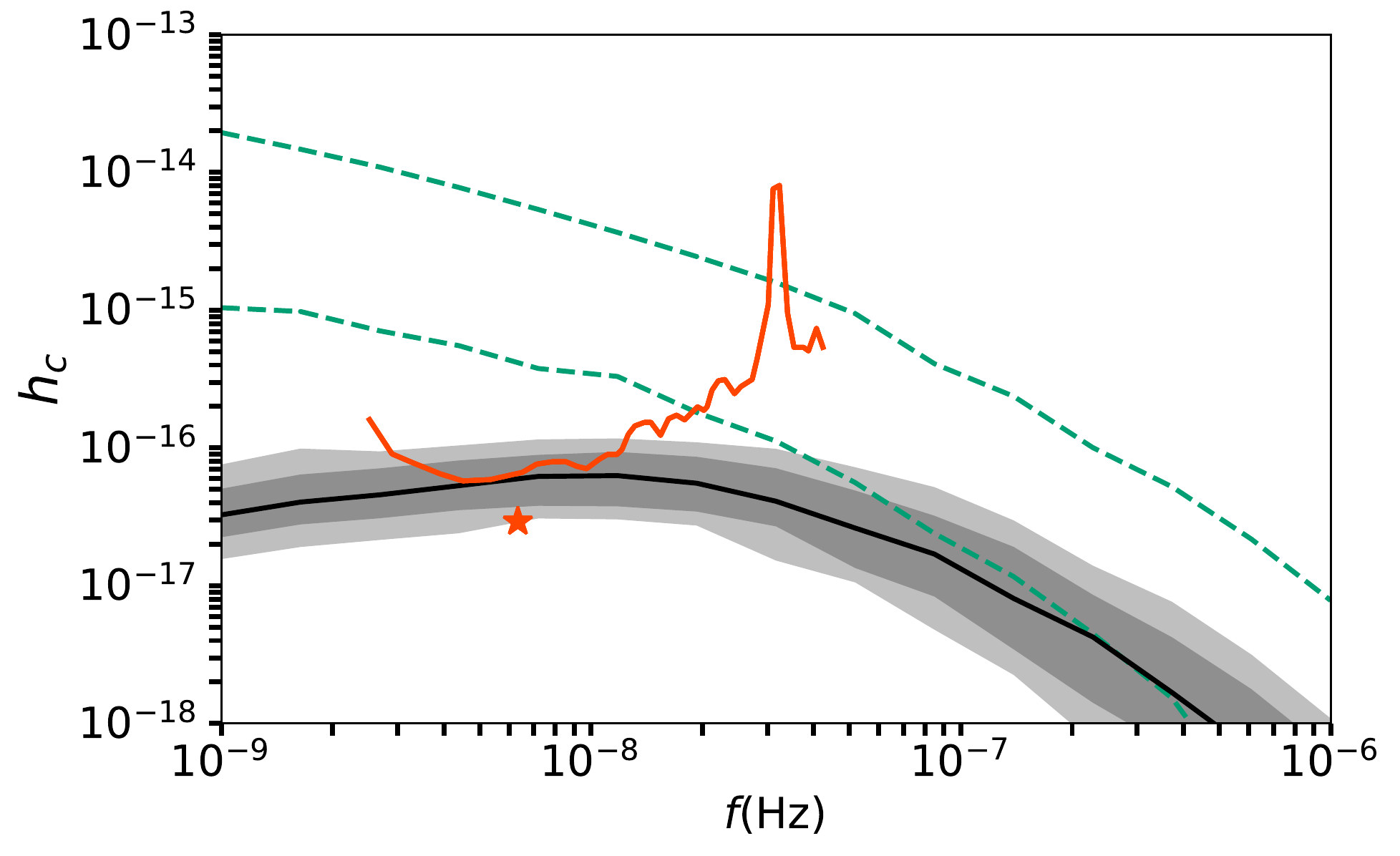} \hspace{1cm}
\includegraphics[width=0.45\textwidth]{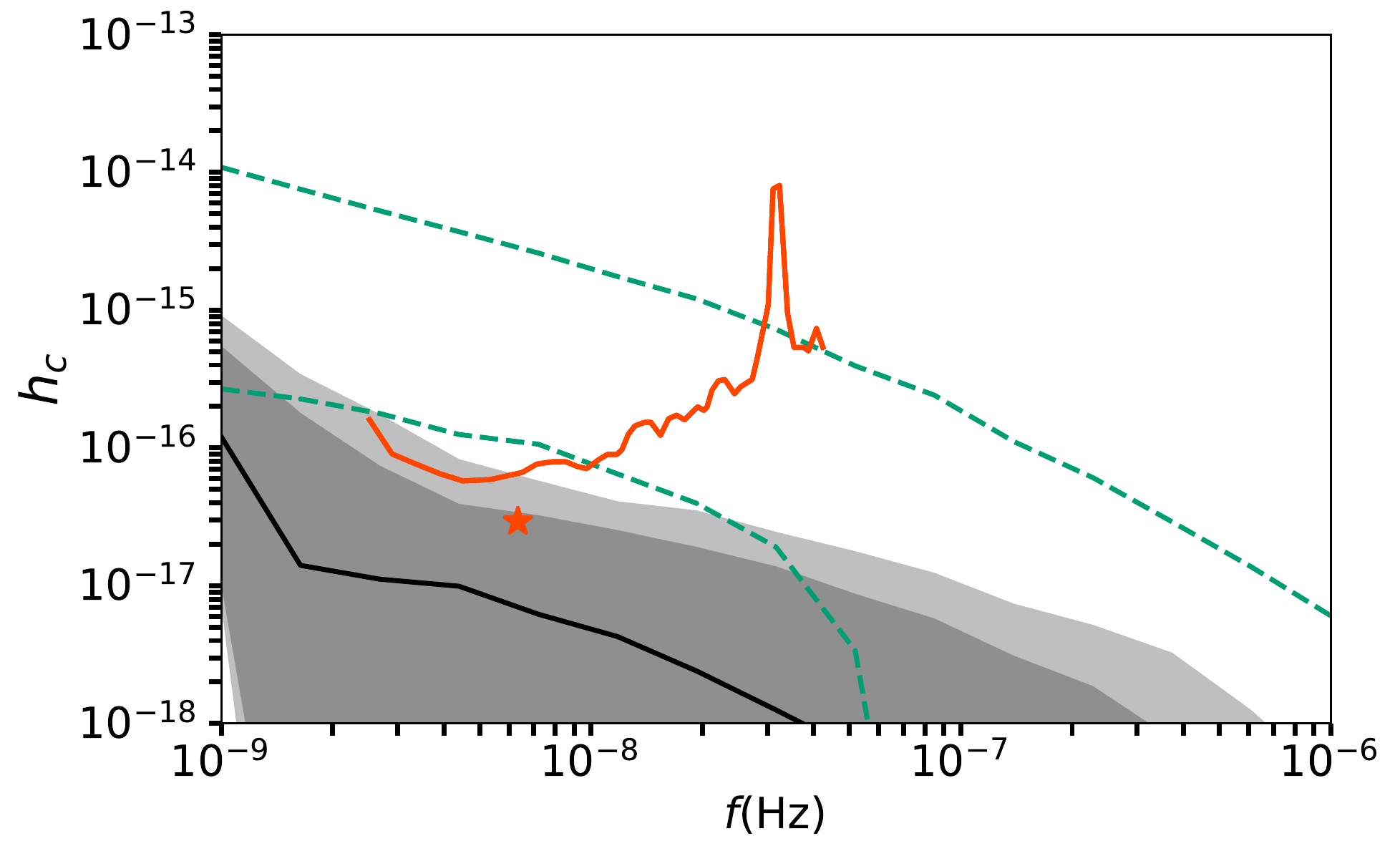} \\
\includegraphics[width=0.45\textwidth]{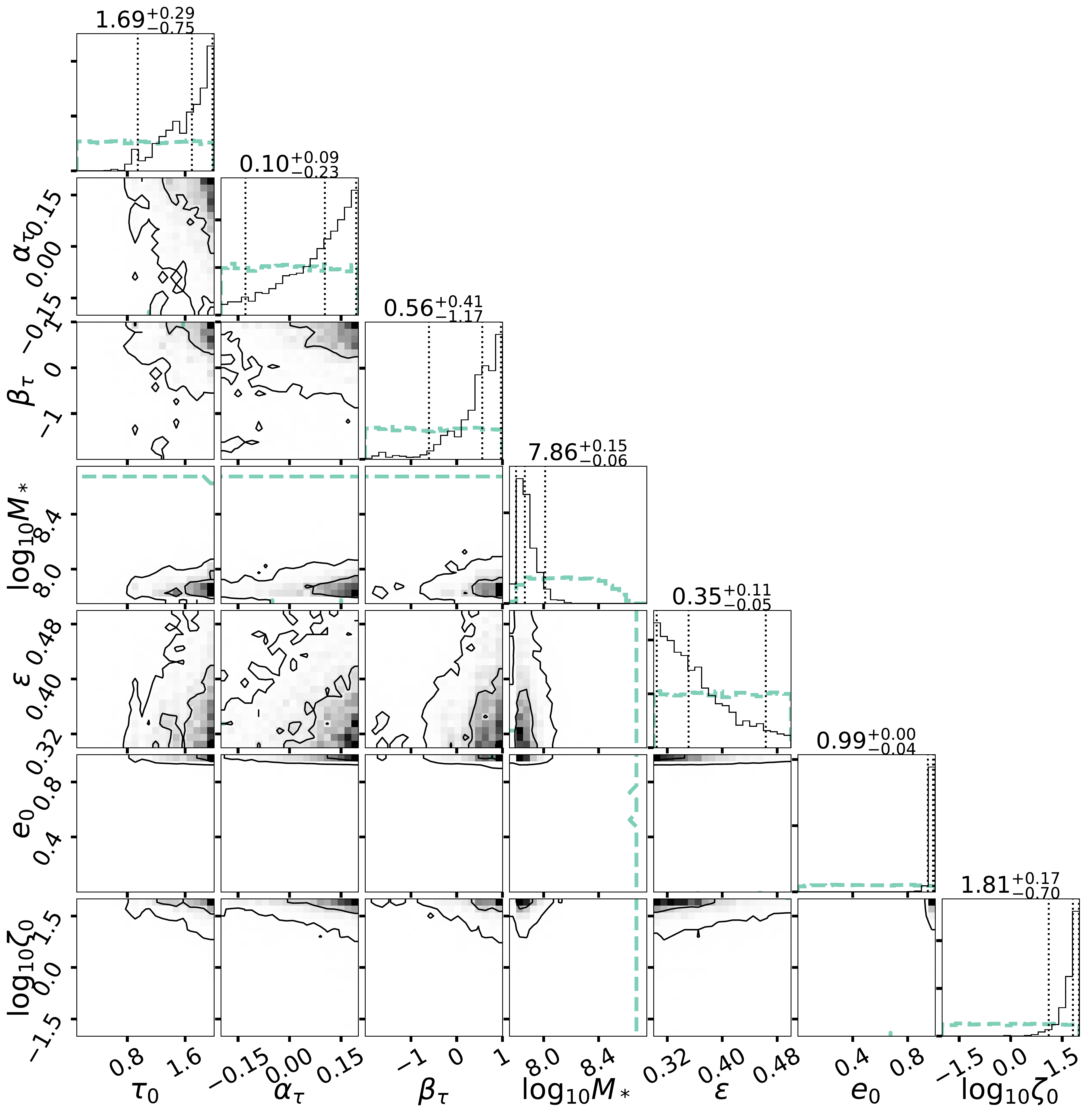} \hspace{1cm}
\includegraphics[width=0.45\textwidth]{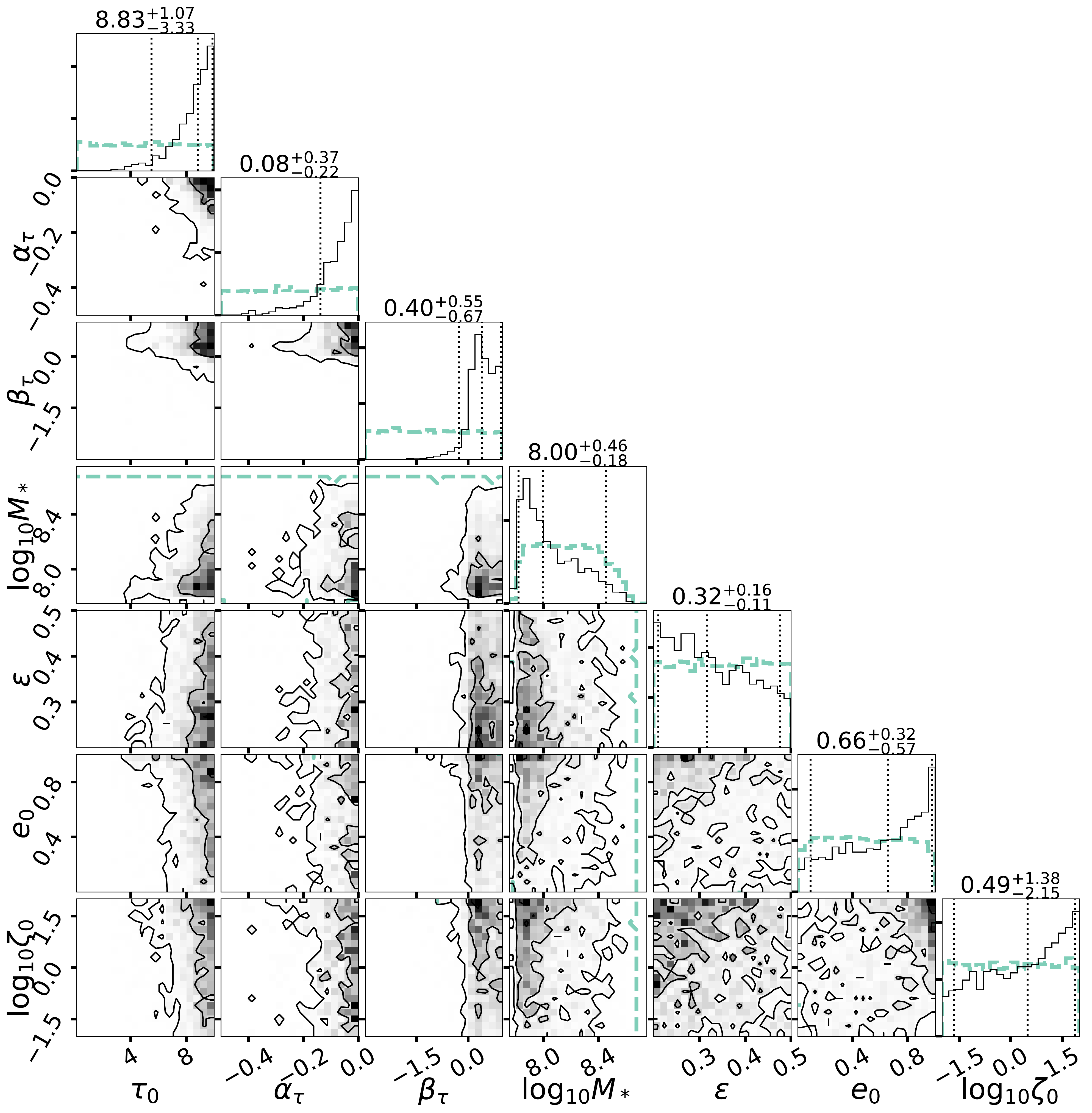} \\
\caption{Implications of a 95\% upper-limit of $A(f=f=1/{\rm yr})= 1\times 10^{-17}$ with standard (left column) and extended (right column) prior distributions. In each column, the top panel shows the posterior of the recovered GWB spectrum with the central 68\% and 90\% confidence regions indicated by progressively lighter shades of grey, and the solid black line marking the median of the posterior. While the prior is indicated by green dashed lines. The bottom corner plots show the (one)two-dimensional posteriors for each model parameter pairs as shaded area, the different levels of shading indicate the 5\%, 50\% and 95\% confidence regions. In each sub-panel, the green dashed lines show the 100\% confidence levels for the prior.}
\label{fig:uplim17}
\end{figure*}

A direct product of combining the GWB model described in Section \ref{sec:Model} to the astrophysical priors presented in Section \ref{sec:Prior} is a robust update to the expected shape and normalization of the signal. Thus, before proceeding with the analysis of our PTA simulations, we present this result. In figure \ref{fig:hccomp} the predicted strain of the GWB using our standard prior is compared to the \textit{ALL} model from \cite{2018NatCo...9..573M}.
The shapes and normalization of the two predictions, shown in the top panel, are fairly consistent. At $f=1/{\rm yr}$ our model predicts $10^{-16} < h_c < 10^{-15}$ at 90\% confidence, which is slightly more restrictive than the \textit{ALL} model. This has to be expected since model \textit{ALL} from \cite{2018NatCo...9..573M} is constructed following the method of \cite{2013CQGra..30v4014S}. The latter, in fact, gave equal credit to all measurements of the galaxy mass function, pair fractions and SMBH -- galaxy scaling relations, without considering any possible correlation between their underlying parameters. Our detailed selection of the prior range takes correlations between different parameters into account (see e.g. figure \ref{fig:gsmfprior}) and is likely more restrictive in terms of galaxy pair fraction.

The bottom panel of figure \ref{fig:hccomp} shows the predicted $h_c$ range assuming circular, GW driven binaries and no high frequency drop, hence producing the standard $f^{-2/3}$ spectral shape. In this simplified case $h_c(f=1/{\rm yr})$ is a factor of $\approx 2$ higher, spanning from $2\times10^{-16}$ to $2\times10^{-15}$. Still, most of the predicted range lies below current PTA upper limits, as well as being consistent with other recent theoretical calculations \citep{2017MNRAS.470.4547D,2017MNRAS.471.4508K,2018MNRAS.477.2599B}.

\subsection{Upper limits}

\subsubsection{Current Upper limit at $A(f=f=1/{\rm yr})= 1\times 10^{-15}$}

Firstly, we discuss the implication of current PTA upper limits. Here, we use the PPTA upper limit, nominally quoted as $A(f=1/{\rm yr}) = 1.0 \times 10^{-15}$, which represents the integrated constraining power over the entire frequency range assuming a $f^{-2/3}$ power-law. As it has been recently pointed out by \cite{2018ApJS..235...37A}, the sensitivity of PTAs has become comparable to the uncertainty in the determination of the solar system ephemeris SSE -- the knowledge of which is required to refer pulse time of arrivals collected at the telescopes to the solar system baricenter. Thus, it has become necessary to include an extra parametrized model of the SSE into the GWB search analysis pipelines. This leads to a more robust albeit higher upper limit, as part of the constraining power is absorbed into the uncertainty of the SSE. A robust upper limit including this effect has recently been placed by the NANOGrav Collaboration at $A(f=1/{\rm yr}) = 1.45 \times 10^{-15}$, which is higher but of the same order as the PPTA upper limit. We therefore consider the PPTA upper limit in this analysis, with the understanding that the recent NANOGrav upper limit would lead to very similar implications. Since the NANOGrav and PPTA upper limits are in fact obtained at each frequency independently, our analysis takes advantage of this by using the constraining power for the GWB spectrum at each frequency separately.

Figure \ref{fig:uplim15} shows that upper limits add very little knowledge to our understanding of the SMBHB population as constrained by the priors on our model parameters. This is in agreement with there being no tension between the current PTA non-detection of the GWB and other astrophysical observations, as extensively discussed in \citep{2018NatCo...9..573M}. The range of characteristic strain of the GWB predicted by the prior ranges of our model $10^{-16} < h_c < 10^{-15}$, shown in the upper left plot of figure \ref{fig:uplim15}, is only mildly reduced by current PTA observations. Therefore, PTAs are starting to probe the interesting, astrophysical region of the parameter space, without yet being able to rule out significant areas, as can be seen in the posterior distribution of the model parameters shown at the bottom of figure \ref{fig:uplim15}. This results into a logarithmic Bayesian evidence $\log_e \mathcal{Z}(10^{-15}) = -0.55$. The evidence is normalized so that an ideal reference model that is unaffected by the measurement has $\log_e = 0$. The log evidence can therefore be directly interpreted as the Bayes factor against such a model. In this specific case, we find $e^{-0.55}=0.58$, indicating that current upper limits do not significantly disfavour the prior range of our astrophysical model. This can also be seen in the bottom row posteriors of figure \ref{fig:uplim15} where the posterior and prior distributions are almost identical, e.g. the effective merger rate (top left histogram) has an upper limit of $n_{\rm eff} \sim 1.4(2.1) \times 10^{-4} \text{Mpc}^{-3} \text{Gyr}^{-1}$ for the posterior(prior) respectively.

\subsubsection{Future Upper limit at $A(f=1/{\rm yr})= 1\times 10^{-16}$}

To investigate what useful information on astrophysical observables can be extracted by future improvements of the PTA sensitivity, we have shifted the upper limit down by an order of magnitude to $A(f=1/{\rm yr}) = 1.0 \times 10^{-16}$, indicative of the possible capabilities in the SKA era \citep{2015aska.confE..37J}.

Results are shown in figure \ref{fig:uplim16}. Unlike the current situation, a future upper limit can put significant constraints on the allowed parameter space, also reflected in value of the Bayesian evidence $\log_e \mathcal{Z}(10^{-16}) = -4.32$. The odds ratio compared to a reference model untouched by the limit is now $e^{-4.32}=0.013$, indicating that our astrophysical prior would be disfavoured a $2.5\sigma$ level. The Bayes factor $\mathcal{B} = \exp \mathcal{Z}(10^{-15}) / \exp \mathcal{Z}(10^{-16}) \approx 43$ provides evidence that there is tension between current constraints on astrophysical observables (defining our prior) and a PTA upper limit of $10^{-16}$ on the GWB level. The top left panel of figure \ref{fig:uplim16} shows that $h_c$ is relegated at the bottom of the allowed prior range, and the top right panel indicates that a low normalization to the $M_{\rm BH}-M_{\rm bulge}$ relation is preferred. The bottom row posteriors in figure \ref{fig:uplim16} show significant updates with respect to their prior distributions. A more restrictive upper limit on the effective merger rate (top left histogram) at $n_{\rm eff} \sim 6.3 \times 10^{-5} \text{Mpc}^{-3} \text{Gyr}^{-1}$ can be placed and the distribution of all parameters defining the merger timescale are skewed towards high values, meaning that longer merger timescales, i.e. fewer mergers within the Hubble time, are preferred. Besides favouring lower merger rates, light SMBH are also required, as shown by the posterior of the $M_*$ parameter. Lastly, there is a slight preference for SMBHBs to be very eccentric and in dense stellar environments, although the whole prior range of these parameters is still possible.

\subsubsection{Ideal Upper limit at $A(f=1/{\rm yr})= 1\times 10^{-17}$}

Pushing the exercise to the extreme, we shift the future upper limit down by another order of magnitude to $A(f=1/{\rm yr}) = 1.0 \times 10^{-17}$, which might be reached in the far future by a post-SKA facility \citep{2015aska.confE..37J}. Nonetheless, this unveils what would be the consequences of a severe non detection, well below the level predicted by current SMBHB population models. Figure \ref{fig:uplim17} compares inference on model parameters for the {\textit PPTA17} run, assuming either standard or extended prior distributions.

If we assume standard priors, constraints are pushed to the extreme compared to those derived in the {\textit PPTA16} case. The Bayesian evidence is now $\log_e \mathcal{Z}(10^{-17}) = -13.69$. The odds ratio compared to a reference model untouched by the limit becomes $e^{-13.69}\approx 10^{-6}$, indicating that our astrophysical prior would be severely disfavoured at a $5\sigma$ level. This would rule out the vast majority of our current constraints on the GSMF, pair fraction, merger timescale and $M_{\rm bulge} - M_{\rm BH}$ relation. Although the effective merger rate is only limited to be smaller than $n_{\rm eff} \sim 5.6 \times 10^{-5} \text{Mpc}^{-3} \text{Gyr}^{-1}$, all other parameters in the bottom row corner plots in figure \ref{fig:uplim17} show rather extreme posterior distributions. Since our standard prior does not allow stalling of low redshift SMBHBs (the maximum normalization of the local merger timescale being 2 Gyrs), skewing the merger timescale to extreme values is not sufficient to explain the non detection. Further, the normalization to the $M_{\rm BH}-M_{\rm bulge}$ is severely pushed to the low end, at $M_* < 10^8 M_\odot$, thus completely ruling out several currently popular relations \citep[e.g.,][]{2013ARA&A..51..511K,2013ApJ...764..184M}. Even with the smallest possible $M_{\rm BH}-M_{\rm bulge}$, a non detection at $A(f=1/{\rm yr}) = 1.0 \times 10^{-17}$ requires a very high frequency turnover of the GWB (see upper left panel of figure \ref{fig:uplim17}), which can be realized only if all binaries have eccentricity $e_0>0.95$ and reside in extremely dense environments (at least a factor of 10 larger than our fiducial Dehnen profile). 

As mentioned above, our standard prior on the total merger timescale (see Section \ref{sec:mtime}), implies that stalling hardly occurs in nature. Although this is backed up by recent progresses in N-body simulations and the theory of SMBHB hardening in stellar environments \citep[see, e.g.,][]{2015MNRAS.454L..66S,2015ApJ...810...49V}, we want to keep all possibilities open and check what happens when arbitrary long merger timescales, and thus stalling, are allowed. We note, however, that such a model is intrinsically inconsistent, because when very long merger timescales are allowed, one should also consider the probable formation of SMBH triplets, due to subsequent galaxy mergers. Triple interactions are not included in our models but they have been shown \citep{2018MNRAS.477.2599B,2018MNRAS.473.3410R} to drive about 1/3 to the stalled SMBHBs to coalescence in less than 1 Gyr. Therefore, we caution that actual constrains on model parameters would likely be more stringent than what described in the following.
The extended prior distributions relaxes the strong evidence of $-13.69$ to $\log_e \mathcal{Z}(10^{-17}) = -4.56$ and the Bayes factor becomes comparable to the {\textit PPTA16}, this is mainly due to allowing binaries to stall as the merger timescale increases to $\tau_0 > 5.5$ Gyr. The extreme constraints on the other parameters are consequently loosened, although posterior distributions of $M_*, e_0, \zeta_0$ indicate that light SMBHBs are favoured, along with large eccentricities and dense environments. The stalling of a substantial fraction of SMBHB pushes the effective merger rate to drop below $n_{\rm eff} \sim 1.1 \times 10^{-5} \text{Mpc}^{-3} \text{Gyr}^{-1}$.

Table \ref{table:ev} summarizes the increasing constraining power as the upper limits are lowered. As they become more restrictive, fewer mergers are allowed. The effective merger rate is therefore pushed to be as low as possible with long merger timescales, low SMBHB masses, large eccentricities and dense environments. Bayes factors comparing the current observational constraints, i.e. the prior ranges, with posterior constraints can be calculated from the evidences. These, however, show that the tension increases from $0.6\sigma$ with the current upper limit of $A(f=1/{\rm yr})= 1\times 10^{-15}$ to $5\sigma$ with an ideal upper limit at $A(f=1/{\rm yr})= 1\times 10^{-17}$. Relaxing the upper bound on the merger time norm and other constraints (see Section \ref{sec:ext}) can alleviate the tension between current observations and such a upper limit to $2.6\sigma$ (although this does not take into account for triple-induced mergers, as mentioned above).

\begin{table}
\begin{center}
\def\arraystretch{1.5}
\begin{tabularx}{0.475\textwidth}{c|XXX}
\hline
parameter & $\log_{10} n_\mathrm{eff}$ & $\tau_0$ & $\log_e \mathcal{Z}$ \\
\hline
standard prior: no upper limit         & $<-3.68$ & $>0.2$  & 0 \\
standard prior: $A = 1\times 10^{-15}$ & $<-3.85$ & $>0.28$ & -0.55 \\
standard prior: $A = 1\times 10^{-16}$ & $<-4.2$  & $>0.75$ & -4.32 \\
standard prior: $A = 1\times 10^{-17}$ & $<-4.25$ & $>0.94$ & -13.69 \\
extended prior: $A = 1\times 10^{-17}$ & $<-4.96$ & $>5.5$  & -4.56 \\
\hline
\end{tabularx}
\caption{List of bounds for selected parameters and evidences for the upper limit cases. The 95\% upper bounds for the effective merger rate, the 95\% lower bounds for the merger time norm and the evidences are reported in the columns. The five rows list the values for the standard prior, current, future and ideal upper limit posteriors from top to bottom.}
\label{table:ev}
\end{center}
\end{table}

\subsection{Simulated detections}

Although it is useful to explore the implication of PTA upper limits, it is more interesting to consider the case of a future detection, which is expected within the next decade \citep{2015MNRAS.451.2417R,2016ApJ...819L...6T,2017MNRAS.471.4508K}. We therefore turn our attention at simulated detections and their potential to put further constraints on the astrophysics of galaxy evolution and SMBHB mergers. To simulate a detection, the GWB strain is computed for a specific set of parameters, i.e. the injected signal, which is detected at the computed values $A_k=h_c(f_k)$ with an uncertainty of $\sigma_{\mathrm{det},k}$ given by equation \eqref{eqrhotot}. As these simulated detections are very ideal, effects that could pollute the strength the GWB detection are mostly neglected. However, we include an empirical term in the computation of $S_n$ (see equations 9 and 10 in paper II) to account for the flattening of the sensitivity at low frequencies.

The amplitude of the simulated GWB is defined by the 16 parameters describing the SMBHB merger rate. We fix those as follows: $(\Phi_0, \Phi_I, M_0, \alpha_0, \alpha_I, f_0, \alpha_f, \beta_f, \gamma_f, \tau_0, \alpha_\tau, \beta_\tau, \gamma_\tau, M_*, \alpha_*, \epsilon) =$ (-2.6, -0.45, $10^{11.25}$, -1.15, -0.1, 0.025, 0.1, 0.8, 0.1, 0.8, -0.1, -2, -0.1, $10^8$, 1, 0.3). The low frequency turnover is defined by the two extra parameters $(e_0, \zeta_0)$. We fix $\zeta_0=1$ and we produce two GWB spectra distinguished solely by the assumed value of the eccentricity: $e_i=0.01$ (circular case) and $e_i=0.9$ (eccentric case). This set of parameters is chosen such that it results in a GWB strain of $h_c = 5.0 \times 10^{-16}$ at $f=1/{\rm yr}$ (i.e. well within current upper limits), whilst being consistent with the current constraints of all the relevant astrophysical observables:

\begin{itemize}

\item GSMF: the values for $(\Phi_0, \Phi_I, M_0, \alpha_0, \alpha_I)$ are chosen, such that they accurately reproduce the currently best measured GSMF, i.e., they are close to the best fit values of the re-parametrisation described in Section \ref{sec:gsmfprior};

\item merger timescale: $\tau_0=0.8$ Gyr is chosen to match the predicted merger timescales found in simulations by \cite{lotz10}, while $\beta_\tau=-2$ is chosen to match the expected redshift evolution of the merger timescale from \cite{2017MNRAS.468..207S};

\item $M_{\rm bulge} - M_{\rm BH}$ relation: $(M_*, \alpha_*, \epsilon)$ have been chosen to produce the injected characteristic strain amplitude, consistent with the allowed prior shown in figure \ref{fig:mbulge}.

\end{itemize}

The other parameters are chosen to be close to the centre of their prior ranges, except for the eccentricity, as mentioned above.

\begin{figure*}
\begin{subfigure}{0.4\textwidth}
\centering
$IPTA30, \ e_t = 0.01$
\includegraphics[width=5.5cm]{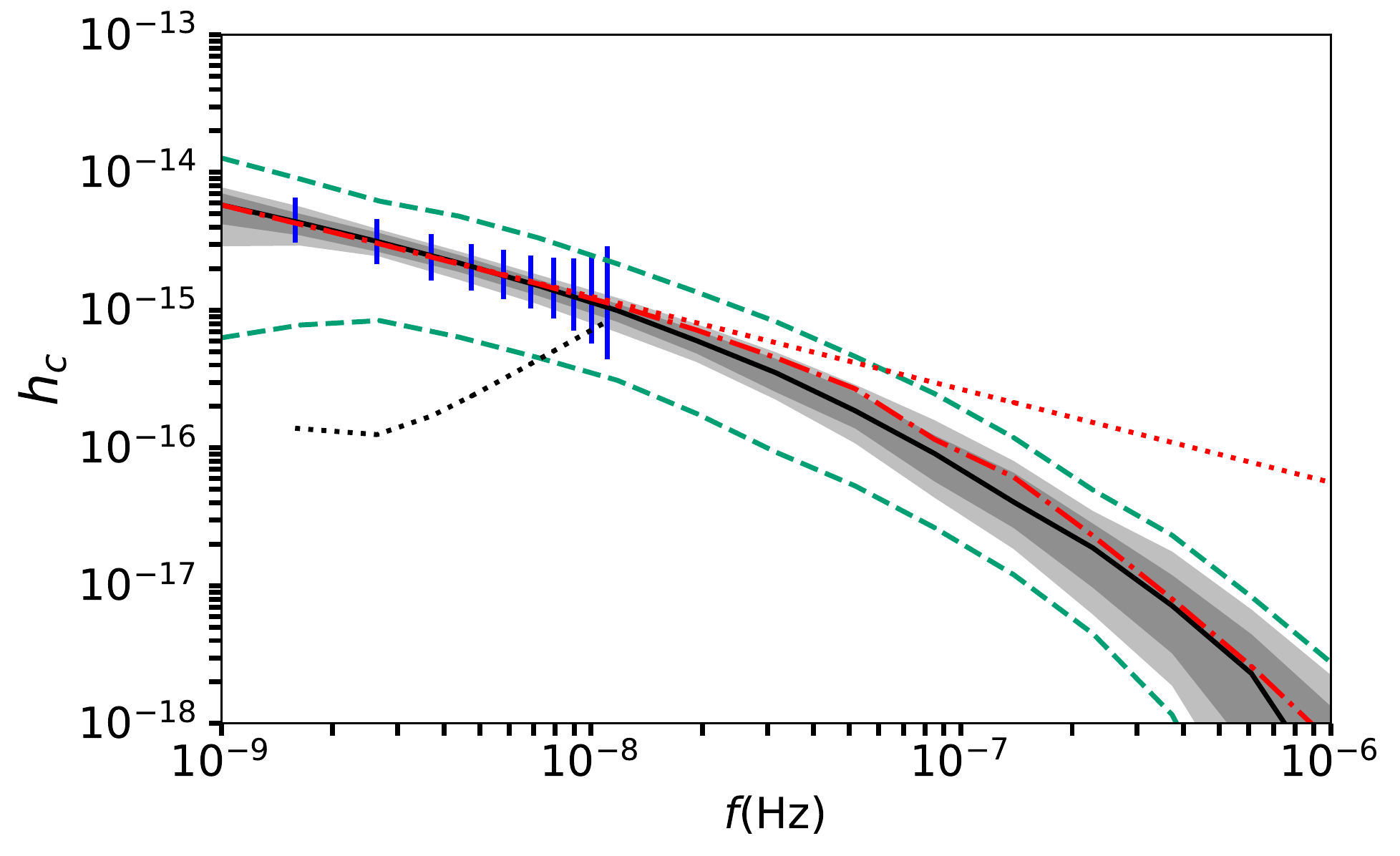}\\
\includegraphics[width=5.5cm]{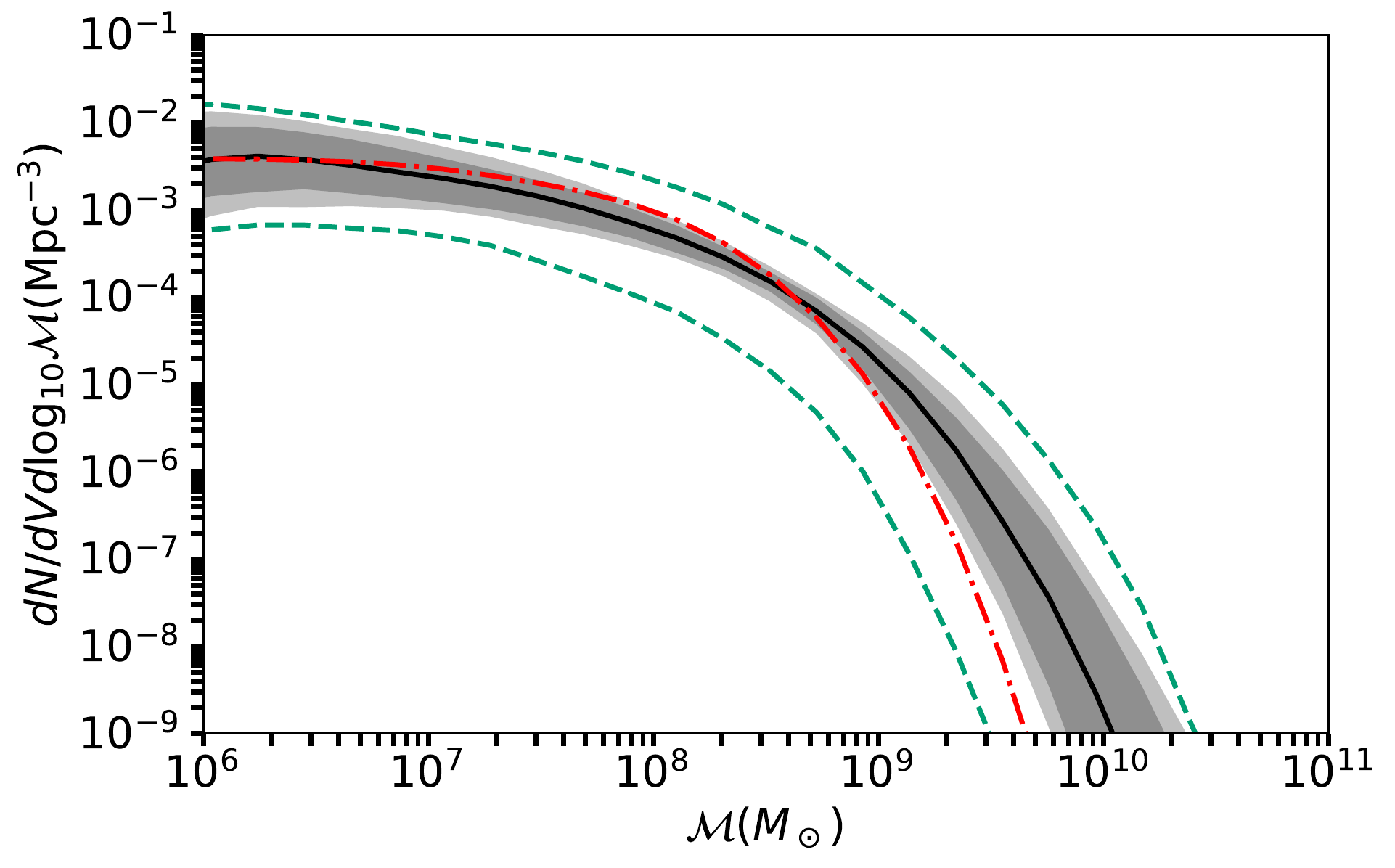}\\
\includegraphics[width=5.5cm]{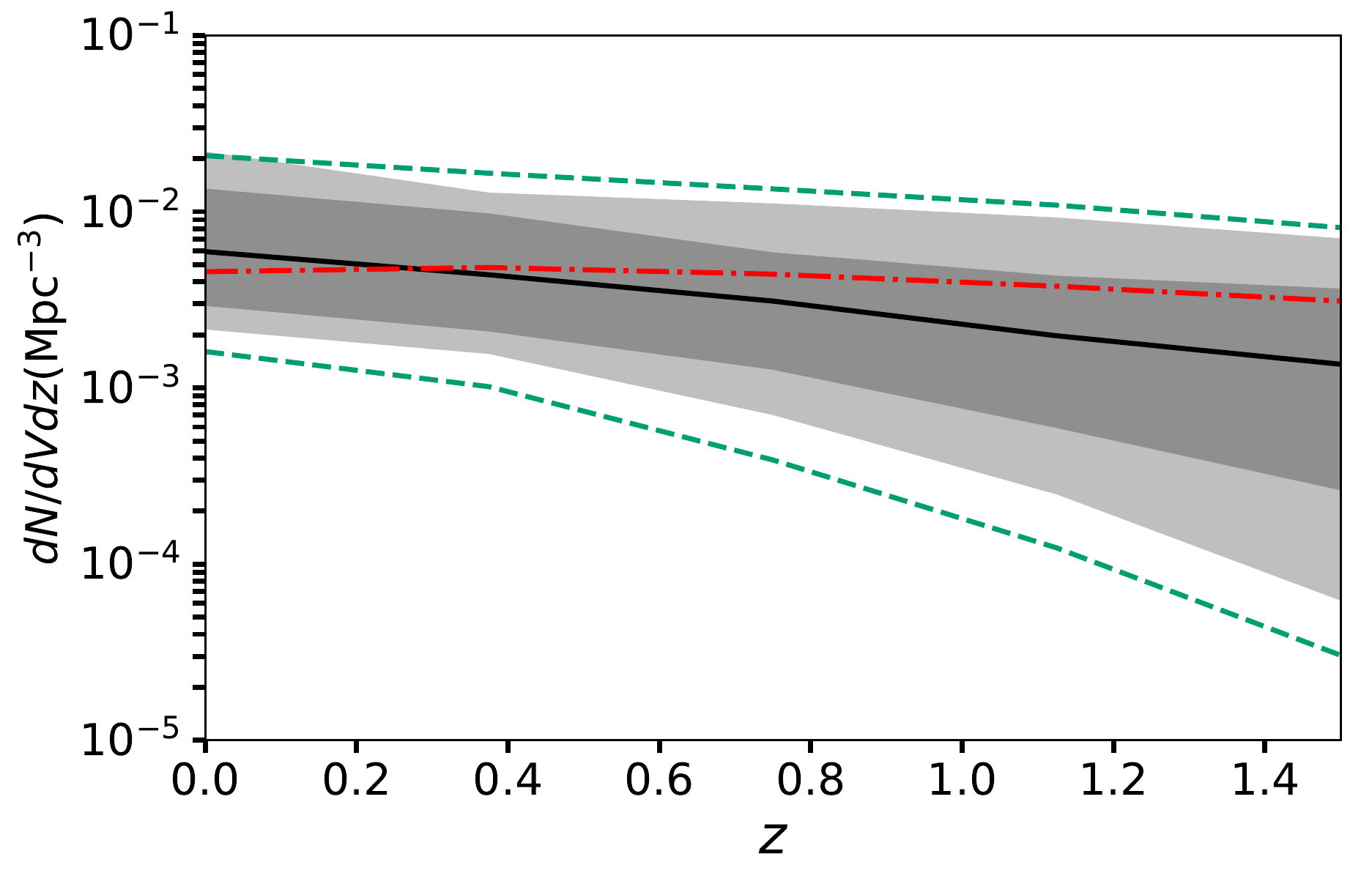}\\
\includegraphics[width=5.5cm]{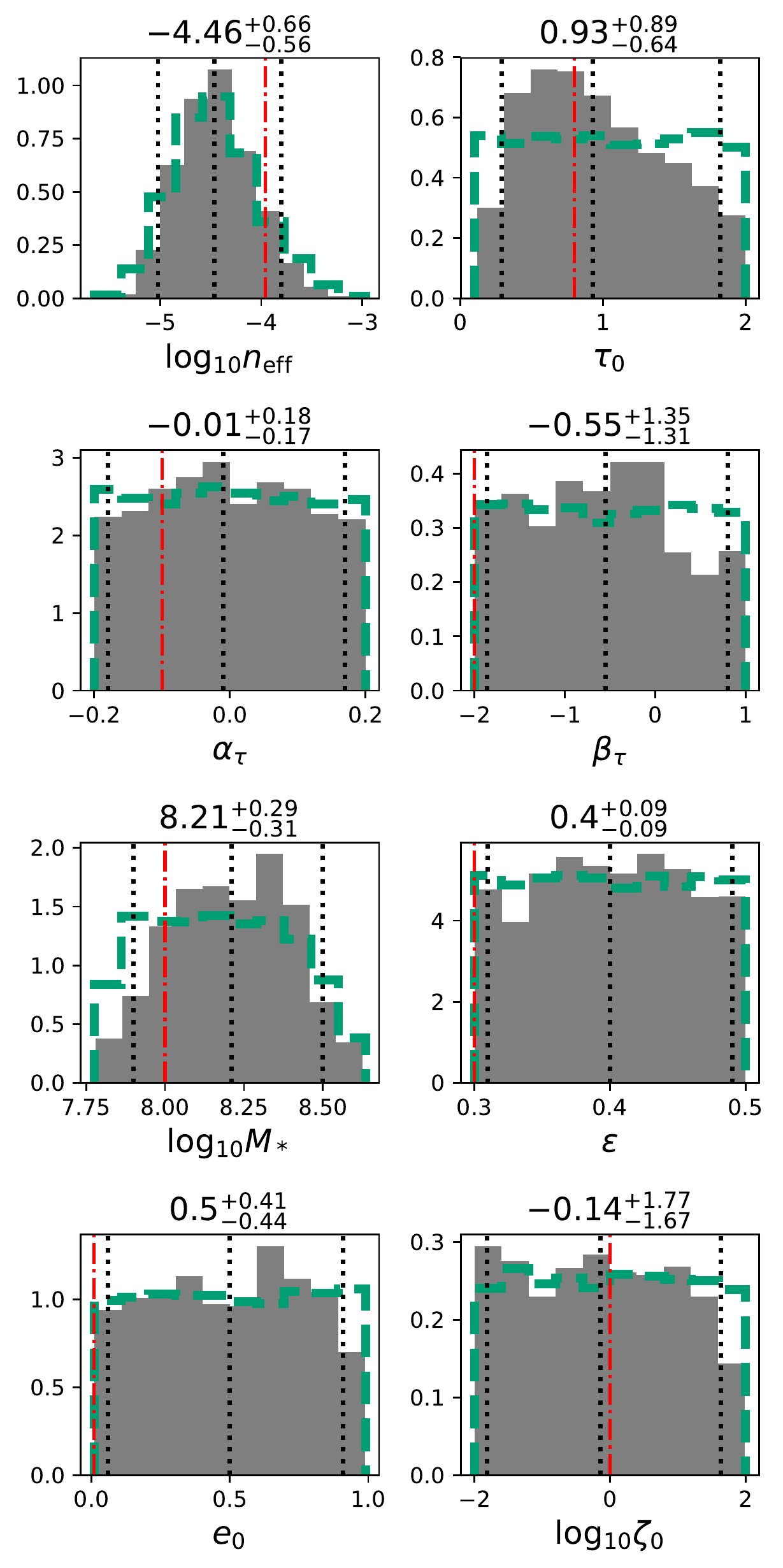}
\end{subfigure} \hspace{1cm}
\begin{subfigure}{0.4\textwidth}
\centering
$SKA20, \ e_t = 0.01$
\includegraphics[width=5.5cm]{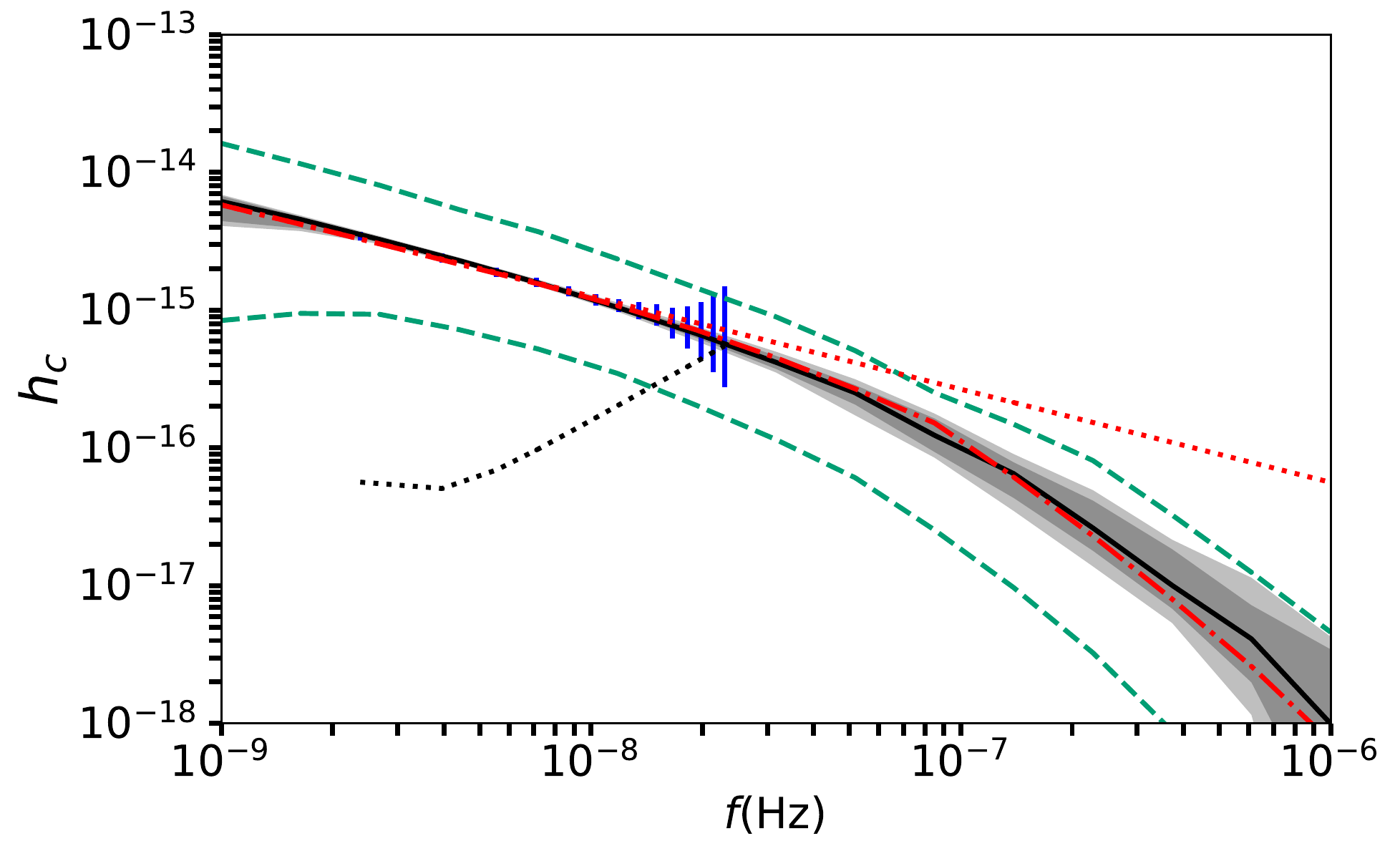}\\
\includegraphics[width=5.5cm]{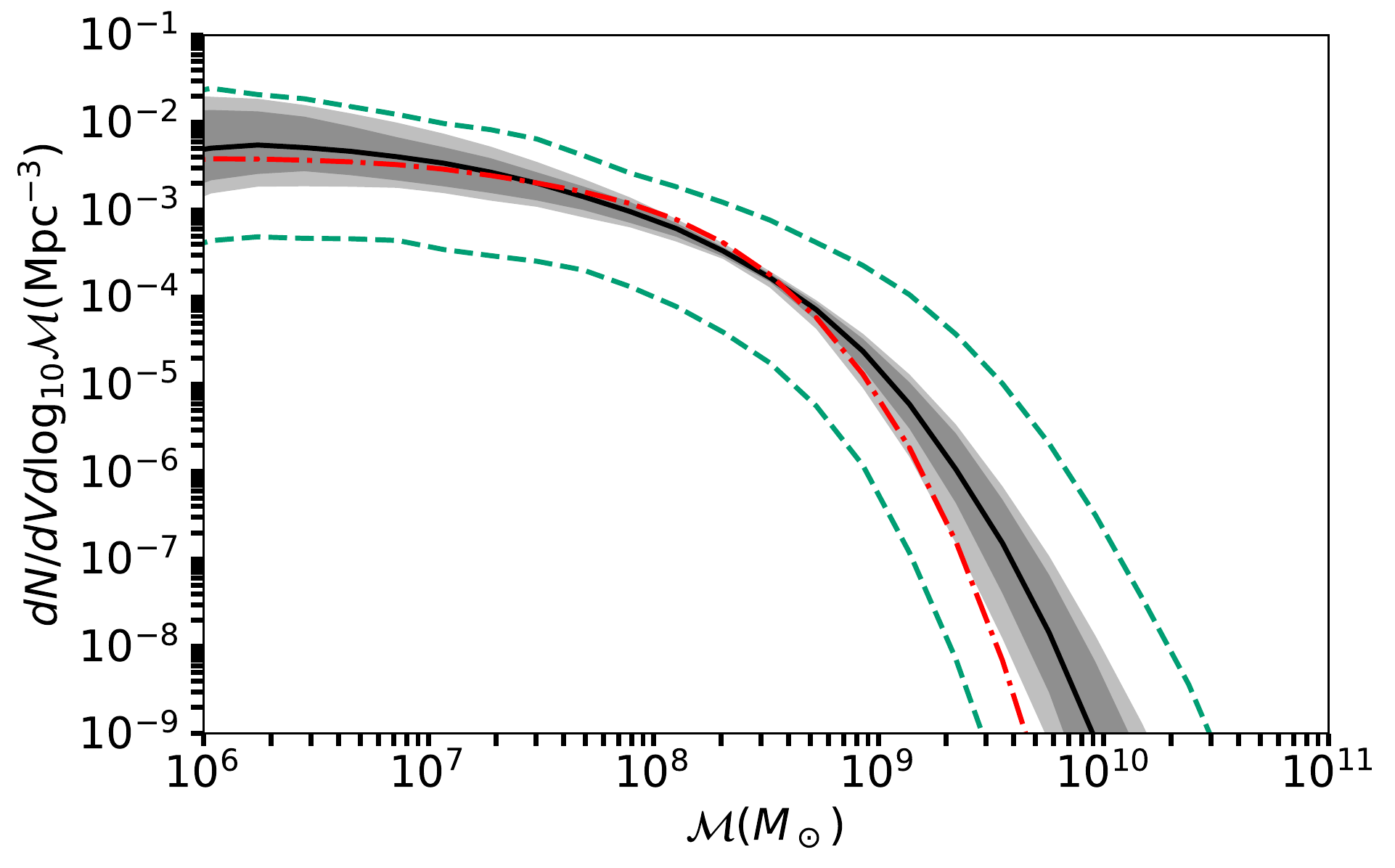}\\
\includegraphics[width=5.5cm]{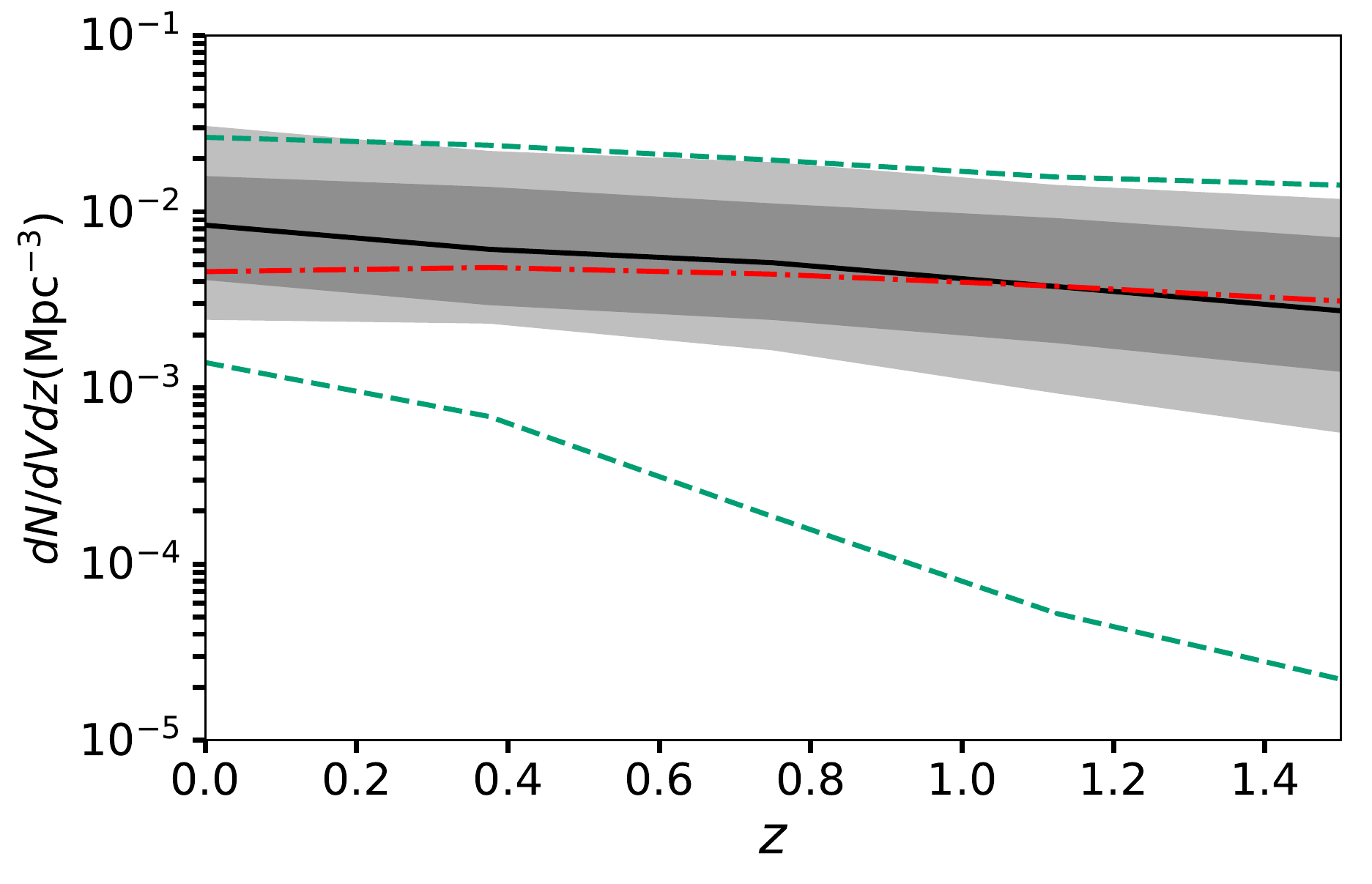}\\
\includegraphics[width=5.5cm]{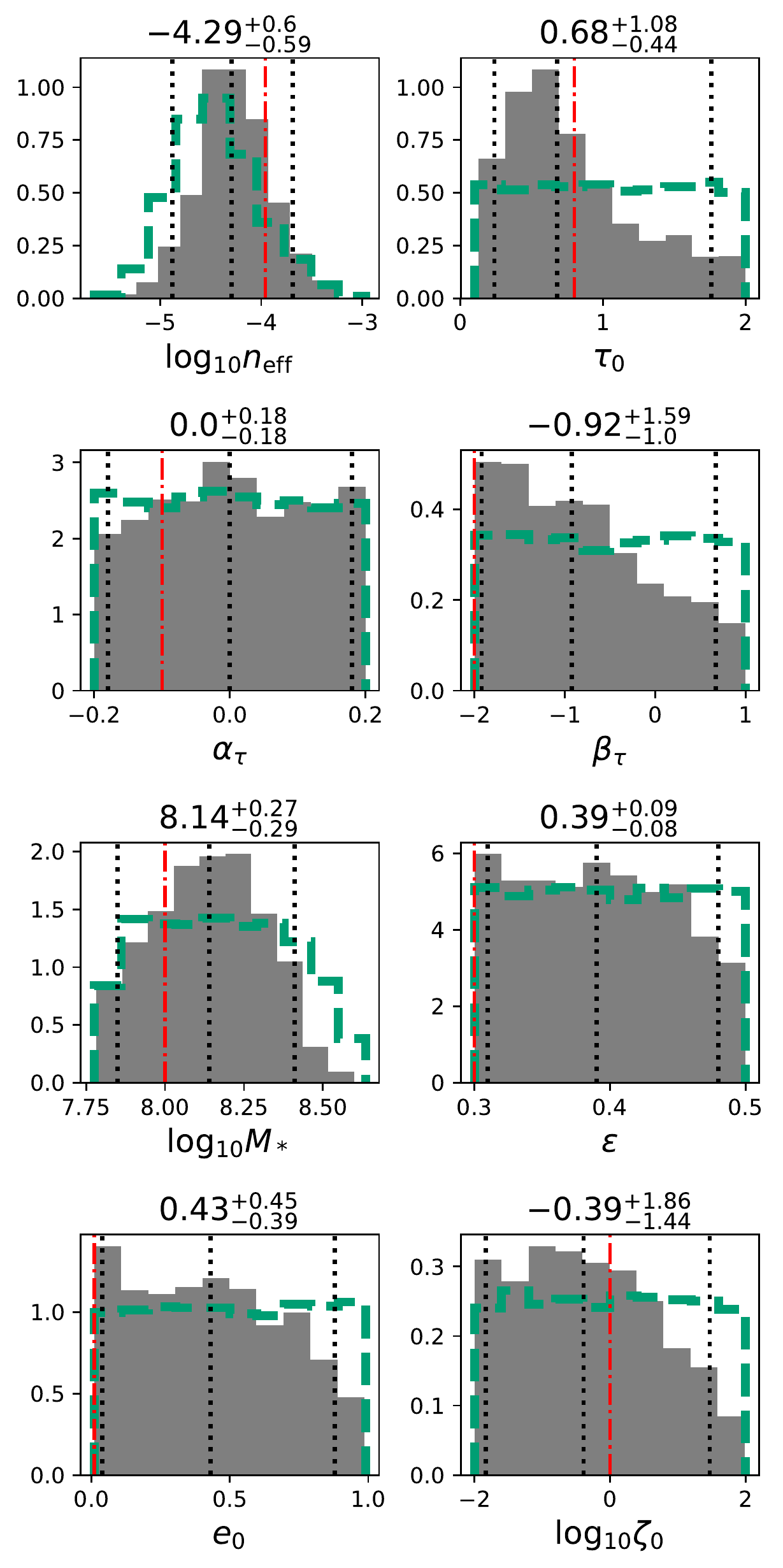}
\end{subfigure} \\
\caption{Implications of a PTA detection at a low (S/N$\approx$20, left column) and moderate (S/N$\approx$100, right column) significance, assuming a SMBHB population with default parameters and almost circular ($e_t=0.01$) at decoupling. As in figure \ref{fig:uplim15}, the posteriors for the spectrum, mass and redshift functions (in descending order from the top) are shown as shaded areas. In each of those panels, the dashed green lines indicate the prior and the dash--dotted red line indicates the injected model. In the top panels the vertical blue bands indicate the 68\% confidence interval of the observed signal amplitude at each frequency bin. The dotted line is the nominal $1\sigma$ sensitivity of the considered PTA. The dotted red line shows the simulated spectrum assuming no drop due to missing sources at high frequencies. The bottom row histograms in shades of grey show the marginalized posteriors for selected model parameters with the prior distributions indicated by dashed green lines. The injected parameter values are marked by red dash--dotted lines.}
\label{fig:detection_circ}
\end{figure*}
\begin{figure*}
\begin{subfigure}{0.4\textwidth}
\centering
$IPTA30, \ e_t = 0.9$
\includegraphics[width=6.0cm]{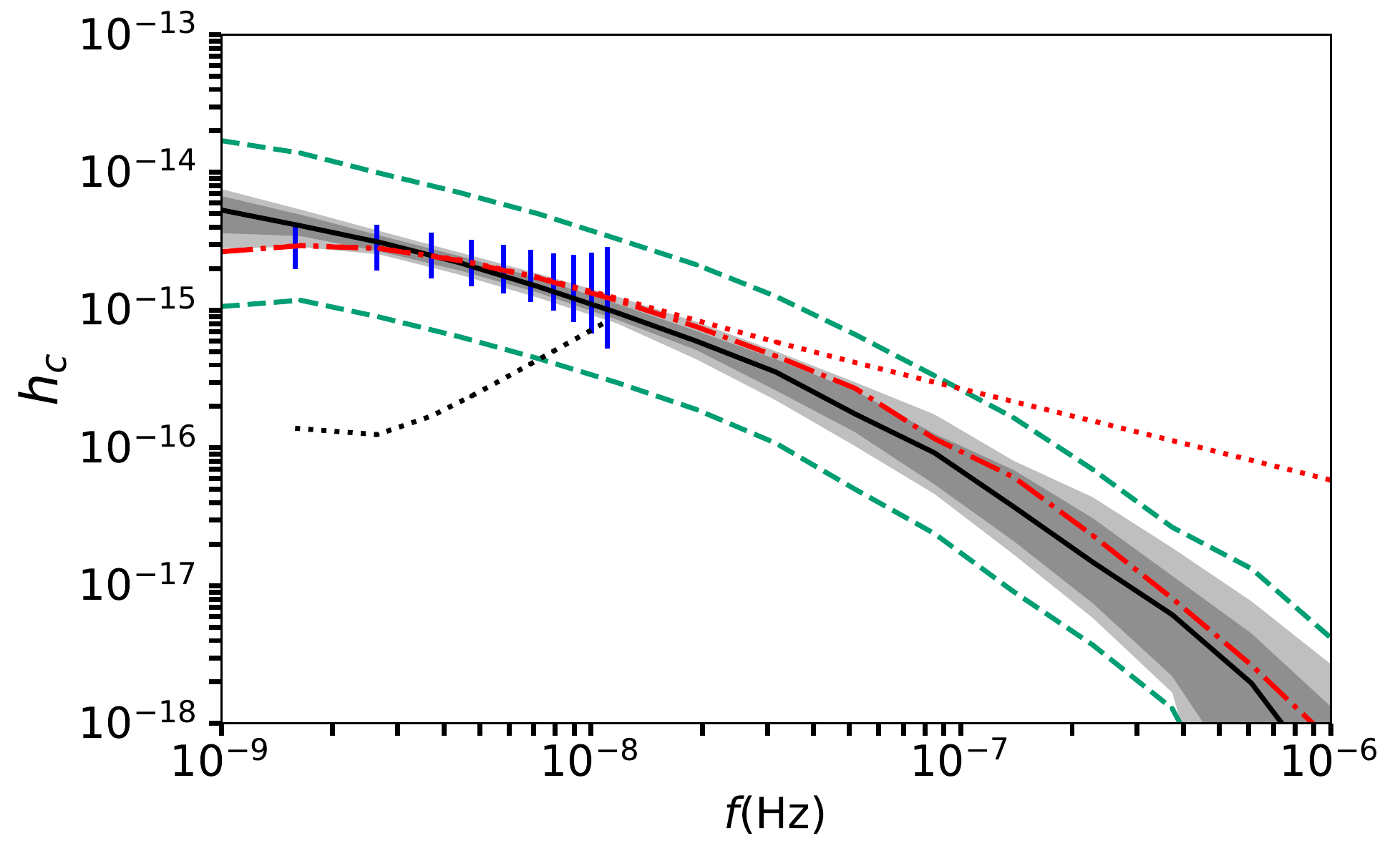}\\
\includegraphics[width=6.0cm]{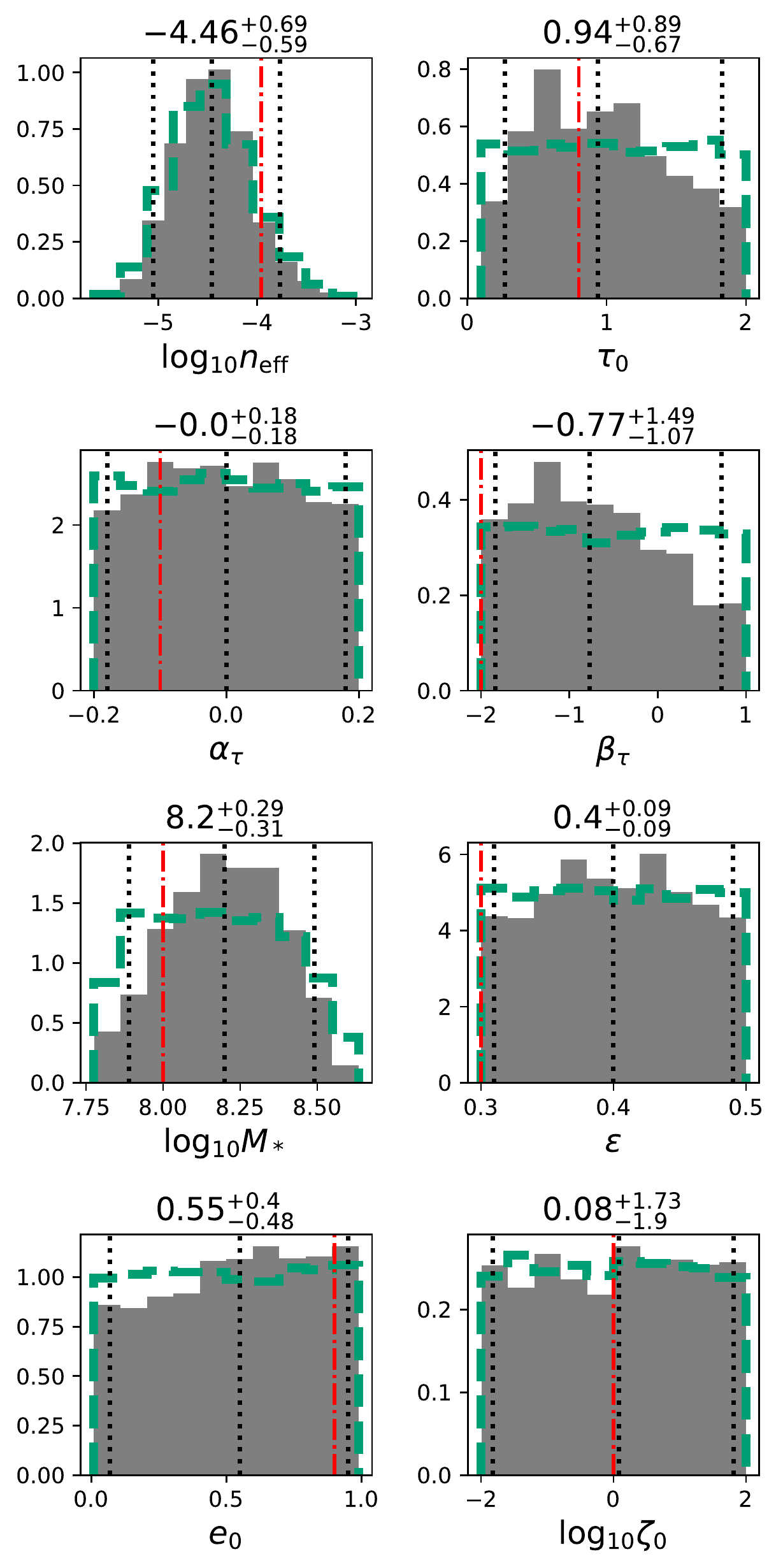}
\end{subfigure} \hspace{1cm}
\begin{subfigure}{0.4\textwidth}
\centering
$SKA20, \ e_t = 0.9$
\includegraphics[width=6.0cm]{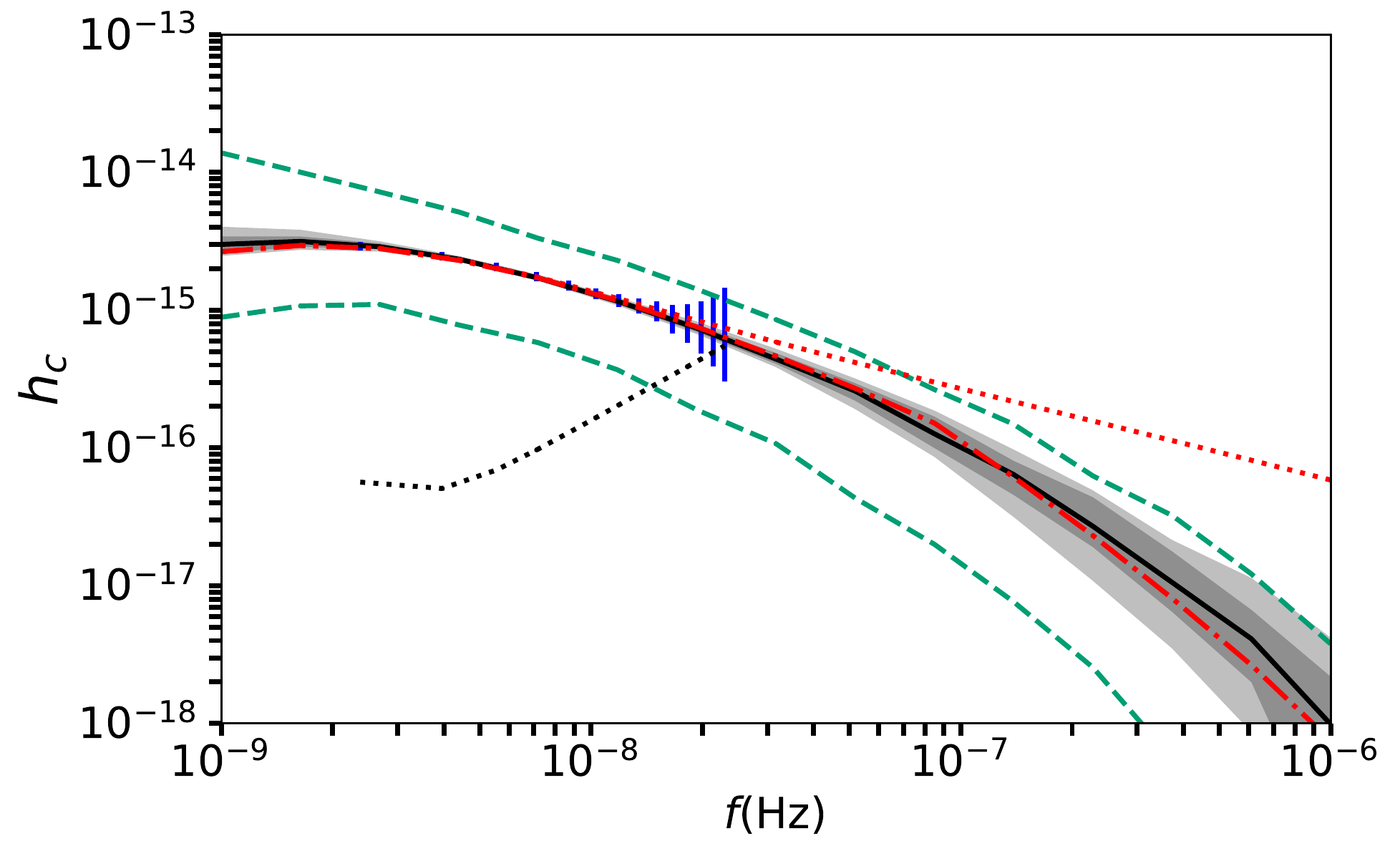}\\
\includegraphics[width=6.0cm]{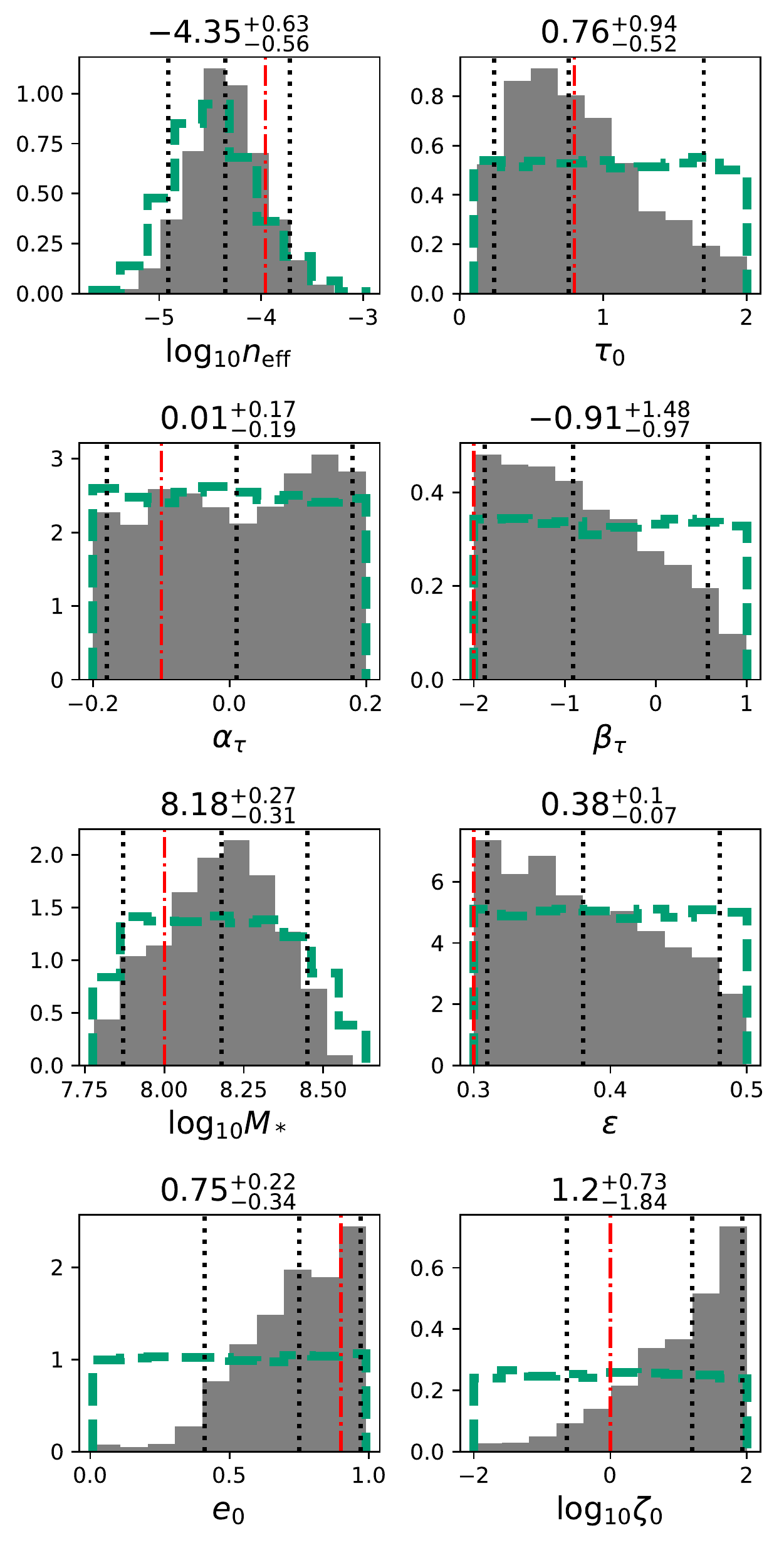}
\end{subfigure} \\
\caption{Same as figure \ref{fig:detection_circ} but assuming decoupling eccentricity of $e_t=0.9$. The posterior panels of the mass and redshift functions are analogue to those in the circular case and thus not shown here.}
\label{fig:detection_ecc}
\end{figure*}

\subsubsection{Circular case}

Figure \ref{fig:detection_circ} shows a comparison of the results of the \textit{IPTA30} (left column) and \textit{SKA20} (right column) setups for the circular case ($e_0=0.01$). In the \textit{IPTA30}(\textit{SKA20}) case the GWB has been detected in 10(14) frequency bins up to frequencies of $\sim 1(2) \times 10^{-8}$ Hz, for a total detection S/N ${\mathcal S}\approx$ 20(100). Qualitatively, both detections provide some extra constraints on selected prior parameters. The injected spectrum, mass and redshift function are recovered increasingly better as the S/N increases. Still, a broad portion of the initial parameter space is allowed, especially for the redshift evolution of the SMBHB merger rate. It should be noted that PTAs have the most constraining power around the bend of the mass function, at the SMBHB chirp mass $\mathcal{M} \approx 3 \times 10^8 \Msol$. The posterior panels at the bottom of figure \ref{fig:detection_circ} show that there is not much additional information gained compared to the prior knowledge for most of the parameters (full corner plots shown in Appendix A, available in electronic form), with three notable exceptions:

\begin{enumerate}

\item merger timescale. $\tau_0$ is marginally constrained around the injected value (0.8 Gyrs) in the \textit{IPTA30} case, the constraint becomes better in the \textit{SKA20} case. $\beta_\tau$ is also skewed towards low values (consistent with the $\beta_\tau=-2$ injection). A clean PTA detection thus potentially allow to constrain the timescale of SMBHB coalescence, which can help in understanding the processes driving the merger;

\item $M_{\rm bulge} - M_{\rm BH}$ relation. The $M_*$ panels show a tightening of the $M_*$ distribution with increasing S/N. A detection would thus also allow to constrain the $M_{\rm bulge} - M_{\rm BH}$ relation;

\item eccentricity and stellar density. The posterior distributions for $e_0$ and $\log_{10} \zeta_0$ show some marginal update. In particular  in the \textit{SKA20} case, extreme eccentricities, above $e_0 > 0.9$ can be safely ruled out. Note that the absence of a low frequency turnover also favours small value of $\zeta_0$, fully consistent with the injected value $\zeta_0=1$.

\end{enumerate}

\subsubsection{Eccentric case}

The results for the \textit{IPTA30} and \textit{SKA20} eccentric cases are shown in figure \ref{fig:detection_ecc}, with full corner plots reported in Appendix A, available in electronic form. In general results are comparable to the circular case shown above, as the only difference is in the injected eccentricity parameter. The left column (\textit{IPTA30} case) of figure \ref{fig:detection_ecc} shows nearly identical posterior distributions to its circular counterpart reported in figure \ref{fig:detection_circ}, this also translates into similar recovered spectrum, mass and redshift functions.

However, in the \textit{SKA20} case, the detection S/N is high enough to allow a clear detection of the spectrum turnover in the lowest frequency bins. Which is not the case for \textit{IPTA30}, as can be seen in the top row spectra plots of figure \ref{fig:detection_ecc}. This has important consequences for astrophysical inference since an observable turnover is only possible if binaries are significantly eccentric and/or evolve in very dense environments. This is shown in the $e_0$ and $\zeta_0$ posterior distributions at the bottom right of figure \ref{fig:detection_ecc}: eccentricities $e_0<0.4$ are excluded and densities higher than what predicted by the fiducial Dehnen model are strongly favoured. The full corner plot \ref{fig:cornerplot_uplim15_full} reported in Appendix A, available in electronic form, also highlights the $e_0 - \zeta_0$ degeneracy, as a low frequency turnover can be caused by either parameters; very eccentric binaries in low density stellar environments pruduce a turnover at the same frequency as more circular binaries in denser stellar environments. Additionally, a large region in the $e_0 - \zeta_0$ plane has been ruled out($e_0 > 0.41$ and $\log_{10} \zeta_0 > -0.63$). This also prompts some extra constrain in the $M_{\rm BH} - M_{\rm bulge}$ relation, as can be seen in the trends in the $\alpha_*$ and $\epsilon$ distributions.

Summarizing, little extra astrophysical information (besides the non-trivial confirmation that SMBHBs actually do merge) can be extracted in the \textit{IPTA30}, whereas many more interesting constrains emerge as more details of the GWB spectrum are unveiled in the \textit{SKA20} case. Although posteriors on most of the parameters remain broad, the typical SMBHB coalescence timescale can be constrained around the injected value; the posterior distributions of $n_{\rm eff}$ and $M_*$ are tightened, providing some extra information on the SMBHB merger rate and on the $M_{\rm BH} - M_{\rm bulge}$ scaling relation; significant constrains onto the SMBHB eccentricity and immediate environment can be placed if a low frequency turnover is detected.

\subsubsection{Ideal case}

\begin{figure*}
\begin{subfigure}{0.4\textwidth}
\centering
$Ideal, \ e_t = 0.01$
\includegraphics[width=7.0cm]{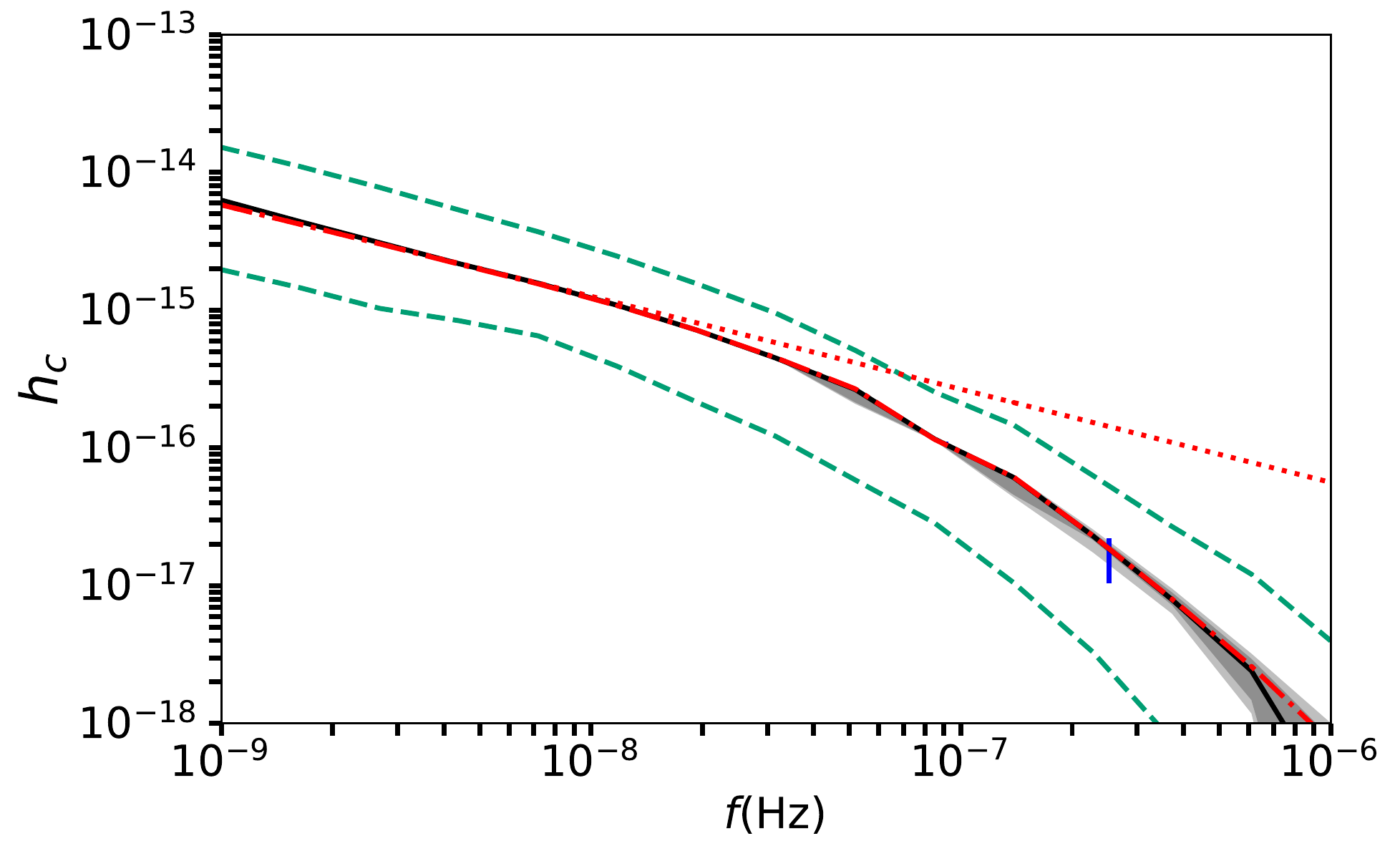}\\
\includegraphics[width=8.0cm]{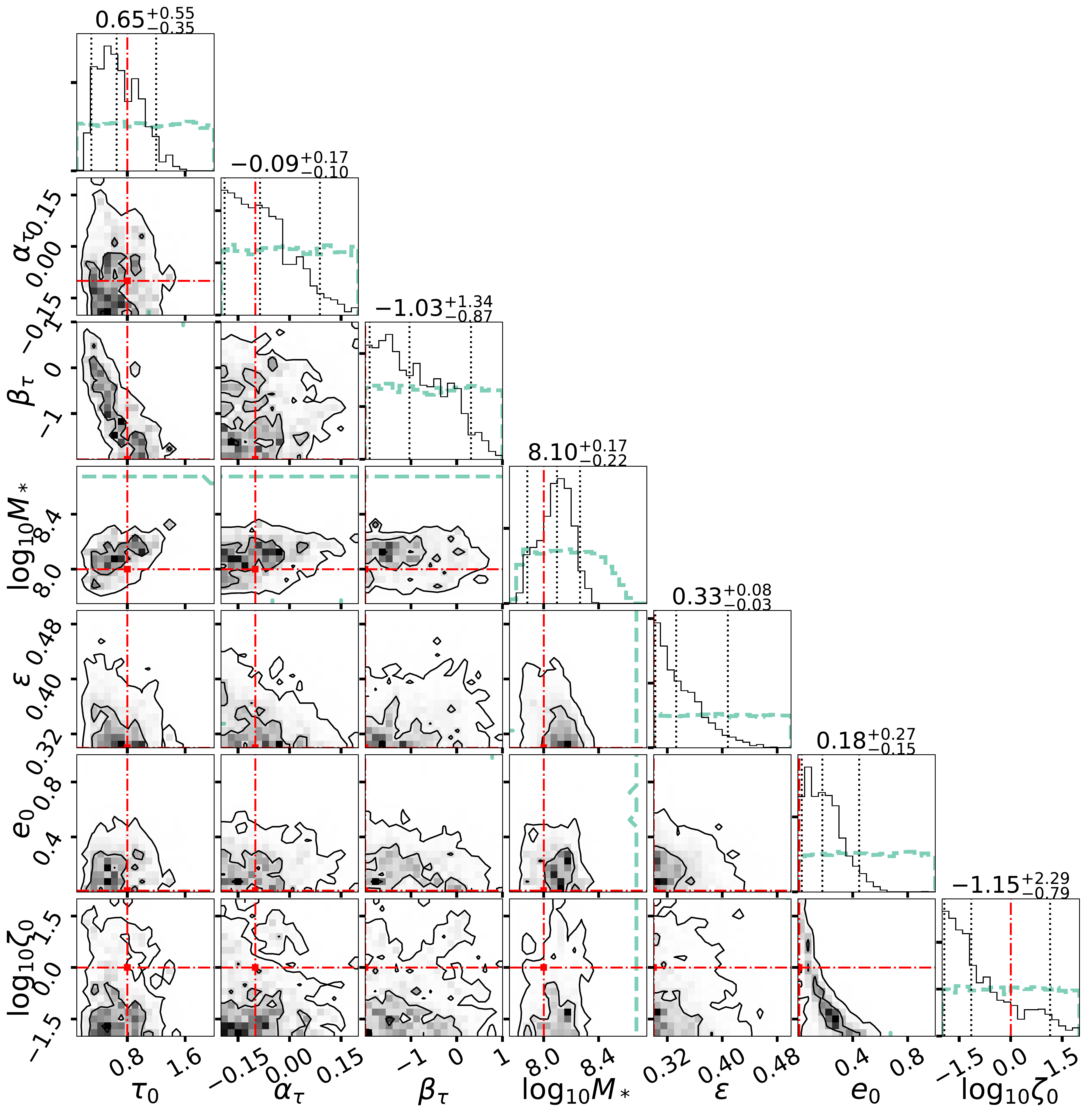}
\end{subfigure} \hspace{1cm}
\begin{subfigure}{0.4\textwidth}
\centering
$Ideal, \ e_t = 0.9$
\includegraphics[width=7.0cm]{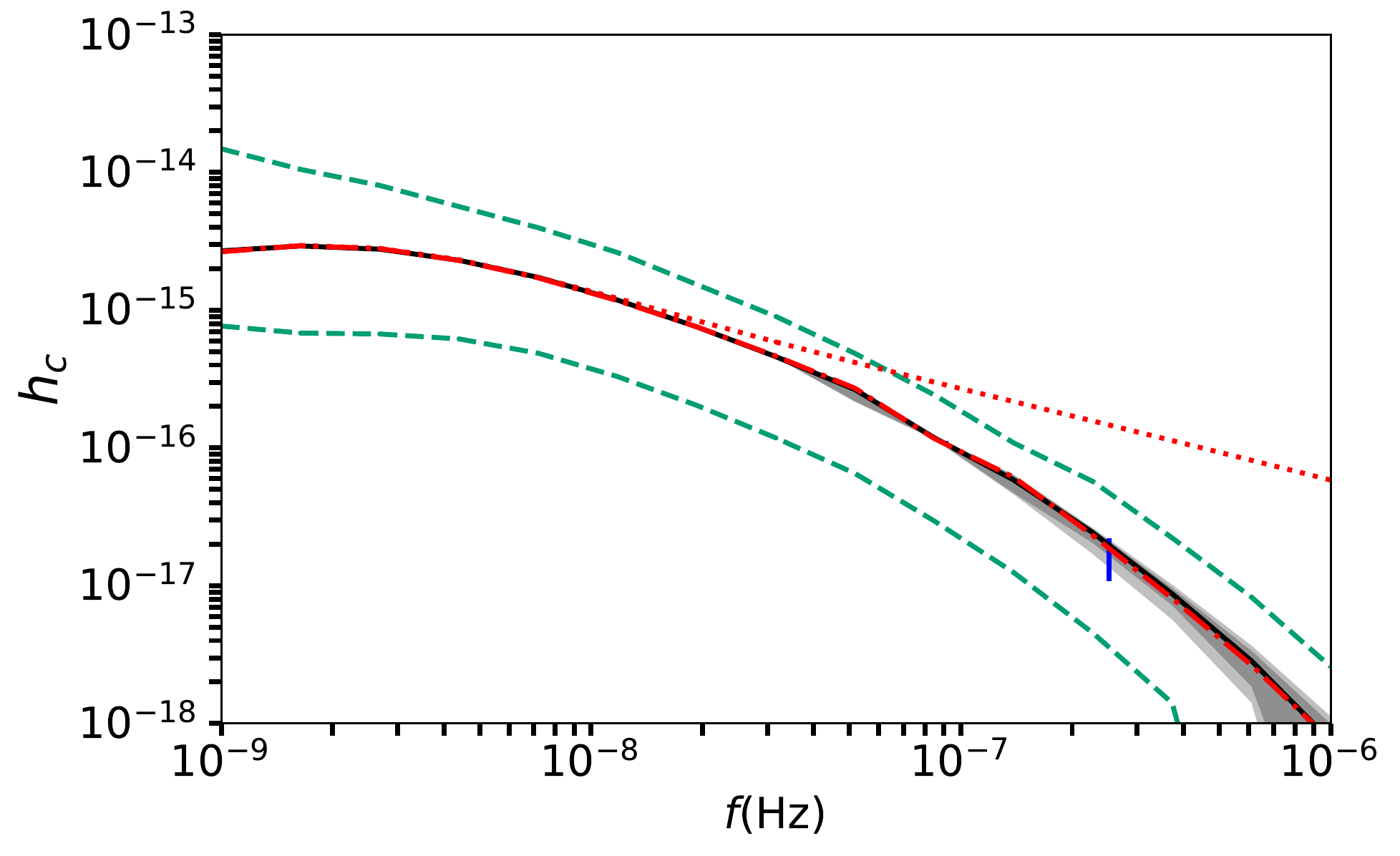}\\
\includegraphics[width=8.0cm]{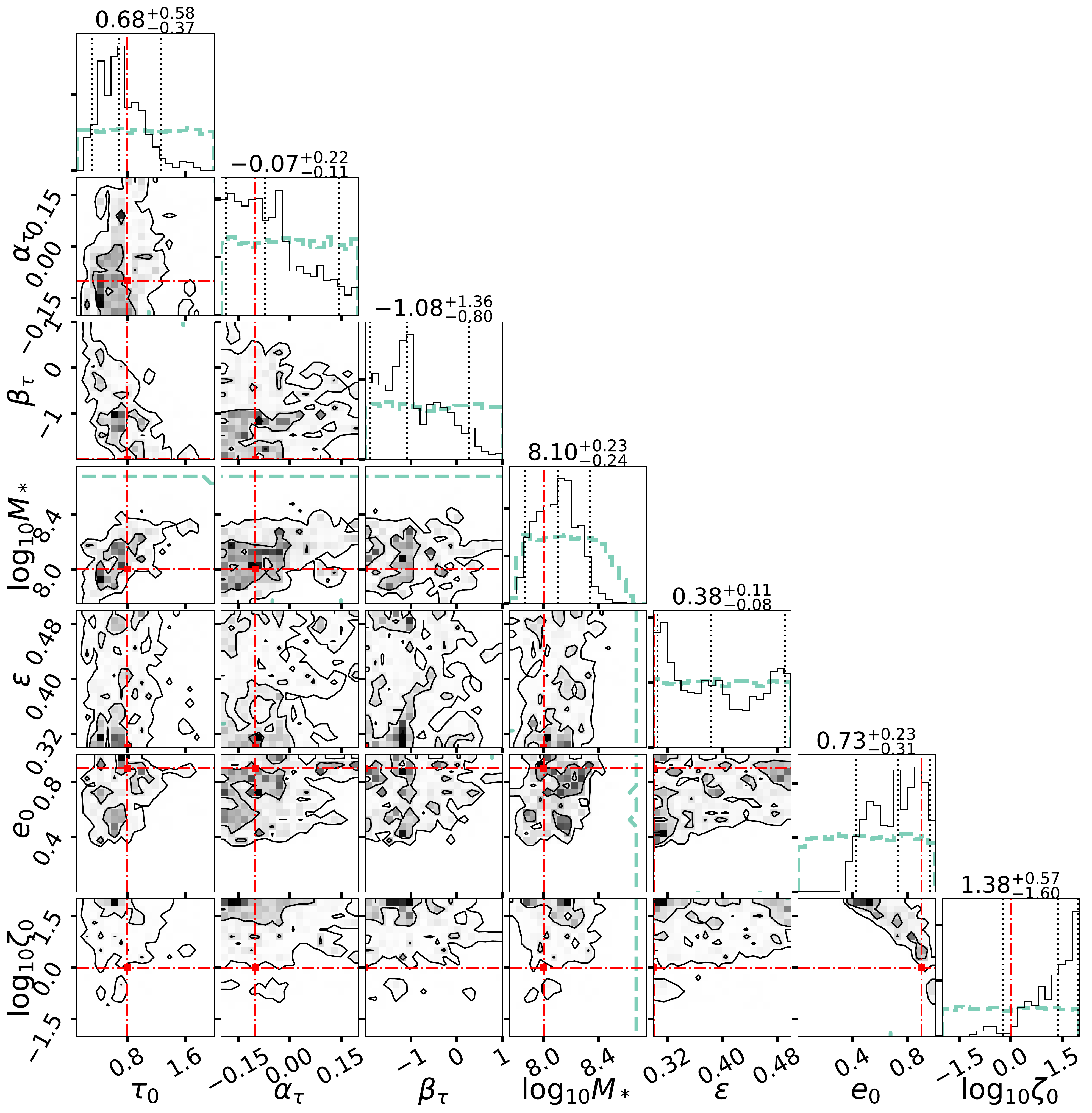}
\end{subfigure} \\
\caption{Implications of an ideal detection with 500 MSPs timed at sub-ns precision for 30 years. The injected model has default parameters with $e_t=0.01$ (left column) and $e_t=0.9$ (right column). Panel sequence and style as in figure \ref{fig:uplim17}.}
\label{fig:detection_ideal}
\end{figure*}

We show ideal detections for both the circular and eccentric cases in figure \ref{fig:detection_ideal}. Although, such detection may not be achievable by PTAs in the foreseeable future, these results show what might be constrained in principle by combining astrophysical prior knowledge to precise measurements of the amplitude and shape of the nano-Hz stochastic GWB.

The spectra, mass and redshift functions (not shown in the figure) are recovered extremely well in both cases. Both corner plots also show interesting constrains on some key parameters. The typical merger timescale $\tau_0$ is correctly measured and constrained within less than 1 Gyr uncertainty, and clear trends in $\alpha_\tau$ and $\beta_\tau$ provide some extra information on the merger timescale evolution with galaxy mass and redshift. Note that those are parameters defining the {\it SMBHB} coalescence time which are unlikely to be measured by any other means. The normalization of the $M_{\rm bulge} - M_{\rm BH}$ relation is also significantly constrained, as shown by the tight $M_*$ posterior distributions. Again we see both in the circular and eccentric cases the degeneracy between eccentricity $e_0$ and stellar density $\zeta_0$, as in the \textit{SKA20} eccentric case above. The posterior regions contain the injected values and exclude a large area from the prior: $e_0 > 0.42$, $\log_{10} \zeta_0 > -0.22$ for the circular and $e_0 < 0.45$, $\log_{10} \zeta_0 < 1.14$ for the eccentric case (95th percentile). Although, the \textit{ideal} eccentric detection has a vastly larger S/N than its \textit{SKA20} analogue, the constraints on $e_0$ and $\log_{10} \zeta_0$ are comparable due to the degeneracy between the two parameters. Table \ref{table:pos} shows the increasing constraining power on selected key parameters as the detection S/N improves for the eccentric case.

\section{Conclusions and outlook}
\label{sec:Conclusions}

\begin{table}
\begin{center}
\def\arraystretch{1.5}
\begin{tabularx}{0.475\textwidth}{X|XXXXX}
\hline
parameter & $\log_{10} n_\mathrm{eff}$ & $\tau_0$ & $\log_{10} M_*$ & $e_0$ & $\log_{10} \zeta_0$\\
\hline
prior         & $-4.47^{+0.79}_{-0.61}$ & $ 1.04^{+0.86}_{-0.85}$ & $ 8.17^{+0.36}_{-0.32}$ & $ 0.50^{+0.44}_{-0.44}$ & $ 0.01^{+1.79}_{-1.80}$\\
{\it IPTA30}  & $-4.46^{+0.70}_{-0.59}$ & $ 0.94^{+0.89}_{-0.67}$ & $ 8.20^{+0.29}_{-0.31}$ & $ 0.55^{+0.40}_{-0.48}$ & $ 0.08^{+1.73}_{-1.90}$\\
{\it SKA20}   & $-4.35^{+0.63}_{-0.56}$ & $ 0.76^{+0.94}_{-0.52}$ & $ 8.18^{+0.26}_{-0.31}$ & $ 0.75^{+0.22}_{-0.34}$ & $ 1.20^{+0.74}_{-1.83}$\\
{\it ideal}   & $-4.24^{+0.38}_{-0.38}$ & $ 0.68^{+0.58}_{-0.37}$ & $ 8.10^{+0.23}_{-0.24}$ & $ 0.73^{+0.23}_{-0.31}$ & $ 1.38^{+0.57}_{-1.60}$\\
injection     & $-4.0$                  & $ 0.8$                  & $ 8.0$                  & $ 0.9$                  & $ 0.0$\\
\hline
\end{tabularx}
\caption{List of credible intervals for selected parameters of the model. Each column reports the median value together with the errors bracketing the 90\% confidence regions for selected parameters. The five rows list the boundaries defined by the prior distributions, the posterior distributions as measured in the {\it IPTA30}, {\it SKA20} and {\it ideal} cases and the injected values from top to bottom.}
\label{table:pos}
\end{center}
\end{table}

We have presented an analytic parametrized model for the SMBHB merger rate in terms of astrophysical observables, including: galaxy stellar mass function, pair fraction, merger timescale and black hole - host galaxy relations. We described each individual ingredient with a simple analytic function and exploited our state of the art knowledge from observations, theory and simulations to define the prior range of each free parameter in the model. We then sampled the allowed parameter space (18 parameters in total) to produce an updated measure of the expected amplitude of the stochastic gravitational wave background across the frequency range. At $f=1\text{yr}^{-1}$ our model with the prior selection from Section \ref{sec:Prior} results in a characteristic strain $10^{-16}<h_c<10^{-15}$, confirming recent findings \citep[e.g.,][]{2018NatCo...9..573M}. We used our model to interpret current and future pulsar timing array upper limits and detections, linking the outcome of PTA observations to constraints on interesting observed quantities describing the cosmic population of merging galaxies and SMBHs. 

Consistent with our previous results \citep{2018NatCo...9..573M}, we find that current PTA upper limits can only add very little to the prior knowledge of the physical parameters as determined by current observations and simulations. However, as the sensitivity of PTA improves over time, upper limits can become stringent enough to probe interesting regions of the prior parameter space. The more stringent the upper limit becomes, the more extreme the conditions for the SMBHB population must be. Longer merger time (maybe even stalling) of binaries, less massive black holes and a spectral turnover at $f>10$ nano-Hz, all contribute to reduce the characteristic strain of the GWB in the PTA observable band. A upper limit at $A(f=1/{\rm yr}) = 1.0 \times 10^{-16}$ indicates moderate tension (at a nominal 2.5$\sigma$ level) between PTA observations and current astrophysical constraints. Pushing it down to $A(f=1/{\rm yr}) = 1.0 \times 10^{-17}$ would imply a strong $5\sigma$ tension with our current knowledge of the process of SMBHB formation and dynamics. Explaining a GWB below this level requires invoking a combination of SMBHB stalling, over-estimate of the SMBH -- host galaxy scaling relations, extreme eccentricities and dense environments. 

Although exploring progressively stringent upper limits is a useful exercise, we are particularly interested in addressing the astrophysical significance of a future PTA detection.
A weak initial detection at S/N ${\mathcal S}\approx 5-10$ will only put marginally better constraints on the underlying astrophysics of galaxies and SMBHs. As the detection significance increases, so do the constraints, as shown in table \ref{table:pos}. A full SKA-type array, detecting the GWB at ${\mathcal S}\approx 30-40$, will enable us to place important constrains on the normalization of the cosmic SMBHB merger rate, the time elapsed between galaxy pairing and SMBHB mergers, the normalization of the SMBH -- host galaxy relations and the dynamical properties of the merging SMBHBs. Since there is limited information in the GWB amplitude and spectral shape, even an ideal detection, reconstructing the GWB almost perfectly, will allow to place constrains only on a sub-set of the 18 parameters of the model. 
In particular, we have identified four quantities that can be well constrained with PTA observations on the GWB: the merger timescale of the SMBHB, the $M_{\rm bulge} - M_{\rm BH}$ relation, the eccentricity - density of the stellar environment and the overall effective merger rate of galaxies. This can be understood in terms of the distinctive features of the GWB spectrum. The observation of a low frequency turnover constrains the dynamics of individual SMBHBs, providing information about their eccentricity and the effectiveness of the hardening mechanism driving the merger process (i.e. the density of the stellar environment). The high frequency drop is determined by the high mass tail of the SMBH mass function, which is directly connected to the $M_{\rm BH} - M_{\rm bulge}$ relation. Whether a GWB is detected or not, immediately put a (loose) constrain on the SMBHB merger timescale. And the general amplitude of the strain allows to refine the measurement of the SMBHB merger timescale as well as determining the overall cosmic merger rate.

We stress that our model is still idealised in many ways. In particular, we employ a deterministic relation between model parameters and GWB spectrum. In reality, the GWB has some intrinsic variance due to the specific statistical realization of the SMBHB population occurring in nature. This is particularly important because the GWB strain is dominated by the most massive SMBHBs in the universe, which are intrinsically rare. Including a self-consistent computation of the variance in the model requires extensive Monte Carlo simulations, making the computation of the likelihood function prohibitively expensive for a direct nested sampling exploration of the parameter space. This difficulty can be overcome in the future by combining targeted simulations, sparsely sampling the parameter space with dedicated interpolation processes, which was demonstrated by \citep{2017PhRvL.118r1102T} on a parameter space of reduced complexity.

Although the introduction of intrinsic variance will likely degrade the inference on astrophysical observables, we also stress that we are still not using all the information encoded in the GW signal. In particular, information extracted from the shape and normalization of the GWB should be complemented with the statistics and properties of individually resolvable sources, which will provide precious extra information about the most massive SMBHBs and their physical properties \citep[e.g. their eccentricity][]{2016ApJ...817...70T}. Likewise, non-stationarity of the GWB will be indicative of highly eccentric binaries, allowing to disentangle eccentricity from extreme environments as the cause of a putative low frequency turnover. A comprehensive inference model from PTA observations will have to simultaneously combine all this information. Although there is still a lot of work to do, this study constitutes an important step forward in this endeavour.

\section*{acknowledgements}
We acknowledge the support of our colleagues in the European Pulsar Timing Array. S.C. thanks the University of Birmingham for their support via the AE Hills studentship. A.S. is supported by a University Research Fellow of the Royal Society.

\bibliographystyle{mnras}
\bibliography{bibliography}

\begin{thebibliography}{}
\makeatletter
\relax
\def\mn@urlcharsother{\let\do\@makeother \do\$\do\&\do\#\do\^\do\_\do\%\do\~}
\def\mn@doi{\begingroup\mn@urlcharsother \@ifnextchar [ {\mn@doi@}
  {\mn@doi@[]}}
\def\mn@doi@[#1]#2{\def\@tempa{#1}\ifx\@tempa\@empty \href
  {http://dx.doi.org/#2} {doi:#2}\else \href {http://dx.doi.org/#2} {#1}\fi
  \endgroup}
\def\mn@eprint#1#2{\mn@eprint@#1:#2::\@nil}
\def\mn@eprint@arXiv#1{\href {http://arxiv.org/abs/#1} {{\tt arXiv:#1}}}
\def\mn@eprint@dblp#1{\href {http://dblp.uni-trier.de/rec/bibtex/#1.xml}
  {dblp:#1}}
\def\mn@eprint@#1:#2:#3:#4\@nil{\def\@tempa {#1}\def\@tempb {#2}\def\@tempc
  {#3}\ifx \@tempc \@empty \let \@tempc \@tempb \let \@tempb \@tempa \fi \ifx
  \@tempb \@empty \def\@tempb {arXiv}\fi \@ifundefined
  {mn@eprint@\@tempb}{\@tempb:\@tempc}{\expandafter \expandafter \csname
  mn@eprint@\@tempb\endcsname \expandafter{\@tempc}}}

\bibitem[\protect\citeauthoryear{{Arzoumanian} et~al.,}{{Arzoumanian}
  et~al.}{2016}]{2016ApJ...821...13A}
{Arzoumanian} Z.,  et~al., 2016, \mn@doi [\apj] {10.3847/0004-637X/821/1/13},
  \href {http://esoads.eso.org/abs/2016ApJ...821...13A} {821, 13}

\bibitem[\protect\citeauthoryear{{Arzoumanian} et~al.,}{{Arzoumanian}
  et~al.}{2018}]{2018ApJS..235...37A}
{Arzoumanian} Z.,  et~al., 2018, \mn@doi [\apjs] {10.3847/1538-4365/aab5b0},
  \href {http://adsabs.harvard.edu/abs/2018ApJS..235...37A} {235, 37}

\bibitem[\protect\citeauthoryear{{Begelman}, {Blandford}  \& {Rees}}{{Begelman}
  et~al.}{1980}]{1980Natur.287..307B}
{Begelman} M.~C.,  {Blandford} R.~D.,   {Rees} M.~J.,  1980, \mn@doi [\nat]
  {10.1038/287307a0}, \href {http://adsabs.harvard.edu/abs/1980Natur.287..307B}
  {287, 307}

\bibitem[\protect\citeauthoryear{{Bernardi}, {Meert}, {Vikram},
  {Huertas-Company}, {Mei}, {Shankar}  \& {Sheth}}{{Bernardi}
  et~al.}{2014}]{2014MNRAS.443..874B}
{Bernardi} M.,  {Meert} A.,  {Vikram} V.,  {Huertas-Company} M.,  {Mei} S.,
  {Shankar} F.,   {Sheth} R.~K.,  2014, \mn@doi [\mnras]
  {10.1093/mnras/stu1106}, \href
  {http://adsabs.harvard.edu/abs/2014MNRAS.443..874B} {443, 874}

\bibitem[\protect\citeauthoryear{{Bonetti}, {Sesana}, {Barausse}  \&
  {Haardt}}{{Bonetti} et~al.}{2018}]{2018MNRAS.477.2599B}
{Bonetti} M.,  {Sesana} A.,  {Barausse} E.,   {Haardt} F.,  2018, \mn@doi
  [\mnras] {10.1093/mnras/sty874}, \href
  {http://adsabs.harvard.edu/abs/2018MNRAS.477.2599B} {477, 2599}

\bibitem[\protect\citeauthoryear{{Chen}, {Middleton}, {Sesana}, {Del Pozzo}  \&
  {Vecchio}}{{Chen} et~al.}{2017a}]{2017MNRAS.468..404C}
{Chen} S.,  {Middleton} H.,  {Sesana} A.,  {Del Pozzo} W.,   {Vecchio} A.,
  2017a, \mn@doi [\mnras] {10.1093/mnras/stx475}, \href
  {http://adsabs.harvard.edu/abs/2017MNRAS.468..404C} {468, 404}

\bibitem[\protect\citeauthoryear{{Chen}, {Sesana}  \& {Del Pozzo}}{{Chen}
  et~al.}{2017b}]{2017MNRAS.470.1738C}
{Chen} S.,  {Sesana} A.,   {Del Pozzo} W.,  2017b, \mn@doi [\mnras]
  {10.1093/mnras/stx1093}, \href
  {http://adsabs.harvard.edu/abs/2017MNRAS.470.1738C} {470, 1738}

\bibitem[\protect\citeauthoryear{{Conselice}, {Bershady}, {Dickinson}  \&
  {Papovich}}{{Conselice} et~al.}{2003}]{2003AJ....126.1183C}
{Conselice} C.~J.,  {Bershady} M.~A.,  {Dickinson} M.,   {Papovich} C.,  2003,
  \mn@doi [\aj] {10.1086/377318}, \href
  {http://adsabs.harvard.edu/abs/2003AJ....126.1183C} {126, 1183}

\bibitem[\protect\citeauthoryear{{Conselice}, {Wilkinson}, {Duncan}  \&
  {Mortlock}}{{Conselice} et~al.}{2016}]{2016ApJ...830...83C}
{Conselice} C.~J.,  {Wilkinson} A.,  {Duncan} K.,   {Mortlock} A.,  2016,
  \mn@doi [\apj] {10.3847/0004-637X/830/2/83}, \href
  {http://adsabs.harvard.edu/abs/2016ApJ...830...83C} {830, 83}

\bibitem[\protect\citeauthoryear{{Dehnen}}{{Dehnen}}{1993}]{1993MNRAS.265..250D}
{Dehnen} W.,  1993, \mn@doi [\mnras] {10.1093/mnras/265.1.250}, \href
  {http://adsabs.harvard.edu/abs/1993MNRAS.265..250D} {265, 250}

\bibitem[\protect\citeauthoryear{{Del Pozzo} \& {Veitch}}{{Del Pozzo} \&
  {Veitch}}{2015}]{cpnest}
{Del Pozzo} W.,  {Veitch} J.,  2015, {CPNest: Parallel nested sampling in
  python}, https://github.com/johnveitch/cpnest

\bibitem[\protect\citeauthoryear{{Desvignes} et~al.,}{{Desvignes}
  et~al.}{2016}]{2016MNRAS.458.3341D}
{Desvignes} G.,  et~al., 2016, \mn@doi [\mnras] {10.1093/mnras/stw483}, \href
  {http://adsabs.harvard.edu/abs/2016MNRAS.458.3341D} {458, 3341}

\bibitem[\protect\citeauthoryear{{Dotti}, {Sesana}  \& {Decarli}}{{Dotti}
  et~al.}{2012}]{2012AdAst2012E...3D}
{Dotti} M.,  {Sesana} A.,   {Decarli} R.,  2012, \mn@doi [Advances in
  Astronomy] {10.1155/2012/940568}, \href
  {http://adsabs.harvard.edu/abs/2012AdAst2012E...3D} {2012, 3}

\bibitem[\protect\citeauthoryear{{Duncan} et~al.,}{{Duncan}
  et~al.}{2019}]{2019ApJ...876..110D}
{Duncan} K.,  et~al., 2019, \mn@doi [\apj] {10.3847/1538-4357/ab148a}, \href
  {https://ui.adsabs.harvard.edu/abs/2019ApJ...876..110D} {876, 110}

\bibitem[\protect\citeauthoryear{{Dvorkin} \& {Barausse}}{{Dvorkin} \&
  {Barausse}}{2017}]{2017MNRAS.470.4547D}
{Dvorkin} I.,  {Barausse} E.,  2017, \mn@doi [\mnras] {10.1093/mnras/stx1454},
  \href {http://adsabs.harvard.edu/abs/2017MNRAS.470.4547D} {470, 4547}

\bibitem[\protect\citeauthoryear{{Event Horizon Telescope Collaboration}
  et~al.,}{{Event Horizon Telescope Collaboration}
  et~al.}{2019}]{2019ApJ...875L...1E}
{Event Horizon Telescope Collaboration} et~al., 2019, \mn@doi [\apjl]
  {10.3847/2041-8213/ab0ec7}, \href
  {http://adsabs.harvard.edu/abs/2019ApJ...875L...1E} {875, L1}

\bibitem[\protect\citeauthoryear{{Foster} \& {Backer}}{{Foster} \&
  {Backer}}{1990}]{1990ApJ...361..300F}
{Foster} R.~S.,  {Backer} D.~C.,  1990, \mn@doi [\apj] {10.1086/169195}, \href
  {http://adsabs.harvard.edu/abs/1990ApJ...361..300F} {361, 300}

\bibitem[\protect\citeauthoryear{{Graham} \& {Scott}}{{Graham} \&
  {Scott}}{2012}]{grahamscott12}
{Graham} A.~W.,  {Scott} N.,  2012, preprint, \href
  {http://adsabs.harvard.edu/abs/2012arXiv1211.3199G} {} (\mn@eprint {arXiv}
  {1211.3199})

\bibitem[\protect\citeauthoryear{{Hellings} \& {Downs}}{{Hellings} \&
  {Downs}}{1983}]{HellingsDowns:1983}
{Hellings} R.~W.,  {Downs} G.~S.,  1983, \mn@doi [The Astrophys. J.l]
  {10.1086/183954}, \href {http://adsabs.harvard.edu/abs/1983ApJ...265L..39H}
  {265, L39}

\bibitem[\protect\citeauthoryear{{Jaffe} \& {Backer}}{{Jaffe} \&
  {Backer}}{2003}]{2003ApJ...583..616J}
{Jaffe} A.~H.,  {Backer} D.~C.,  2003, \mn@doi [\apj] {10.1086/345443}, \href
  {http://adsabs.harvard.edu/abs/2003ApJ...583..616J} {583, 616}

\bibitem[\protect\citeauthoryear{{Janssen} et~al.,}{{Janssen}
  et~al.}{2015}]{2015aska.confE..37J}
{Janssen} G.,  et~al., 2015, Advancing Astrophysics with the Square Kilometre
  Array (AASKA14), \href {http://adsabs.harvard.edu/abs/2015aska.confE..37J}
  {p.~37}

\bibitem[\protect\citeauthoryear{{Jenet} et~al.,}{{Jenet}
  et~al.}{2006}]{JenetEtAl:2006}
{Jenet} F.~A.,  et~al., 2006, \mn@doi [The Astrophys. J.] {10.1086/508702},
  \href {http://adsabs.harvard.edu/abs/2006ApJ...653.1571J} {653, 1571}

\bibitem[\protect\citeauthoryear{{Keenan} et~al.,}{{Keenan}
  et~al.}{2014}]{2014ApJ...795..157K}
{Keenan} R.~C.,  et~al., 2014, \mn@doi [\apj] {10.1088/0004-637X/795/2/157},
  \href {http://adsabs.harvard.edu/abs/2014ApJ...795..157K} {795, 157}

\bibitem[\protect\citeauthoryear{{Kelley}, {Blecha}  \& {Hernquist}}{{Kelley}
  et~al.}{2017a}]{2017MNRAS.464.3131K}
{Kelley} L.~Z.,  {Blecha} L.,   {Hernquist} L.,  2017a, \mn@doi [\mnras]
  {10.1093/mnras/stw2452}, \href
  {http://adsabs.harvard.edu/abs/2017MNRAS.464.3131K} {464, 3131}

\bibitem[\protect\citeauthoryear{{Kelley}, {Blecha}, {Hernquist}, {Sesana}  \&
  {Taylor}}{{Kelley} et~al.}{2017b}]{2017MNRAS.471.4508K}
{Kelley} L.~Z.,  {Blecha} L.,  {Hernquist} L.,  {Sesana} A.,   {Taylor} S.~R.,
  2017b, \mn@doi [\mnras] {10.1093/mnras/stx1638}, \href
  {http://adsabs.harvard.edu/abs/2017MNRAS.471.4508K} {471, 4508}

\bibitem[\protect\citeauthoryear{{Khan}, {Berentzen}, {Berczik}, {Just},
  {Mayer}, {Nitadori}  \& {Callegari}}{{Khan}
  et~al.}{2012}]{2012ApJ...756...30K}
{Khan} F.~M.,  {Berentzen} I.,  {Berczik} P.,  {Just} A.,  {Mayer} L.,
  {Nitadori} K.,   {Callegari} S.,  2012, \mn@doi [\apj]
  {10.1088/0004-637X/756/1/30}, \href
  {http://adsabs.harvard.edu/abs/2012ApJ...756...30K} {756, 30}

\bibitem[\protect\citeauthoryear{{Kitzbichler} \& {White}}{{Kitzbichler} \&
  {White}}{2008}]{kit08}
{Kitzbichler} M.~G.,  {White} S.~D.~M.,  2008, \mn@doi [Mon. Not. R. Astron.
  Soc.] {10.1111/j.1365-2966.2008.13873.x}, \href
  {http://adsabs.harvard.edu/abs/2008MNRAS.391.1489K} {391, 1489}

\bibitem[\protect\citeauthoryear{{Kormendy} \& {Ho}}{{Kormendy} \&
  {Ho}}{2013}]{2013ARA&A..51..511K}
{Kormendy} J.,  {Ho} L.~C.,  2013, \mn@doi [\araa]
  {10.1146/annurev-astro-082708-101811}, \href
  {http://adsabs.harvard.edu/abs/2013ARA%26A..51..511K} {51, 511}

\bibitem[\protect\citeauthoryear{{Kormendy}, {Fisher}, {Cornell}  \&
  {Bender}}{{Kormendy} et~al.}{2009}]{2009ApJS..182..216K}
{Kormendy} J.,  {Fisher} D.~B.,  {Cornell} M.~E.,   {Bender} R.,  2009, \mn@doi
  [\apjs] {10.1088/0067-0049/182/1/216}, \href
  {http://adsabs.harvard.edu/abs/2009ApJS..182..216K} {182, 216}

\bibitem[\protect\citeauthoryear{{Lentati} et~al.,}{{Lentati}
  et~al.}{2015}]{2015MNRAS.453.2576L}
{Lentati} L.,  et~al., 2015, \mn@doi [\mnras] {10.1093/mnras/stv1538}, \href
  {http://esoads.eso.org/abs/2015MNRAS.453.2576L} {453, 2576}

\bibitem[\protect\citeauthoryear{{Li}, {Ho}  \& {Wang}}{{Li}
  et~al.}{2011}]{2011ApJ...742...33L}
{Li} Y.-R.,  {Ho} L.~C.,   {Wang} J.-M.,  2011, \mn@doi [\apj]
  {10.1088/0004-637X/742/1/33}, \href
  {http://adsabs.harvard.edu/abs/2011ApJ...742...33L} {742, 33}

\bibitem[\protect\citeauthoryear{{Lotz}, {Jonsson}, {Cox}  \& {Primack}}{{Lotz}
  et~al.}{2010}]{lotz10}
{Lotz} J.~M.,  {Jonsson} P.,  {Cox} T.~J.,   {Primack} J.~R.,  2010, \mn@doi
  [Mon. Not. R. Astron. Soc.] {10.1111/j.1365-2966.2010.16268.x}, \href
  {http://adsabs.harvard.edu/abs/2010MNRAS.404..575L} {404, 575}

\bibitem[\protect\citeauthoryear{{Lotz}, {Jonsson}, {Cox}, {Croton}, {Primack},
  {Somerville}  \& {Stewart}}{{Lotz} et~al.}{2011}]{2011ApJ...742..103L}
{Lotz} J.~M.,  {Jonsson} P.,  {Cox} T.~J.,  {Croton} D.,  {Primack} J.~R.,
  {Somerville} R.~S.,   {Stewart} K.,  2011, \mn@doi [\apj]
  {10.1088/0004-637X/742/2/103}, \href
  {http://adsabs.harvard.edu/abs/2011ApJ...742..103L} {742, 103}

\bibitem[\protect\citeauthoryear{{McConnell} \& {Ma}}{{McConnell} \&
  {Ma}}{2013}]{2013ApJ...764..184M}
{McConnell} N.~J.,  {Ma} C.-P.,  2013, \mn@doi [\apj]
  {10.1088/0004-637X/764/2/184}, \href
  {http://adsabs.harvard.edu/abs/2013ApJ...764..184M} {764, 184}

\bibitem[\protect\citeauthoryear{{Middleton}, {Del Pozzo}, {Farr}, {Sesana}  \&
  {Vecchio}}{{Middleton} et~al.}{2016}]{2016MNRAS.455L..72M}
{Middleton} H.,  {Del Pozzo} W.,  {Farr} W.~M.,  {Sesana} A.,   {Vecchio} A.,
  2016, \mn@doi [\mnras] {10.1093/mnrasl/slv150}, \href
  {http://esoads.eso.org/abs/2016MNRAS.455L..72M} {455, L72}

\bibitem[\protect\citeauthoryear{{Middleton}, {Chen}, {Del Pozzo}, {Sesana}  \&
  {Vecchio}}{{Middleton} et~al.}{2018}]{2018NatCo...9..573M}
{Middleton} H.,  {Chen} S.,  {Del Pozzo} W.,  {Sesana} A.,   {Vecchio} A.,
  2018, \mn@doi [Nature Communications] {10.1038/s41467-018-02916-7}, \href
  {http://adsabs.harvard.edu/abs/2018NatCo...9..573M} {9, 573}

\bibitem[\protect\citeauthoryear{{Mingarelli} et~al.,}{{Mingarelli}
  et~al.}{2017}]{2017NatAs...1..886M}
{Mingarelli} C.~M.~F.,  et~al., 2017, \mn@doi [Nature Astronomy]
  {10.1038/s41550-017-0299-6}, \href
  {http://adsabs.harvard.edu/abs/2017NatAs...1..886M} {1, 886}

\bibitem[\protect\citeauthoryear{{Mirza}, {Tahir}, {Khan}, {Holley-Bockelmann},
  {Baig}, {Berczik}  \& {Chishtie}}{{Mirza} et~al.}{2017}]{2017MNRAS.470..940M}
{Mirza} M.~A.,  {Tahir} A.,  {Khan} F.~M.,  {Holley-Bockelmann} H.,  {Baig}
  A.~M.,  {Berczik} P.,   {Chishtie} F.,  2017, \mn@doi [\mnras]
  {10.1093/mnras/stx1248}, \href
  {http://adsabs.harvard.edu/abs/2017MNRAS.470..940M} {470, 940}

\bibitem[\protect\citeauthoryear{{Moore}, {Taylor}  \& {Gair}}{{Moore}
  et~al.}{2015}]{2015CQGra..32e5004M}
{Moore} C.~J.,  {Taylor} S.~R.,   {Gair} J.~R.,  2015, \mn@doi [Classical and
  Quantum Gravity] {10.1088/0264-9381/32/5/055004}, \href
  {http://adsabs.harvard.edu/abs/2015CQGra..32e5004M} {32, 055004}

\bibitem[\protect\citeauthoryear{{Mortlock} et~al.,}{{Mortlock}
  et~al.}{2015}]{2015MNRAS.447....2M}
{Mortlock} A.,  et~al., 2015, \mn@doi [\mnras] {10.1093/mnras/stu2403}, \href
  {http://adsabs.harvard.edu/abs/2015MNRAS.447....2M} {447, 2}

\bibitem[\protect\citeauthoryear{{Mundy}, {Conselice}, {Duncan}, {Almaini},
  {H{\"a}u{\ss}ler}  \& {Hartley}}{{Mundy} et~al.}{2017}]{2017MNRAS.470.3507M}
{Mundy} C.~J.,  {Conselice} C.~J.,  {Duncan} K.~J.,  {Almaini} O.,
  {H{\"a}u{\ss}ler} B.,   {Hartley} W.~G.,  2017, \mn@doi [\mnras]
  {10.1093/mnras/stx1238}, \href
  {http://adsabs.harvard.edu/abs/2017MNRAS.470.3507M} {470, 3507}

\bibitem[\protect\citeauthoryear{{Peters} \& {Mathews}}{{Peters} \&
  {Mathews}}{1963}]{PetersMathews:1963}
{Peters} P.~C.,  {Mathews} J.,  1963, \mn@doi [Physical Review]
  {10.1103/PhysRev.131.435}, \href
  {http://adsabs.harvard.edu/abs/1963PhRv..131..435P} {131, 435}

\bibitem[\protect\citeauthoryear{{Rajagopal} \& {Romani}}{{Rajagopal} \&
  {Romani}}{1995}]{1995ApJ...446..543R}
{Rajagopal} M.,  {Romani} R.~W.,  1995, \mn@doi [\apj] {10.1086/175813}, \href
  {http://adsabs.harvard.edu/abs/1995ApJ...446..543R} {446, 543}

\bibitem[\protect\citeauthoryear{{Ravi}, {Wyithe}, {Hobbs}, {Shannon},
  {Manchester}, {Yardley}  \& {Keith}}{{Ravi} et~al.}{2012}]{RaviEtAl:2012}
{Ravi} V.,  {Wyithe} J.~S.~B.,  {Hobbs} G.,  {Shannon} R.~M.,  {Manchester}
  R.~N.,  {Yardley} D.~R.~B.,   {Keith} M.~J.,  2012, \mn@doi [\apj]
  {10.1088/0004-637X/761/2/84}, \href
  {http://adsabs.harvard.edu/abs/2012ApJ...761...84R} {761, 84}

\bibitem[\protect\citeauthoryear{{Reardon} et~al.,}{{Reardon}
  et~al.}{2016}]{2016MNRAS.455.1751R}
{Reardon} D.~J.,  et~al., 2016, \mn@doi [\mnras] {10.1093/mnras/stv2395}, \href
  {http://adsabs.harvard.edu/abs/2016MNRAS.455.1751R} {455, 1751}

\bibitem[\protect\citeauthoryear{{Robotham} et~al.,}{{Robotham}
  et~al.}{2014}]{2014MNRAS.444.3986R}
{Robotham} A.~S.~G.,  et~al., 2014, \mn@doi [\mnras] {10.1093/mnras/stu1604},
  \href {http://adsabs.harvard.edu/abs/2014MNRAS.444.3986R} {444, 3986}

\bibitem[\protect\citeauthoryear{{Rosado}, {Sesana}  \& {Gair}}{{Rosado}
  et~al.}{2015}]{2015MNRAS.451.2417R}
{Rosado} P.~A.,  {Sesana} A.,   {Gair} J.,  2015, \mn@doi [\mnras]
  {10.1093/mnras/stv1098}, \href
  {http://adsabs.harvard.edu/abs/2015MNRAS.451.2417R} {451, 2417}

\bibitem[\protect\citeauthoryear{{Ryu}, {Perna}, {Haiman}, {Ostriker}  \&
  {Stone}}{{Ryu} et~al.}{2018}]{2018MNRAS.473.3410R}
{Ryu} T.,  {Perna} R.,  {Haiman} Z.,  {Ostriker} J.~P.,   {Stone} N.~C.,  2018,
  \mn@doi [\mnras] {10.1093/mnras/stx2524}, \href
  {http://adsabs.harvard.edu/abs/2018MNRAS.473.3410R} {473, 3410}

\bibitem[\protect\citeauthoryear{{Salviander}, {Shields}  \&
  {Bonning}}{{Salviander} et~al.}{2015}]{2015ApJ...799..173S}
{Salviander} S.,  {Shields} G.~A.,   {Bonning} E.~W.,  2015, \mn@doi [\apj]
  {10.1088/0004-637X/799/2/173}, \href
  {http://adsabs.harvard.edu/abs/2015ApJ...799..173S} {799, 173}

\bibitem[\protect\citeauthoryear{{Schulze} \& {Wisotzki}}{{Schulze} \&
  {Wisotzki}}{2014}]{2014MNRAS.438.3422S}
{Schulze} A.,  {Wisotzki} L.,  2014, \mn@doi [\mnras] {10.1093/mnras/stt2457},
  \href {http://adsabs.harvard.edu/abs/2014MNRAS.438.3422S} {438, 3422}

\bibitem[\protect\citeauthoryear{{Sesana}}{{Sesana}}{2013a}]{2013CQGra..30v4014S}
{Sesana} A.,  2013a, \mn@doi [Classical and Quantum Gravity]
  {10.1088/0264-9381/30/22/224014}, \href
  {http://adsabs.harvard.edu/abs/2013CQGra..30v4014S} {30, 224014}

\bibitem[\protect\citeauthoryear{{Sesana}}{{Sesana}}{2013b}]{Sesana:2013}
{Sesana} A.,  2013b, \mn@doi [\mnras] {10.1093/mnrasl/slt034}, \href
  {http://adsabs.harvard.edu/abs/2013MNRAS.433L...1S} {433, L1}

\bibitem[\protect\citeauthoryear{{Sesana} \& {Khan}}{{Sesana} \&
  {Khan}}{2015}]{2015MNRAS.454L..66S}
{Sesana} A.,  {Khan} F.~M.,  2015, \mn@doi [\mnras] {10.1093/mnrasl/slv131},
  \href {http://adsabs.harvard.edu/abs/2015MNRAS.454L..66S} {454, L66}

\bibitem[\protect\citeauthoryear{{Sesana}, {Vecchio}  \& {Colacino}}{{Sesana}
  et~al.}{2008}]{SesanaVecchioColacino:2008}
{Sesana} A.,  {Vecchio} A.,   {Colacino} C.~N.,  2008, \mn@doi [\mnras]
  {10.1111/j.1365-2966.2008.13682.x}, \href
  {http://adsabs.harvard.edu/abs/2008MNRAS.390..192S} {390, 192}

\bibitem[\protect\citeauthoryear{{Sesana}, {Vecchio}  \& {Volonteri}}{{Sesana}
  et~al.}{2009}]{SesanaVecchioVolonteri:2009}
{Sesana} A.,  {Vecchio} A.,   {Volonteri} M.,  2009, \mn@doi [\mnras]
  {10.1111/j.1365-2966.2009.14499.x}, \href
  {http://adsabs.harvard.edu/abs/2009MNRAS.394.2255S} {394, 2255}

\bibitem[\protect\citeauthoryear{{Sesana}, {Shankar}, {Bernardi}  \&
  {Sheth}}{{Sesana} et~al.}{2016}]{2016MNRAS.463L...6S}
{Sesana} A.,  {Shankar} F.,  {Bernardi} M.,   {Sheth} R.~K.,  2016, \mn@doi
  [\mnras] {10.1093/mnrasl/slw139}, \href
  {http://adsabs.harvard.edu/abs/2016MNRAS.463L...6S} {463, L6}

\bibitem[\protect\citeauthoryear{{Shankar} et~al.,}{{Shankar}
  et~al.}{2016}]{2016MNRAS.460.3119S}
{Shankar} F.,  et~al., 2016, \mn@doi [\mnras] {10.1093/mnras/stw678}, \href
  {http://adsabs.harvard.edu/abs/2016MNRAS.460.3119S} {460, 3119}

\bibitem[\protect\citeauthoryear{{Shannon} et~al.,}{{Shannon}
  et~al.}{2015}]{2015Sci...349.1522S}
{Shannon} R.~M.,  et~al., 2015, \mn@doi [Science] {10.1126/science.aab1910},
  \href {http://esoads.eso.org/abs/2015Sci...349.1522S} {349, 1522}

\bibitem[\protect\citeauthoryear{{Simon} \& {Burke-Spolaor}}{{Simon} \&
  {Burke-Spolaor}}{2016}]{2016ApJ...826...11S}
{Simon} J.,  {Burke-Spolaor} S.,  2016, \mn@doi [\apj]
  {10.3847/0004-637X/826/1/11}, \href
  {http://esoads.eso.org/abs/2016ApJ...826...11S} {826, 11}

\bibitem[\protect\citeauthoryear{Skilling}{Skilling}{2004}]{Skilling2004a}
Skilling J.,  2004, in American Institute of Physics Conference Series. pp
  395--405

\bibitem[\protect\citeauthoryear{{Snyder}, {Lotz}, {Rodriguez-Gomez},
  {Guimar{\~a}es}, {Torrey}  \& {Hernquist}}{{Snyder}
  et~al.}{2017}]{2017MNRAS.468..207S}
{Snyder} G.~F.,  {Lotz} J.~M.,  {Rodriguez-Gomez} V.,  {Guimar{\~a}es}
  R.~d.~S.,  {Torrey} P.,   {Hernquist} L.,  2017, \mn@doi [\mnras]
  {10.1093/mnras/stx487}, \href
  {http://adsabs.harvard.edu/abs/2017MNRAS.468..207S} {468, 207}

\bibitem[\protect\citeauthoryear{{Taylor}, {Huerta}, {Gair}  \&
  {McWilliams}}{{Taylor} et~al.}{2016a}]{2016ApJ...817...70T}
{Taylor} S.~R.,  {Huerta} E.~A.,  {Gair} J.~R.,   {McWilliams} S.~T.,  2016a,
  \mn@doi [\apj] {10.3847/0004-637X/817/1/70}, \href
  {http://adsabs.harvard.edu/abs/2016ApJ...817...70T} {817, 70}

\bibitem[\protect\citeauthoryear{{Taylor}, {Vallisneri}, {Ellis}, {Mingarelli},
  {Lazio}  \& {van Haasteren}}{{Taylor} et~al.}{2016b}]{2016ApJ...819L...6T}
{Taylor} S.~R.,  {Vallisneri} M.,  {Ellis} J.~A.,  {Mingarelli} C.~M.~F.,
  {Lazio} T.~J.~W.,   {van Haasteren} R.,  2016b, \mn@doi [\apjl]
  {10.3847/2041-8205/819/1/L6}, \href
  {http://adsabs.harvard.edu/abs/2016ApJ...819L...6T} {819, L6}

\bibitem[\protect\citeauthoryear{{Taylor}, {Simon}  \& {Sampson}}{{Taylor}
  et~al.}{2017}]{2017PhRvL.118r1102T}
{Taylor} S.~R.,  {Simon} J.,   {Sampson} L.,  2017, \mn@doi [Physical Review
  Letters] {10.1103/PhysRevLett.118.181102}, \href
  {http://adsabs.harvard.edu/abs/2017PhRvL.118r1102T} {118, 181102}

\bibitem[\protect\citeauthoryear{{Vasiliev}, {Antonini}  \&
  {Merritt}}{{Vasiliev} et~al.}{2015}]{2015ApJ...810...49V}
{Vasiliev} E.,  {Antonini} F.,   {Merritt} D.,  2015, \mn@doi [\apj]
  {10.1088/0004-637X/810/1/49}, \href
  {http://adsabs.harvard.edu/abs/2015ApJ...810...49V} {810, 49}

\bibitem[\protect\citeauthoryear{{Veitch} et~al.,}{{Veitch}
  et~al.}{2015}]{LALInference}
{Veitch} J.,  et~al., 2015, \mn@doi [\prd] {10.1103/PhysRevD.91.042003}, \href
  {http://adsabs.harvard.edu/abs/2015PhRvD..91d2003V} {91, 042003}

\bibitem[\protect\citeauthoryear{{Verbiest} et~al.,}{{Verbiest}
  et~al.}{2016}]{2016IPTA}
{Verbiest} J.~P.~W.,  et~al., 2016, \mn@doi [\mnras] {10.1093/mnras/stw347},
  \href {http://adsabs.harvard.edu/abs/2016MNRAS.458.1267V} {458, 1267}

\bibitem[\protect\citeauthoryear{{White} \& {Rees}}{{White} \&
  {Rees}}{1978}]{1978MNRAS.183..341W}
{White} S.~D.~M.,  {Rees} M.~J.,  1978, \mn@doi [\mnras]
  {10.1093/mnras/183.3.341}, \href
  {http://adsabs.harvard.edu/abs/1978MNRAS.183..341W} {183, 341}

\bibitem[\protect\citeauthoryear{{Xu}, {Zhao}, {Scoville}, {Capak}, {Drory}  \&
  {Gao}}{{Xu} et~al.}{2012}]{xu12}
{Xu} C.~K.,  {Zhao} Y.,  {Scoville} N.,  {Capak} P.,  {Drory} N.,   {Gao} Y.,
  2012, \mn@doi [Astrophys. J.] {10.1088/0004-637X/747/2/85}, \href
  {http://adsabs.harvard.edu/abs/2012ApJ...747...85X} {747, 85}

\makeatother
\end{thebibliography}

\label{lastpage}

\appendix
\section{Full corner plots}
\label{app:corner}
\begin{figure*}
\centering
{\large \bf 1e-15 Upper limit}
\includegraphics[width=\textwidth]{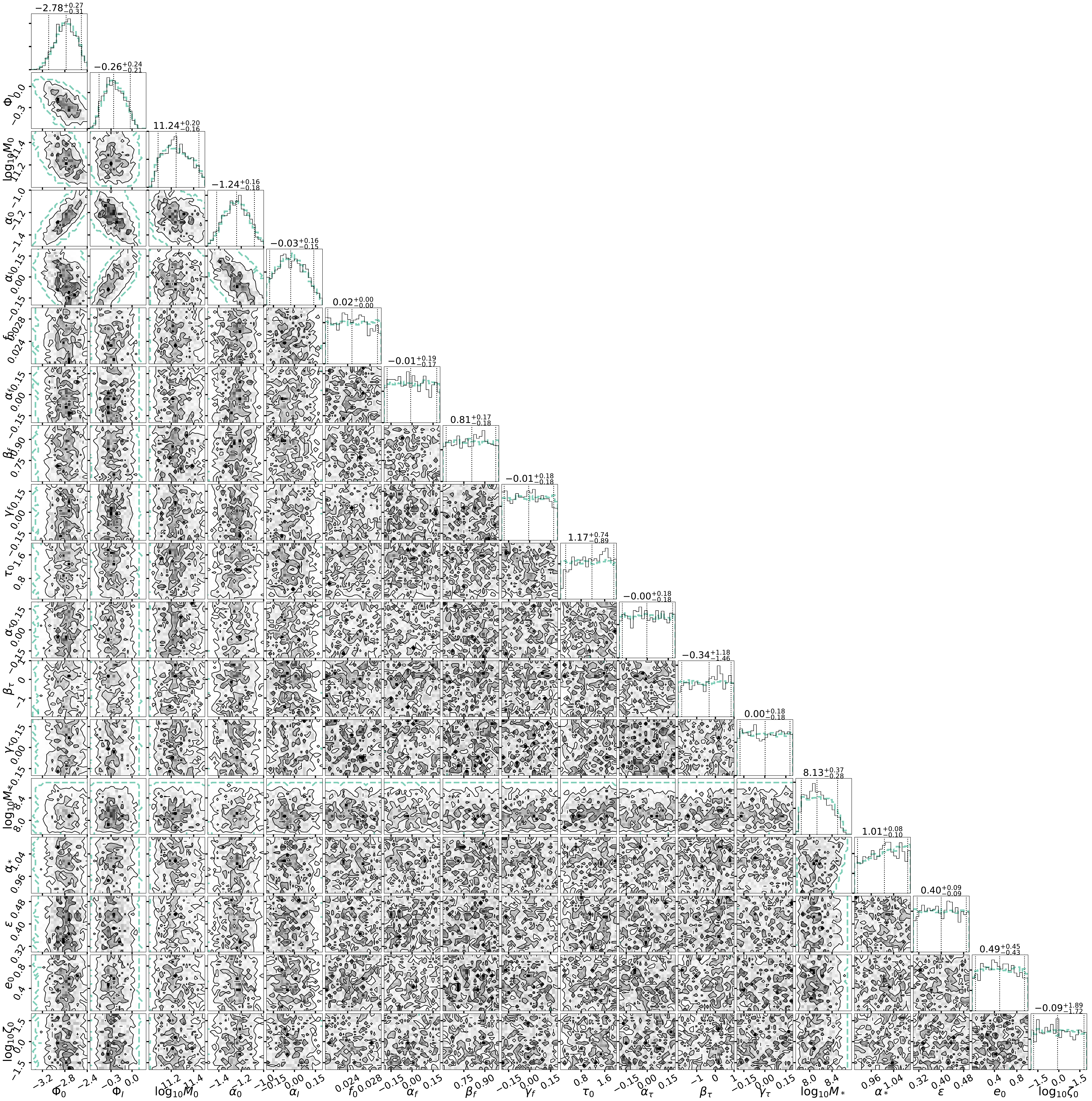}
\caption{Full corner plot of the 18 parameters for the PPTA15 case}
\label{fig:cornerplot_uplim15_full}
\end{figure*}

\begin{figure*}
\centering
{\large \bf 1e-16 Upper limit}
\includegraphics[width=\textwidth]{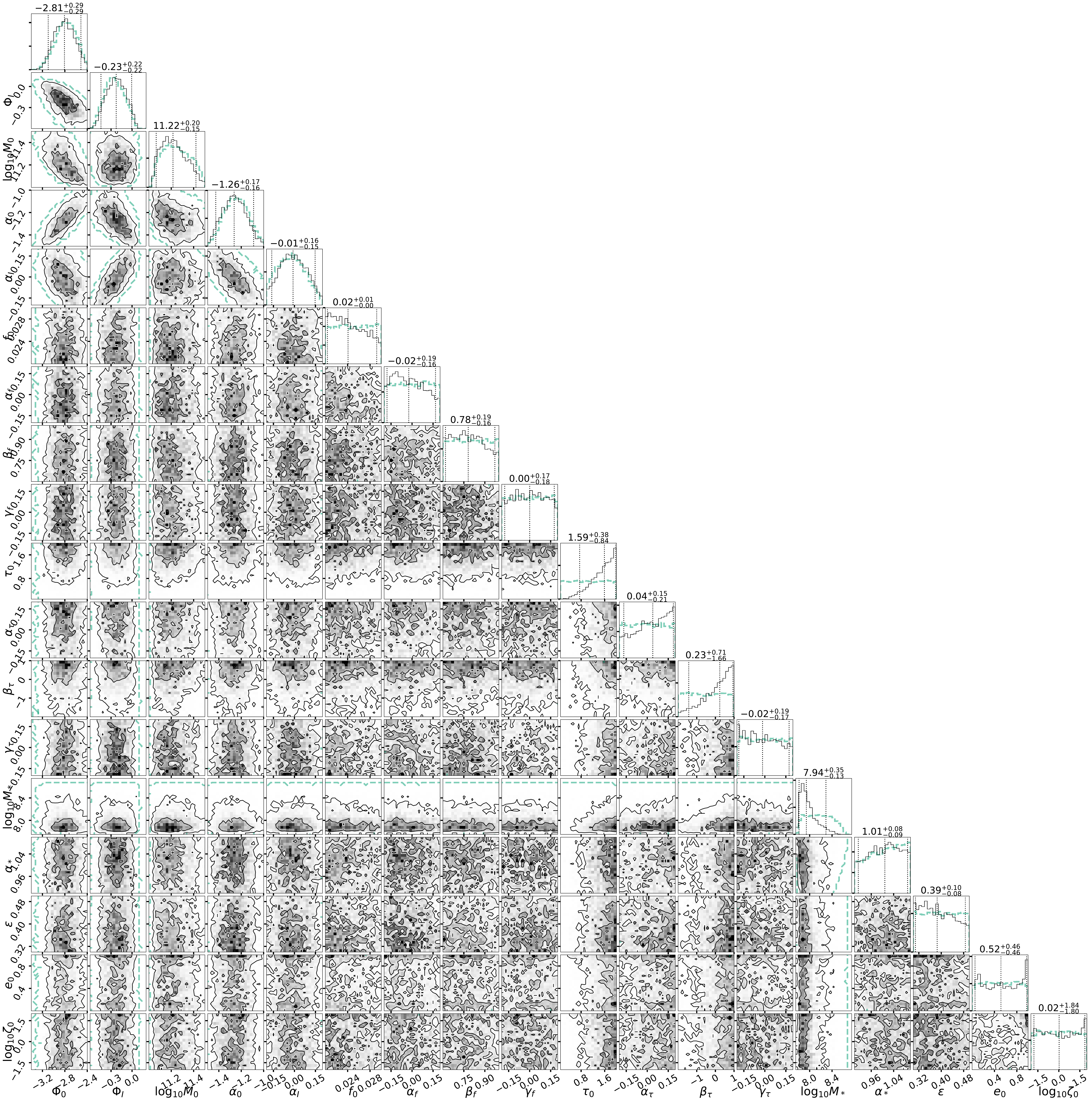}
\caption{Full corner plot of the 18 parameters for the PPTA16 case}
\label{fig:cornerplot_uplim15_full}
\end{figure*}

\begin{figure*}
\centering
{\large \bf 1e-17 Upper limit}
\includegraphics[width=\textwidth]{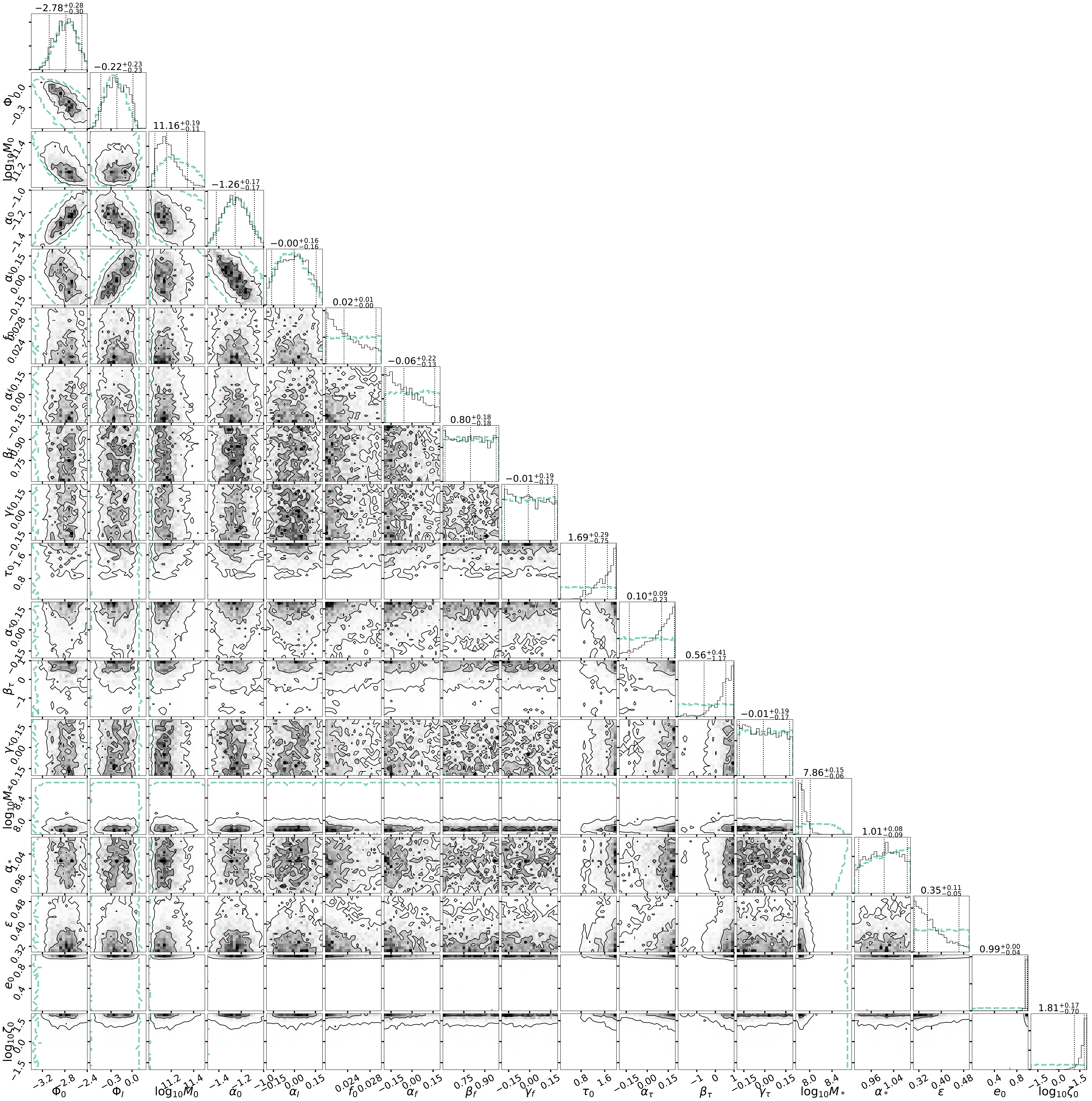}
\caption{Full corner plot of the 18 parameters for the PPTA17 case}
\label{fig:cornerplot_uplim15_full}
\end{figure*}

\begin{figure*}
\centering
{\large \bf 1e-17 Upper limit Extended Prior}
\includegraphics[width=\textwidth]{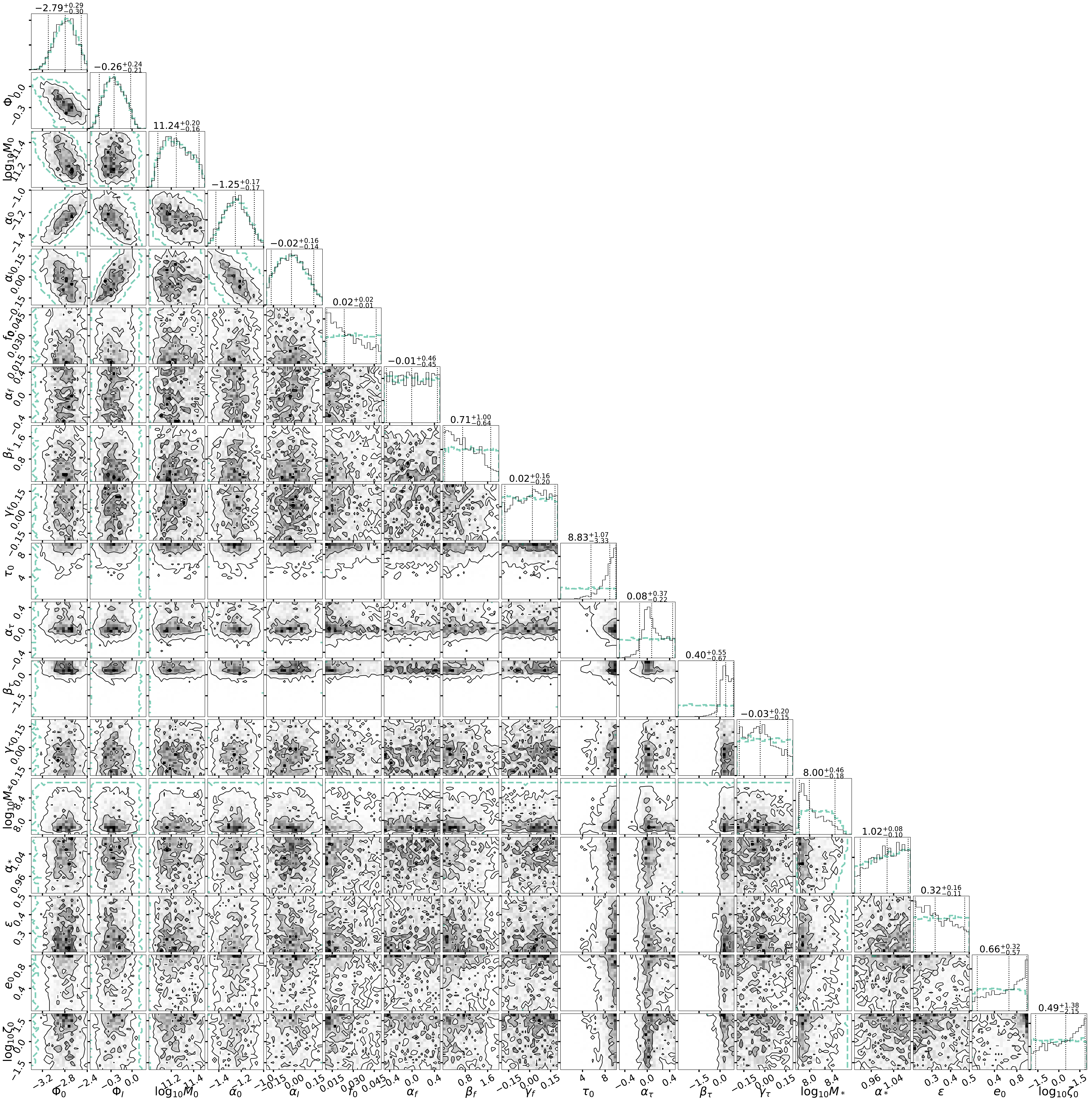}
\caption{Full corner plot of the 18 parameters for the PPTA17 extended prior case}
\label{fig:cornerplot_uplim15_full}
\end{figure*}

\begin{figure*}
\centering
{\large \bf Circular case IPTA30}
\includegraphics[width=\textwidth]{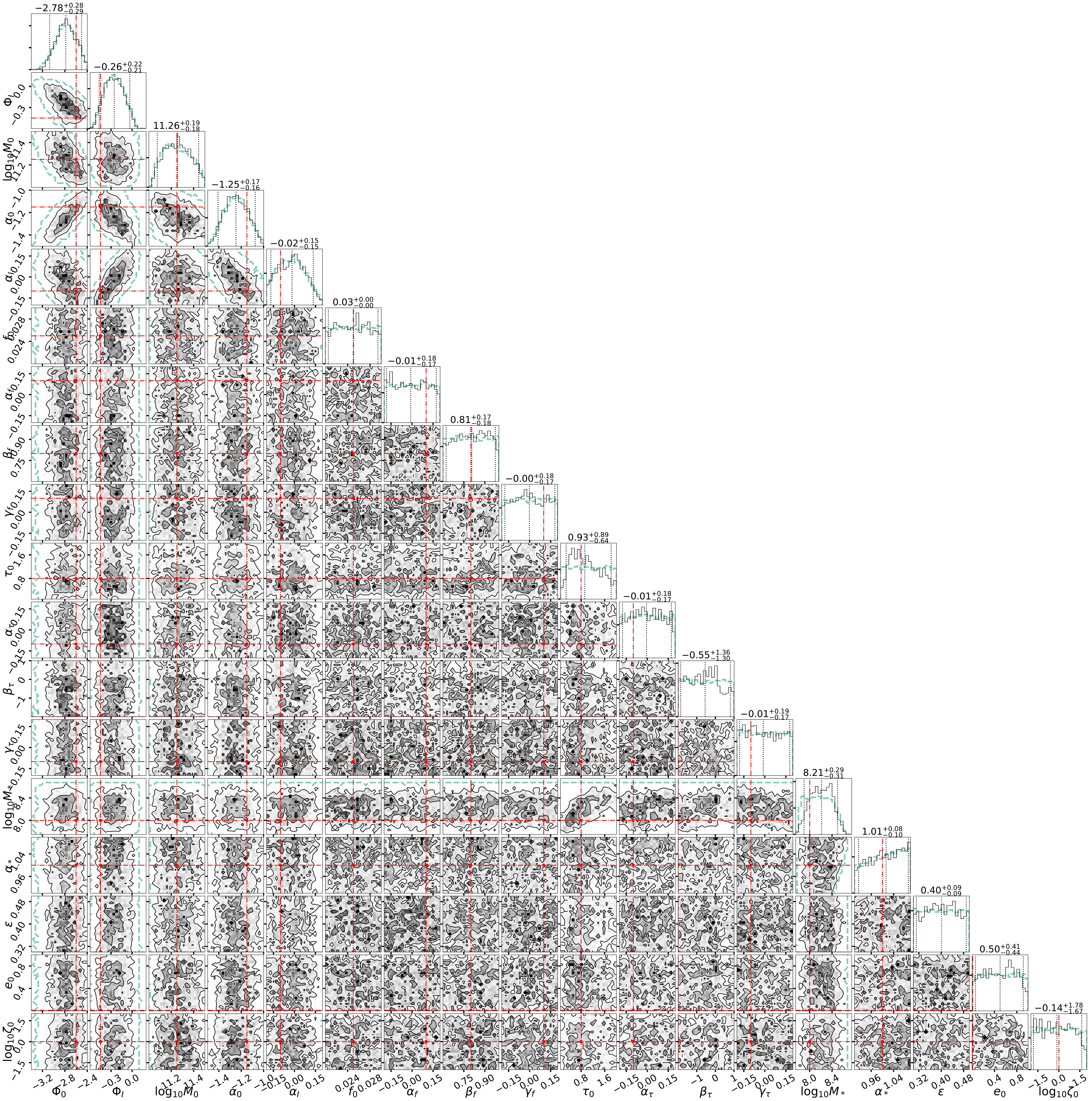}
\caption{Full corner plot of the 18 parameters for the IPTA30 Circular case}
\label{fig:cornerplot_uplim15_full}
\end{figure*}

\begin{figure*}
\centering
{\large \bf Circular case SKA20}
\includegraphics[width=\textwidth]{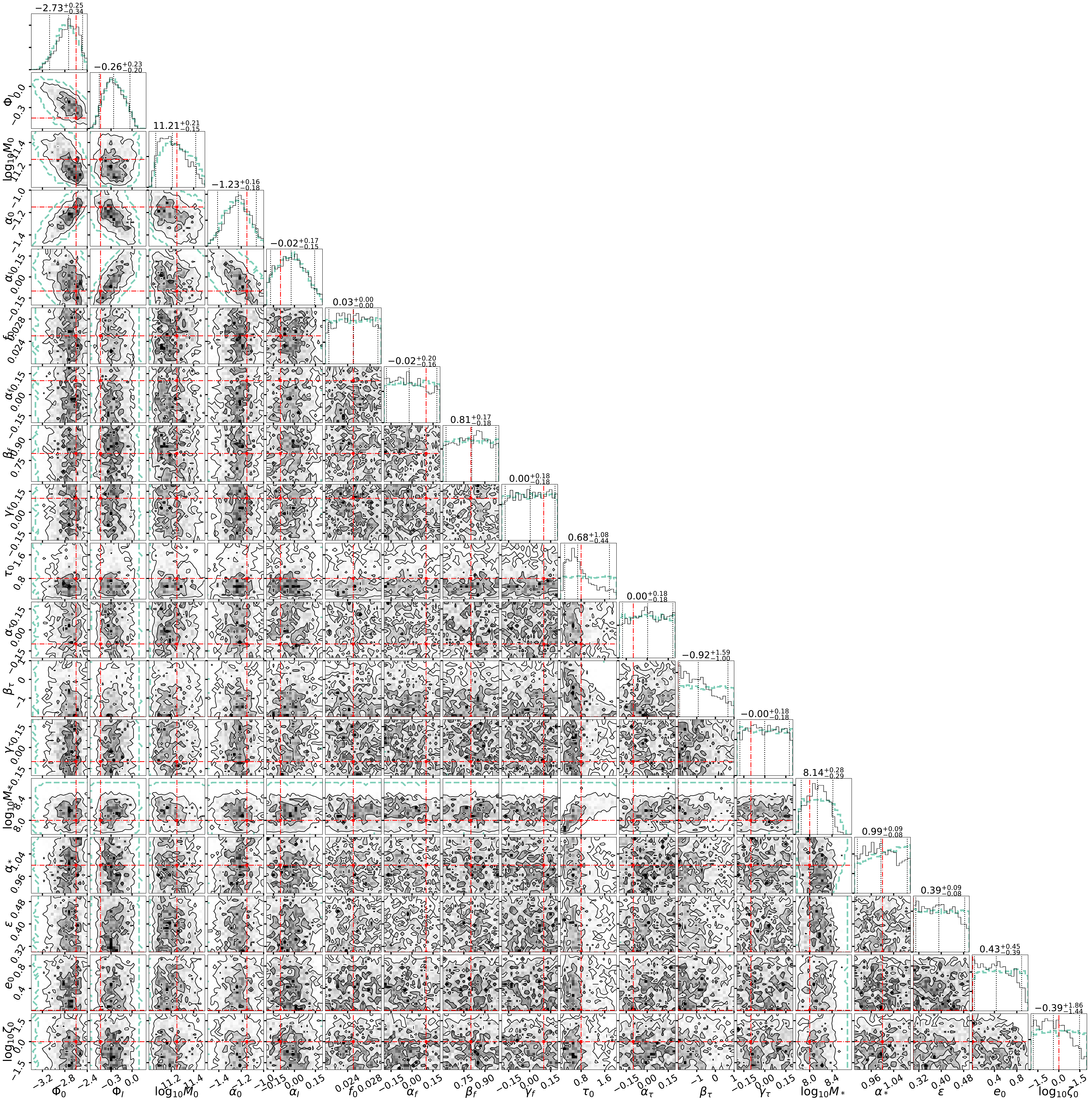}
\caption{Full corner plot of the 18 parameters for the SKA20 Circular case}
\label{fig:cornerplot_uplim15_full}
\end{figure*}

\begin{figure*}
\centering
{\large \bf Eccentric case IPTA30}
\includegraphics[width=\textwidth]{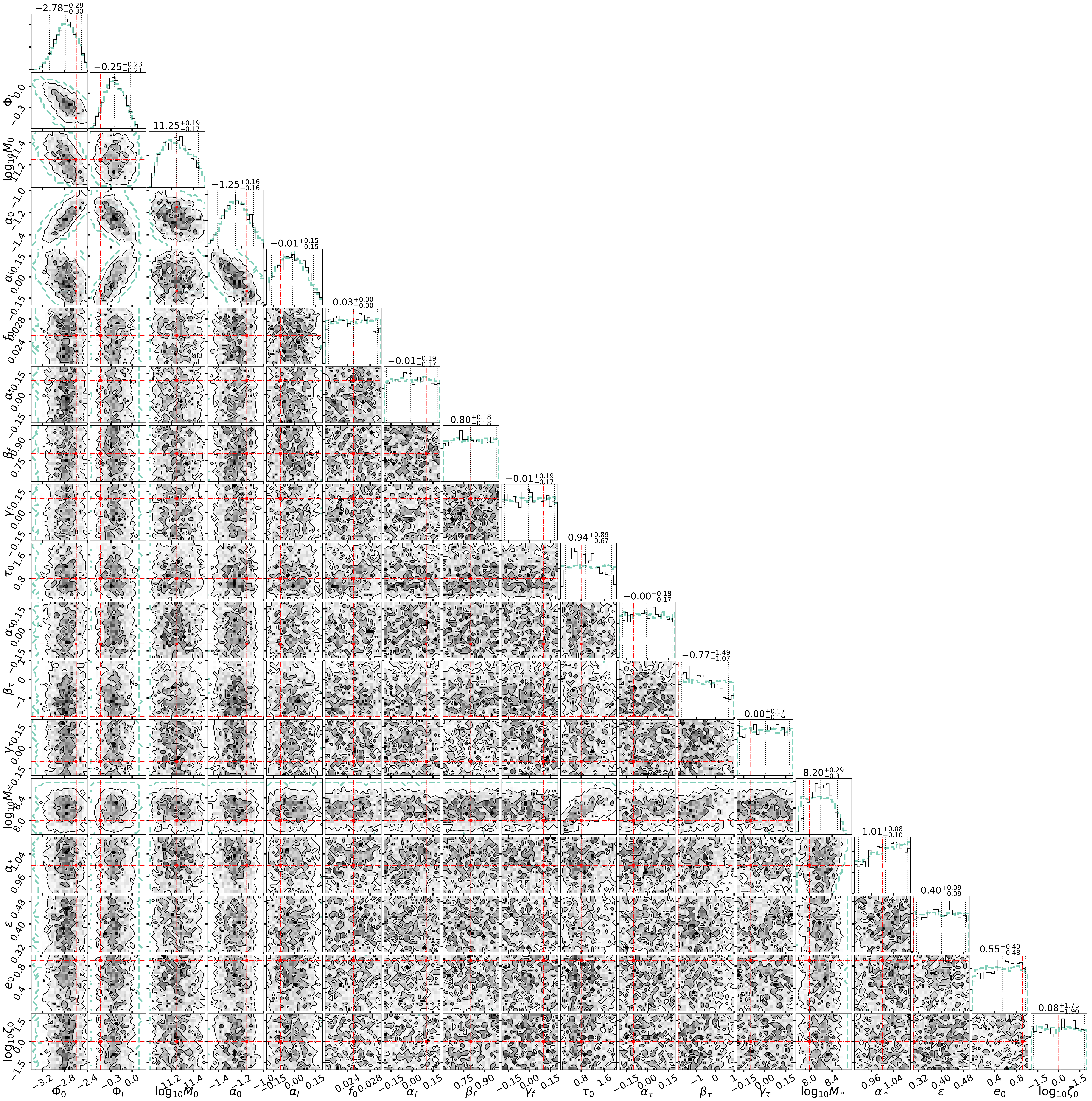}
\caption{Full corner plot of the 18 parameters for the IPTA30 Eccentric case}
\label{fig:cornerplot_uplim15_full}
\end{figure*}

\begin{figure*}
\centering
{\large \bf Eccentric case SKA20}
\includegraphics[width=\textwidth]{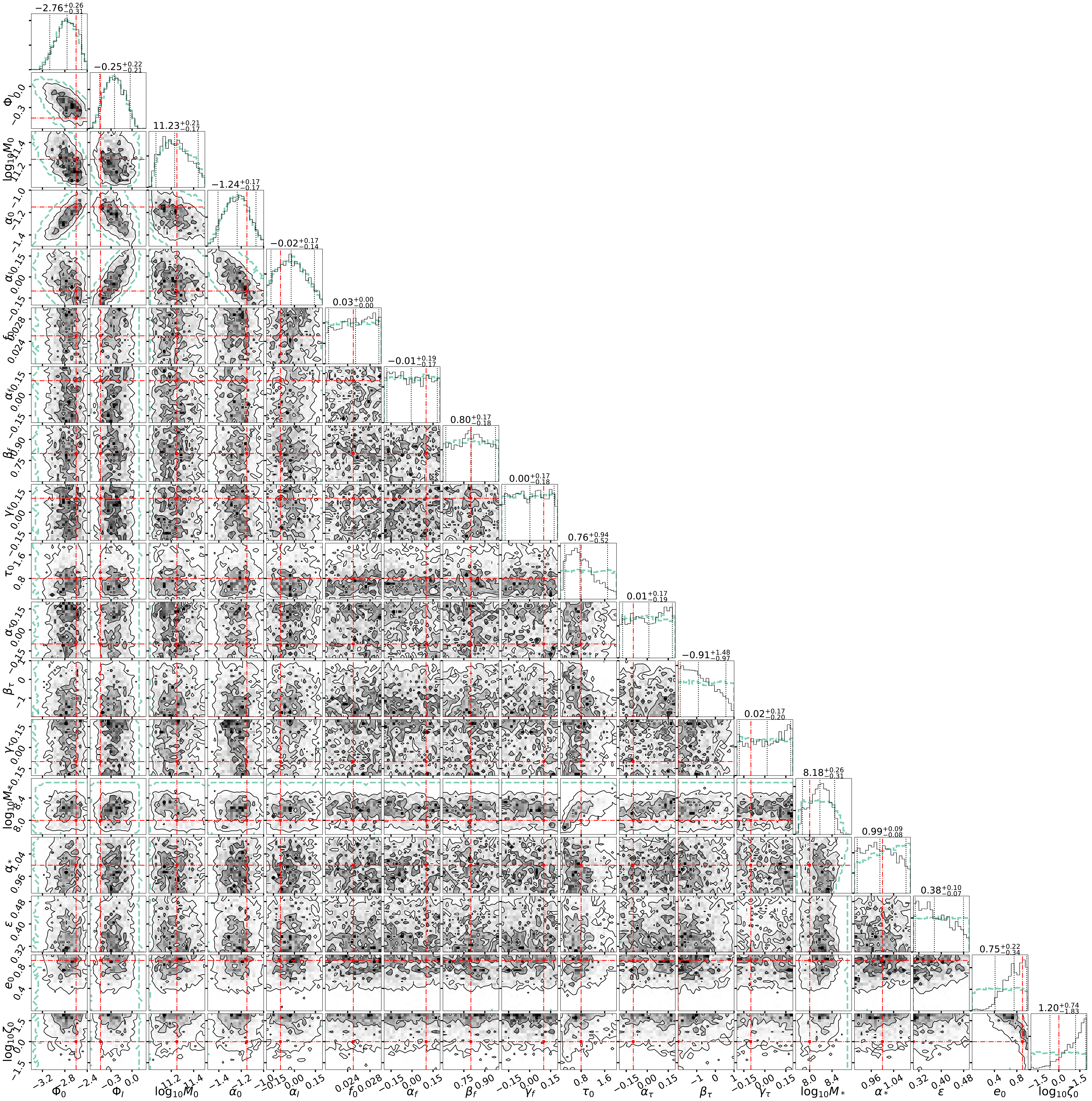}
\caption{Full corner plot of the 18 parameters for the SKA20 Eccentric case}
\label{fig:cornerplot_uplim15_full}
\end{figure*}

\begin{figure*}
\centering
{\large \bf Circular case Ideal}
\includegraphics[width=\textwidth]{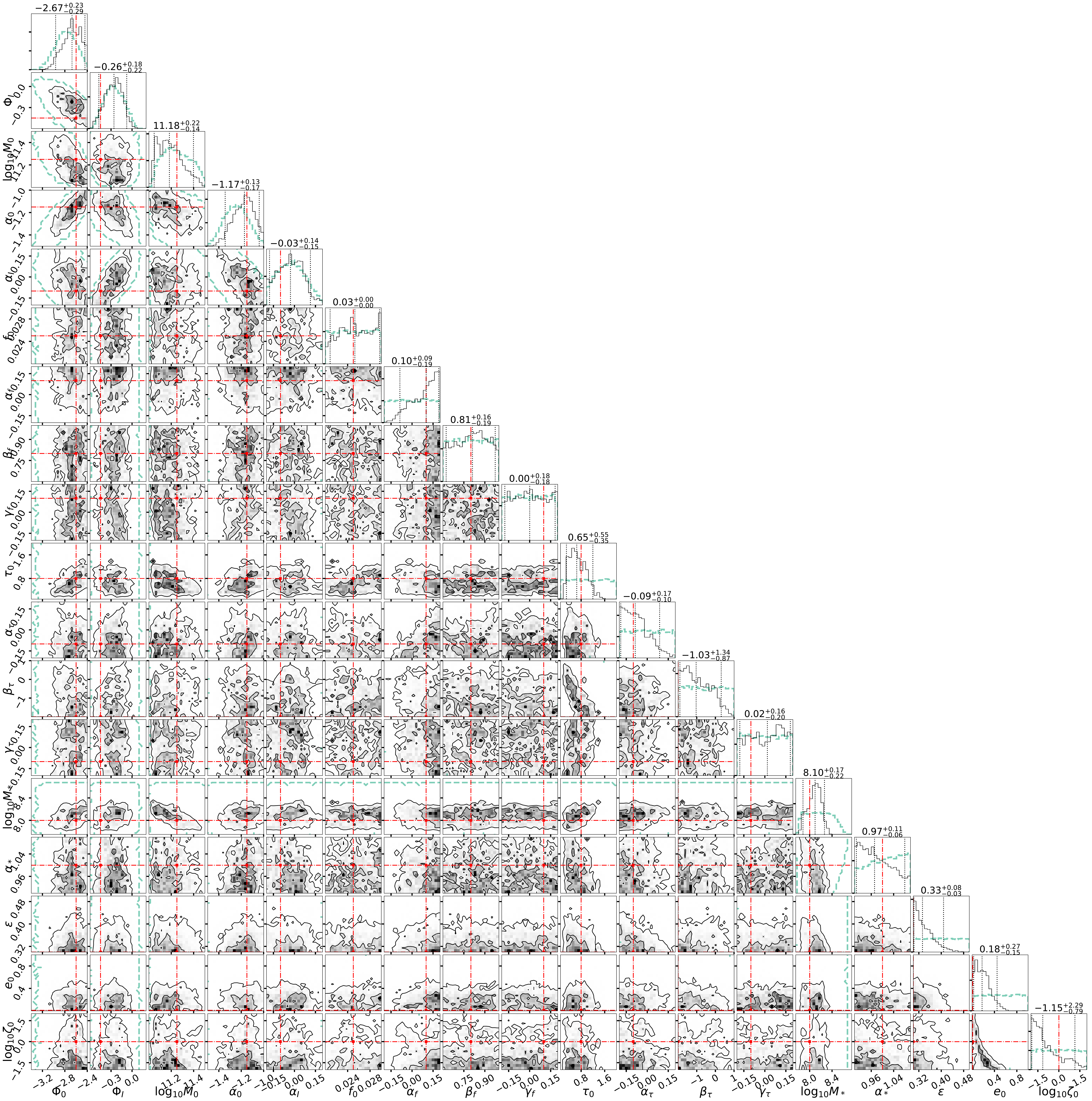}
\caption{Full corner plot of the 18 parameters for the Ideal Circular case}
\label{fig:cornerplot_uplim15_full}
\end{figure*}

\begin{figure*}
\centering
{\large \bf Eccentric case Ideal}
\includegraphics[width=\textwidth]{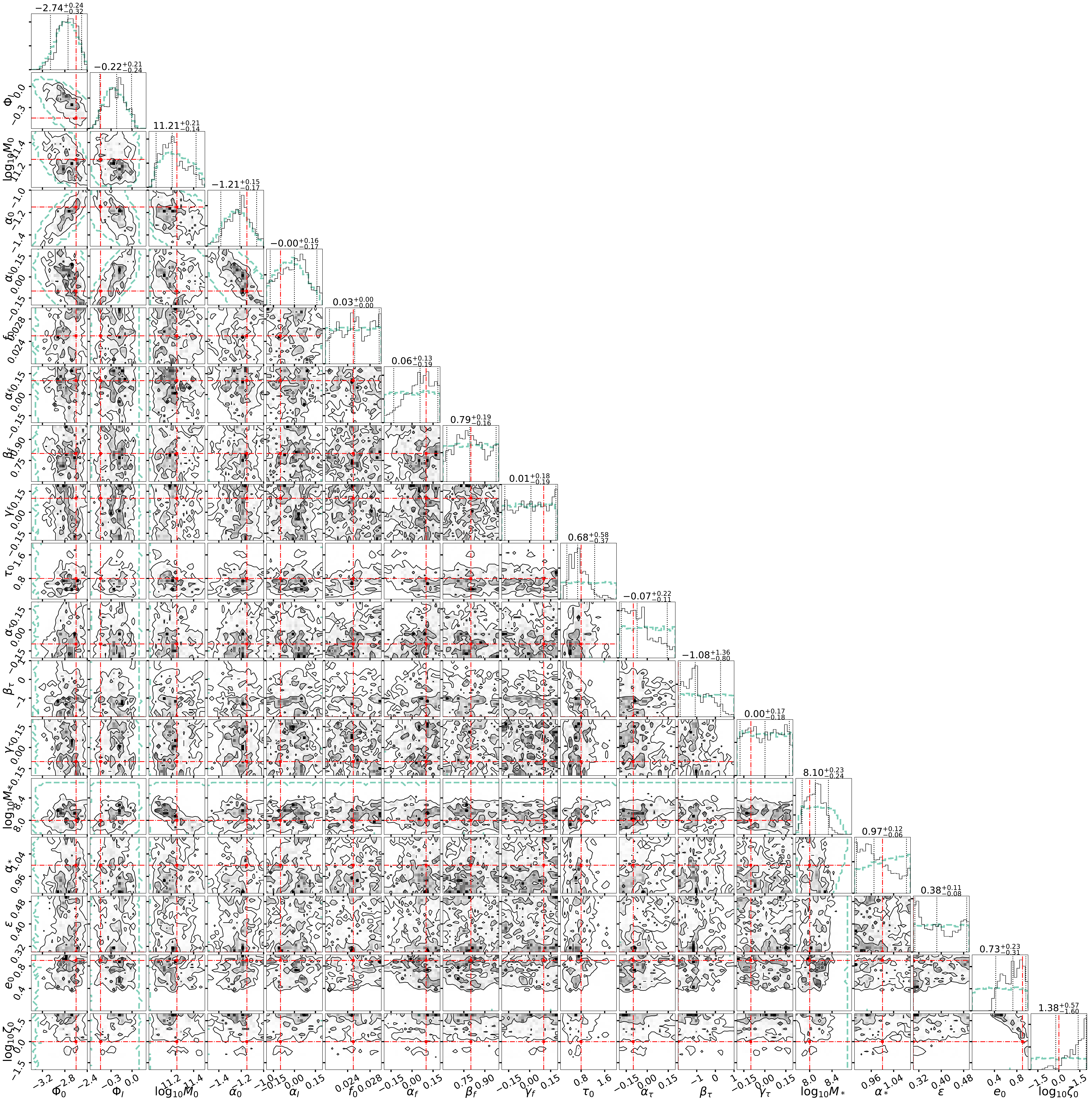}
\caption{Full corner plot of the 18 parameters for the Ideal Eccentric case}
\label{fig:cornerplot_uplim15_full}
\end{figure*}

\end{document}